\documentclass[aps,prb,floatfix,twocolumn,showpacs]{revtex4}%
\usepackage{amsmath}
\usepackage{graphicx}
\usepackage{amsfonts}
\usepackage{color}
\usepackage{amssymb}%
\providecommand{\U}[1]{\protect\rule{.1in}{.1in}}
\providecommand{\U}[1]{\protect\rule{.1in}{.1in}}
\providecommand{\U}[1]{\protect\rule{.1in}{.1in}}
\providecommand{\U}[1]{\protect\rule{.1in}{.1in}}

\begin{document}
\title{Theory of interactions between cavity photons induced by a mesoscopic circuit}
\author{Audrey Cottet$^{1}$, Zaki Leghtas$^{2,1}$ and Takis Kontos$^{1}$}
\affiliation{$^{1}$\textit{Laboratoire de Physique de l'Ecole normale sup\'{e}rieure, ENS,
Universit\'{e} PSL, CNRS, Sorbonne Universit\'{e}, Universit\'{e} de Paris,
F-75005 Paris, France}}
\affiliation{$^{2}$\textit{Centre Automatique et Syst\`{e}mes, Mines-ParisTech, PSL
Research University, 60, bd Saint-Michel, 75006 Paris, France}}

\date{\today}

\begin{abstract}
We use a quantum path integral approach to describe the behavior of a
microwave cavity coupled to a dissipative mesoscopic circuit.\ We integrate
out the mesoscopic electronic degrees of freedom to obtain a cavity effective
action at fourth order in the light/matter coupling. By studying the structure
of this action, we establish sufficient conditions in which the
cavity dynamics can be described with a Lindblad equation. This equation
depends on effective parameters set by electronic correlation
functions. It reveals that the mesoscopic circuit induces an
effective Kerr interaction and two-photon dissipative processes. We use our
method to study the effective dynamics of a cavity coupled to a double quantum
dot with normal metal reservoirs. If the cavity is driven at twice its
frequency, the double dot circuit generates photonic squeezing and
non-classicalities visible in the cavity Wigner function. In particular, we
find a counterintuitive situation where mesoscopic dissipation enables the
production of photonic Schr\"{o}dinger cats. These effects can occur for
realistic circuit parameters. Our method can be generalized straightforwardly
to more complex circuit geometries with, for instance, multiple quantum dots,
and other types of fermionic reservoirs such as superconductors and ferromagnets.

\end{abstract}
\maketitle

\section{Introduction}

Embedding nonlinear Josephson circuits into microwave cavities has enabled
impressive progress in the quantum control of microwave
light\cite{Wallraff:2004}. Indeed, the field of circuit Quantum
Electrodynamics (QED) offers many functionalities. For instance, squeezed
photonic states, where the uncertainty of one quadrature is reduced below the
zero-point level, can be obtained by embedding a nonlinear circuit such as a
Superconducting Quantum Interference Device (SQUID) array into a microwave
cavity\cite{Mallet:2011}. A classical cavity state can evolve into a quantum
superposition of coherent states due to an effective Kerr interaction provided
by a superconducting quantum bit\cite{Kirchmair:2013}. One can also generate
arbitrary quantum superpositions of Fock states by using the time-dependent
coupling of a superconducting qubit to a microwave
resonator\cite{Hofheinz:2008,Hofheinz:2009}. For most quantum protocols
implemented so far, cavity damping is a spurious effect. However, it has been
demonstrated experimentally that in a nonlinear circuit QED setup driven with
microwaves, photon-number dependent losses can be used to prepare photonic
Schr\"{o}dinger cat states\cite{Leghtas:2015,Touzard:2018} and stabilize
autonomously Fock states \cite{Holland:2015}. This result contributes to a
research field called \textquotedblleft reservoir
engineering\textquotedblright, which promotes the idea that, contrarily to the
common belief, dissipation is not always harmful for the quantumness of a
system\cite{Poyatos:1996,Sarlette:2012,Aron:2014,Roy:2015}. Thanks to this
rich phenomenology, nonlinear microwave cavities offer many possibilities of
applications, from sensing to quantum information and communication. For
example, squeezed states of light offer a powerful resource for
quantum-enhanced sensing\cite{Caves:1980,Giovannetti:2004}. More recently,
quantum computing schemes have been suggested, where quantum information would
be encoded in a manifold of cavity \begin{figure}[ptb]
\includegraphics[width=1.\linewidth]{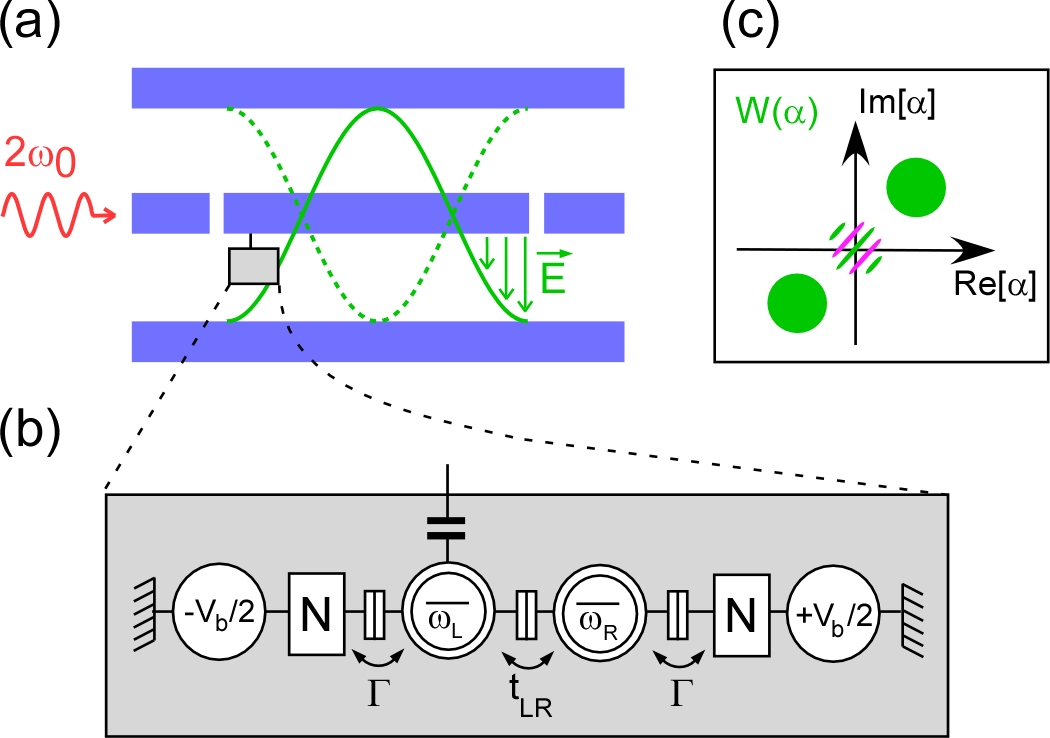}\label{Fig0}\caption{ Example of
Mesoscopic QED\ device. Panel (a): Microwave cavity ac driven at twice the
cavity frequency $\omega_{0}$. The nanocircuit (in grey) is coupled
capacitively to the cavity central conductor at an electric field node. Panel
(b): Double quantum dot coupled to normal metal reservoirs N with a tunnel
rate $\Gamma$. The dots are tunnel coupled with a hopping strength $t_{LR}$.
The normal metal reservoirs have a voltage bias $V_{b}$. Panel (c): Schematic
representation of the cavity Wigner function as a function of the field
quadratures, measured by performing the cavity tomography. }%
\end{figure}states stabilized autonomously by two-photon
dissipation\cite{Mirrahimi: 2013}. In this context, the photonic Wigner
function is a widely measured quantity to characterize the joint statistics of
the cavity field quadratures\cite{Haroche: 2006}. It is obtained
experimentally by performing the cavity tomography\cite{Hofheinz:2009}.

In standard Circuit QED experiments, the Josephson circuits coupled to
microwave cavities are exclusively made of superconducting metals and
Josephson junctions. However, due to the versatility of microwave fabrication
techniques, the connection between Circuit QED and mesoscopic physics is
naturally growing\cite{Reulet:1995,Cottet:2017}. Recently, circuits enclosing
a single\cite{Delbecq:2011} or a double\cite{Frey:2012} quantum dot and
normal\cite{Delbecq:2011,Frey:2012},
ferromagnetic\cite{Viennot:2015,Cubaynes:2019} or superconducting
reservoirs\cite{Bruhat:2016,Bruhat:2016b} have been coupled to microwave
cavities. In the experiments performed so far, microwave cavities have
appeared as a powerful means to characterize the electronic spectrum and
dynamics of mesoscopic circuits. However, the scope of Mesoscopic QED could go
far beyond. Indeed, mesoscopic circuits are intrinsically nonlinear due to
their anharmonic energy spectrum. Besides, fermionic reservoirs represent a
specific source of dissipation which involves electrically controlled quantum
transport. It is therefore appealing to investigate the potentialities of
Mesoscopic QED for producing quantum cavity states. In this direction,
entangled light/matter states due to a strong
charge/photon\cite{Bruhat:2016b,Mi:2017,Stockklauser:2017} or
spin/photon\cite{Mi:20118,Samkharadze:2018,Landig:2018,Cubaynes:2019} coupling
have been obtained in recent experiments, using double quantum dots circuits.
However, many more situations remain to be explored.

On the theory side, the effect of dissipative fermionic reservoirs in
Mesoscopic QED setups has been mostly investigated in the semiclassical regime
where the number of cavity photons is so large that quantum fluctuations in
the photon number can be
disregarded\cite{Schiro:2014,Dmytruk:2015,Dartiailh:2017,Bruhat:2016,Dmytruk:2016}%
. Otherwise, a sequential tunneling description of quantum transport has been
used, which is valid only for very small tunnel rates
\cite{PeiQing:2011,Kulkarni:2014,Stockklauser:2015,Viennot:2014,Jin:2012,Jin:2013}%
. A general quantum description of Mesoscopic QED is lacking. One needs to
develop a theory which describes the cavity quantum dynamics in the presence
of dissipative mesoscopic transport. This description must apply to complex
circuit configurations with arbitrary tunnel couplings to voltage-biased
fermionic reservoirs. It is also important to take into account the nonlinear
photonic effects inherited from the light/matter interactions, which have been
eluded so far in the theory of Mesoscopic QED, and offer a vast field of
investigation. This requires to work beyond the second order treatment of the
light/matter coupling.

In this work, we fill these gaps by employing a quantum path integral
technique along the Keldysh contour, which is particularly convenient to
integrate out electronic degrees of freedom and obtain an effective
description of the cavity nonlinear behavior\cite{Kamenev}. We consider a
cavity with frequency $\omega_{0}$ coupled to a mesoscopic circuit and excited
with a microwave tone at frequency $2\omega_{0}$ with a moderate amplitude
$\varepsilon_{p}$ (i.e. $\varepsilon_{p}$ can be treated to first order). We
note $g$ the order of magnitude of the light/matter coupling in the mesoscopic
QED device. We expand the effective quantum action of the cavity up
to fourth order in the light/matter coupling. The expansion parameter is
described in appendix H, and for conciseness, is hereafter referred to as
$g$. The cavity effective action depends on electronic
correlation functions of the mesoscopic circuit, which we express in terms of
Keldysh Green's functions. It reveals that the cavity is subject to
photon-photon interactions mediated by the mesoscopic circuit. We establish
 sufficient conditions on mesoscopic correlators\textbf{ }for
having a description of the cavity dynamics with a Lindblad
equation. In this case, the $2\omega_{0}$ drive produces, at
third order in $g$, a coherent two-photon drive\cite{Millburn} and a
 less usual dissipative squeezing process\cite{Didier:2014,Lu:2015}%
. Additionally, the mesoscopic circuit induces, at fourth order
in $g$, Kerr photon-photon interaction as well as stochastic two-photon losses
and gains. Importantly, our results are valid for tunnel couplings rates to
the reservoirs of the mesoscopic circuit smaller as well as larger than the
electronic temperature since no sequential tunneling hypothesis is required.
We make the realistic assumption that the cavity has a large quality factor
and a dressed linewidth much smaller than the mesoscopic resonances linewidth.
We finally disregard Coulomb interactions in the mesoscopic circuit.

We use our method to study the quantum dynamics of a microwave cavity coupled
to a non-interacting double quantum dot (DQD) with normal metal contacts
biased with a voltage $V_{b}$. We identify two situations where the effective
dynamics of the cavity is\ described \ by a Lindblad
equation, which includes non-linear light/matter interaction
effects. The first situation is the limit of a low light/matter coupling
($g\sim0.01\omega_{0}$). In this case, we derive an effective Lindblad
equation description of the cavity behavior to third order in $g$, from which
we obtain an analytic expression of the cavity Wigner function in stationary
conditions. The $2\omega_{0}$ drive produces a coherent injection/withdrawal
of photon pairs in the cavity\cite{Millburn} and an less usual squeezing
dissipative process\cite{Didier:2014,Lu:2015}. This
leads to a squeezing of the cavity vacuum, which depends non trivially on the
system parameters\cite{Mendes:2015,Mendes:2016,Grimsmo:2016}%
. The second  Lindbladian situation is
when the double dot is resonant with $2\omega_{0}$ and has moderate interdot
hopping and tunnel couplings to its reservoirs, and the light/matter coupling
is moderate ($g\sim0.1\omega_{0}$). In this case, a description to fourth
order in $g$ is necessary to describe the cavity dynamics. In this limit, we
find that, in the absence of a cavity drive ($\varepsilon_{p}=0$), dissipative
transport in the double dot circuit can enable the stochastic absorption
and/or emission of photon pairs in the cavity, depending on the value of
$V_{b}$. When the cavity is ac driven ($\varepsilon_{p}\neq0$) with $V_{b}=0$,
we show, with numerical simulations of the photonic Lindblad equation, that
the DQD circuit can be used to produce photonic Schr\"{o}dinger cat states.
This effect is expected for realistic circuit parameters. It is due to a
combination of the two-photon drive in $\varepsilon_{p}g^{3}/\omega_{0}^{3}$
and the photon pair damping in $g^{4}/\omega_{0}^{3}$. Hence,
counterintuitively, mesoscopic dissipation enables the generation of a quantum
superposition of cavity states.\textbf{ }In the same vein, recent
experiments with Josephson circuits have shown that the combination of a
two-photon drive with a Kerr photon-photon
interaction\cite{Puri:2017,Grimm:2019,Puri:2019} or two-photon
losses\cite{Leghtas:2015,Lescanne:2019} can be used to prepare autonomously
Schr\"{o}dinger cat states and protect these cats against some types of
decoherence. This represents an important research direction in the context of
the development of a bosonic encoding of quantum information with autonomous
quantum error correction. Our work suggests that mesoscopic QED devices could
offer interesting possibilities in this context.

Thanks to its generality, our approach could be used to explore many more
circuit geometries and protocols. One can consider circuits with
single\cite{Delbecq:2011} or multiple quantum dots\cite{Frey:2012}. One can
also consider extended nanoconductors such as nanowires with a strong-orbit
coupling, which raise a lot of attention in the search for Majorana
quasiparticles\cite{Mourik:2012,Das:2012,Deng:2012,Churchill:2013,Albrecht:2018,Zhang:2018}%
, and which have been recently coupled to microwave cavities\cite{Tosi:2019}.
For this purpose, the nanoconductor can be discretized into various internal
sites by using a Hubbard model\cite{Trif:2012,
Schmidt:2013a,Cottet:2013,Dmytruk:2015,Dartiailh:2017}. Finally, different
types of fermionic reservoirs can be considered, such as normal
metals\cite{Delbecq:2011,Frey:2012},
ferromagnets\cite{Viennot:2015,Cubaynes:2019} or
superconductors\cite{Bruhat:2016,Bruhat:2016b}. These Mesoscopic QED\ devices
could find applications in quantum information science, with for instance spin
quantum bits\cite{Mi:20118,Samkharadze:2018,Landig:2018,Cubaynes:2019} or
Cooper pair splitters\cite{Cottet:2012,Cottet:2014,Mantovani:2019}, in quantum
optics, with for instance lasing generated by mesoscopic
circuits\cite{Jin:2011,Liu:2015,Rastelli:2019}, but also in condensed matter
science, with the simulation of the Kondo effect in quantum
dots\cite{Desjardins:2017}, or the simulation of Anderson-Holstein
problem\cite{Hewson:2001}. Our approach could be instrumental for the study of
these many configurations in the nonlinear quantum regime.

This article is organized as follows. Section II introduces the Mesoscopic QED
Hamiltonian and discusses a direct density matrix description of Mesoscopic
QED and its drawbacks. Section III presents the general description of
Mesoscopic QED with the path integral approach. It also explains how the
cavity effective action leads to a Lindblad description, at third order in $g$
for any parameters, or at fourth order in $g$ provided some mesoscopic
correlation functions fulfill a Lindbladian condition.
Section IV applies the results of section III to the example of a microwave
cavity coupled to a double quantum dot with normal metal contacts. In
particular, it shows how the double dot can be used to squeeze the cavity
vacuum or to produce photonic Schr\"{o}dinger cats. Section V puts
ours results in perspective with other recent works and section VI
concludes. Appendix A gives details on the derivation of the
cavity effective action at fourth order in $g$. Appendix B1 gives a direct
calculation of the possible semiclassical values of the cavity photonic
amplitude at fourth order in $g$ (without using the path integral approach).
This enables a semiclassical interpretation of some of the parameters which
occur in the cavity effective action. Appendix B2 shows an alternative way to
determine the possible semiclassical values of the cavity photonic amplitude,
by considering the saddle points of the cavity action. The agreement between
the results of Appendix B1 and Appendix B2 at fourth order in $g$ provides an
important sanity check for our approach. Appendix C explains how to derive the
action associated to a Lindblad equation. Appendix D establishes a
quantitative equivalence at order 2\ in $g$ between the Lindblad equation
arising from a direct density matrix approach and the Lindblad equation
arising from the path integral approach. Appendix E gives details on the
calculation of the cavity Wigner function. Appendix F gives details on the
dependence of the photonic squeezing effect on the double dot parameters.
Appendix G gives a simple analytical expression of the linear
charge susceptibility of a mesoscopic circuit (i.e. to second order in $g$) in
the sequential tunneling limit, to illustrate the regularization of our
description by dissipative tunneling. Finally, Appendix H shows the
calculation of the generalized charge susceptibilities of the mesoscopic
circuit up to $8^{th}$ order in $g$. This serves as a basis for discussing the
regime of validity of our approach at $4^{th}$ order in $g$.  One needs
sufficiently large tunneling rates to the fermionic reservoirs of the circuit
on top of a small enough coupling $g$ and cavity drive $\varepsilon_{p}$. It
is difficult to give a simple analytic criterion for delimiting this regime.
However, the evaluation of higher order charge susceptibilities given in
Appendix H represents a suitable numerical check for the validity of our
development. 

\section{Description of Mesoscopic QED with a direct density matrix approach}

\subsection{System Hamiltonian}

We consider a cavity with bare frequency $\omega_{0}$ excited by a microwave
drive $\varepsilon_{ac}(t)$, and coupled to a mesoscopic circuit. This circuit
contains $N$ discrete orbitals with index $d$, coupled to fermionic reservoirs
with a continuum of states with index $k$. The mesoscopic circuit can be for
instance a quantum dot circuit, in which case the orbitals $d$ are located in
the dots\cite{Delbecq:2011,Frey:2012,Cottet:2017}. Each orbital $d$ is coupled
to the electric quadrature of the cavity field with a constant $g_{d}$ (see
Ref.\cite{Cottet:2015} for a first-principles description of this effect and a
microscopic expression of $g_{d}$). The resulting Mesoscopic QED device can be
described with the Hamiltonian%
\begin{align}
\hat{H}_{tot}  &  =\omega_{0}\hat{a}^{\dag}\hat{a}+\varepsilon_{ac}(t)\left(
\hat{a}^{\dag}+\hat{a}\right)  +\hat{h}_{b}\label{Htot}\\
&  +\hat{H}_{meso}+%
{\textstyle\sum\limits_{d}}
g_{d}(\hat{a}^{\dag}+\hat{a})\hat{c}_{d}^{\dag}\hat{c}_{d}\nonumber
\end{align}
with%
\begin{align}
\hat{H}_{meso}  &  =%
{\textstyle\sum\limits_{d}}
\omega_{d}\hat{c}_{d}^{\dag}\hat{c}_{d}+%
{\textstyle\sum\limits_{d<d^{\prime}}}
\left(  t_{d^{\prime},d}\hat{c}_{d^{\prime}}^{\dag}\hat{c}_{d}+H.c.\right)
\nonumber\\
&  +%
{\textstyle\sum\limits_{k,d}}
\left(  t_{k,d}\hat{c}_{k}^{\dag}\hat{c}_{d}+H.c.\right)  +%
{\textstyle\sum\limits_{k}}
\omega_{k}\hat{c}_{k}^{\dag}\hat{c}_{k}\text{.} \label{HHmeso}%
\end{align}
Above, $\hat{a}^{\dag}$ is the cavity photon creation operator, $\hat{c}%
_{d}^{\dag}$ the electron creation operator in the discrete orbital
$d\in\lbrack1,N]$ and $\hat{c}_{k}^{\dag}$ an electron creation operator in a
level $k$ of one of the fermionic reservoirs. In the general case, the indices
$k$ and $d$ include the spin degree of freedom. We do not specify the exact
Mesoscopic circuit geometry for the moment. The tunnel hopping strength
between two orbitals $d$ and $d^{\prime}[k]$ located in neighboring sites of
the circuit is noted $t_{d^{\prime}[k],d}$. We use $\hbar=1$. Intrinsic cavity
damping is described by the Hamiltonian $\hat{h}_{b}$ which we do not specify
here. In most cases, the orbital energy $\omega_{d}$ of site $d$ can be finely
tuned with an electrostatic gate, and bias voltages can be applied to the
fermionic reservoirs to induce electronic transport. Note that we
disregard the coupling between the cavity field and the reservoirs levels $k$.
This is relevant for most Mesoscopic QED experiments where the coupling
between discrete internal levels $d$ and the cavity field is dominant due to
the use of ac gates which connect levels $d$ to the cavity central conductor.
In the following, we assume that an ac drive%
\begin{equation}
\varepsilon_{ac}(t)=(\varepsilon_{p}e^{-i2\omega_{0}t}+\varepsilon_{p}^{\ast
}e^{i2\omega_{0}t})/2 \label{eac}%
\end{equation}
is applied to the cavity. We will see that both components in $e^{-i2\omega
_{0}t}$ and $e^{i2\omega_{0}t}$ contribute to the the cavity response through
higher order processes (effect in $g^{3}$ at least). For simplicity, we do not
describe explicitly the microwave inputs and outputs of the cavity but this
can be added straightforwardly by using the input/output
theory\cite{Millburn,Schiro:2014,Benito:2017}.

\subsection{Direct density matrix approach and its drawbacks\label{direct}}

The most commonly used description of Circuit QED is the density matrix
approach which consists in expressing directly the time evolution of the
system density matrix. Here we will shortly discuss this approach to point out
its weaknesses and the interest of the path integral approach in the context
of nonlinear Mesoscopic QED.

We assume that the interaction term $\hat{V}$ is a perturbation in the system
Hamiltonian, in comparison with the cavity contribution in $\omega_{0}$ and
mesoscopic contribution $\hat{H}_{meso}$. For simplicity, in this section, we
also assume that there is no cavity drive ($\varepsilon_{p}=0$) and no cavity
intrinsic dissipation (i.e. $\hat{h}_{b}$ is negligible). In these conditions,
it is convenient to use the interaction picture, where the density matrix
$\rho^{I}(t)=e^{i\omega_{0}\hat{a}^{\dag}\hat{a}t+i\hat{H}_{meso}t}%
\rho(t)e^{-i\omega_{0}\hat{a}^{\dag}\hat{a}t-i\hat{H}_{meso}t}$ of the full
mesoscopic QED device (cavity+mesoscopic circuit) has an evolution equation
\begin{equation}
\frac{\partial\rho^{I}(t)}{\partial t}=-i[\hat{V}(t),\rho^{I}(t)] \label{rt}%
\end{equation}
with%
\begin{equation}
\hat{V}(t)=\hat{N}(t)\left(  \hat{a}e^{-i\omega_{0}t}+\hat{a}^{\dag}%
e^{i\omega_{0}t}\right)  \text{,}%
\end{equation}%
\begin{equation}
\hat{N}(t)=%
{\textstyle\sum\limits_{d}}
g_{d}\hat{n}_{d}(t) \label{NNN2}%
\end{equation}
and%
\begin{equation}
\hat{n}_{d}(t)=e^{i\hat{H}_{meso}t}\hat{c}_{d}^{\dag}\hat{c}_{d}e^{-i\hat
{H}_{meso}t}\text{.} \label{ndI}%
\end{equation}
Note that $\hat{H}_{meso}$ and $\hat{c}_{d}^{\dag}\hat{c}_{d}$ do not commute
due to dot-dot and dot-reservoir tunneling. Hence, from Eqs. (\ref{NNN2}) and
(\ref{ndI}), $\hat{N}(t)$ depends on time.

We now discuss the expression of the cavity dynamics at second order in $g$.
The integration of Eq.(\ref{rt}) gives%
\begin{equation}
\rho^{I}(t)=\rho^{I}(t_{0})-i%
{\textstyle\int\limits_{t_{0}}^{t}}
dt_{1}[V(t_{1}),\rho^{I}(t_{1})] \label{hihi}%
\end{equation}
with $t_{0}$ a reference time far in the past. Inserting this equation back in
Eq.(\ref{rt}) gives%
\begin{equation}
\frac{\partial\rho^{I}(t)}{\partial t}=-i[\hat{V}(t),\rho^{I}(t_{0})]-%
{\textstyle\int\limits_{t_{0}}^{t}}
dt_{1}[\hat{V}(t),[V(t_{1}),\rho^{I}(t_{1})]]\text{.} \label{hihihi}%
\end{equation}
In the limit where the mesoscopic system has a correlation time $\tau$ which
is much shorter than the cavity characteristic timescale of evolution $T$,
only the times $t_{1}$ such that $t-t_{1}\lesssim\tau$ will contribute in the
above integral\cite{Blum:2012}. Accordingly, one can assume that the
mesoscopic system is constantly at equilibrium, i.e.
\begin{equation}
\rho^{I}(t_{1})=\rho_{meso}^{0}\otimes\rho_{cav}^{I}(t_{1}) \label{rhoTot}%
\end{equation}
with $\rho_{meso}^{0}$ the equilibrium density matrix of the mesoscopic
circuit for $g_{d}=0$. Finally, since $\tau\ll T$, one can use $\rho^{I}%
(t_{1})=\rho_{meso}^{0}\otimes\rho_{cav}^{I}(t)$ in the above integral,
Performing the trace $\underset{k,d}{\mathrm{Tr}}$ on the mesoscopic degrees
of freedom, one finally gets%
\begin{align}
&  \frac{\partial\rho_{cav}^{I}(t)}{\partial t}\nonumber\\
&  =-i\underset{k,d}{\mathrm{Tr}}\left[  [V(t),\rho_{meso}^{0}\otimes
\rho_{cav}^{I}(t_{0})]\right] \nonumber\\
&  -%
{\textstyle\int\limits_{t_{0}}^{t}}
dt_{1}~\underset{k,d}{\mathrm{Tr}}\left[  [V(t),[V(t_{1}),\rho_{meso}%
^{0}\otimes\rho_{cav}^{I}(t)]]\right]  \text{.} \label{eqroeff}%
\end{align}
A reorganization of Eq.(\ref{eqroeff}) gives, keeping only resonant terms and
considering a stationary situation,%
\begin{align}
\frac{\partial\rho_{cav}^{I}(t)}{\partial t}  &  =-2\operatorname{Im}[\chi
_{B}(\omega_{0})]\mathcal{D}_{\hat{a}}(\rho_{cav}^{I}(t))\nonumber\\
&  -2\operatorname{Im}[\chi_{A}(\omega_{0})]\mathcal{D}_{\hat{a}^{\dag}}%
(\rho_{cav}^{I}(t))\nonumber\\
&  -i\operatorname{Re}[\chi_{B}(\omega_{0})-\chi_{A}(\omega_{0})]~[\hat
{a}^{\dag}\hat{a},\rho_{cav}^{I}(t)]+o(\check{g}^{2})\text{.} \label{rohPed}%
\end{align}
Above,%
\begin{equation}
\mathcal{D}_{\hat{L}_{j}}(\rho_{cav}^{I})=\hat{L}_{j}\rho_{cav}^{I}\hat{L}%
_{j}^{\dag}-\frac{1}{2}\{\hat{L}_{j}^{\dag}\hat{L}_{j},\rho_{cav}^{I}\}
\label{Lj}%
\end{equation}
is the Lindblad superoperator associated to the jump operator $\hat{L}_{j}$.
We have disregarded the first order term in\textbf{ }$g$ which is non-resonant
with the cavity. The mesoscopic correlators
\begin{equation}
\chi_{A}(t)=-i\theta(t)\left\langle \hat{N}(0)\hat{N}(t)\right\rangle
\label{xiA}%
\end{equation}
and%
\begin{equation}
\chi_{B}(t)=-i\theta(t)\left\langle \hat{N}(t)\hat{N}(0)\right\rangle
\label{xiB}%
\end{equation}
whose Fourier transforms $\chi_{A[B]}(\omega)=%
{\textstyle\int}
dt~\chi_{A[B]}(t)e^{i\omega t}$ appear in Eq.(\ref{rohPed}), have to be
evaluated to second order in the light/matter interaction. More precisely,
from Eq.(\ref{NNN2}), one can use $\left\langle \hat{N}(t^{\prime})\hat
{N}(t)\right\rangle =\sum\nolimits_{d,d^{\prime}}g_{d}g_{d^{\prime}%
}A_{d^{\prime},d}(t^{\prime},t)$ and $A_{d^{\prime},d}(t^{\prime
},t)=\left\langle \hat{c}_{d^{\prime}}^{\dag}(t^{\prime})\hat{c}_{d^{\prime}%
}(t^{\prime})\hat{c}_{d}^{\dag}(t)\hat{c}_{d}(t)\right\rangle _{0}$ where
$\left\langle {}\right\rangle _{0}$ denotes a statistical average calculated
for $g_{d}=0$ for any $d$, i.e. $A_{d^{\prime},d}(t^{\prime},t)=\mathrm{Tr}%
\left[  \rho_{meso}^{0}\hat{c}_{d^{\prime}}^{\dag}(t^{\prime})\hat
{c}_{d^{\prime}}(t^{\prime})\hat{c}_{d}^{\dag}(t)\hat{c}_{d}(t)\right]  $ . In
the absence of Coulomb interactions, the evaluation of $A_{d,d^{\prime}}$ can
be done straightforwardly by using the Wick theorem (see for instance
Ref.\cite{Zamoum:2016}).

To describe the dynamics of $\rho_{cav}^{I}$ beyond the second order in $g$,
one straightforward idea is to start with Eq.(\ref{hihihi}) and iterate the
substitution of $\rho^{I}(t)$ by the right member of Eq.(\ref{hihi}). This gives:%

\begin{align}
&  \frac{\partial\rho_{cav}^{I}(t)}{\partial t}\label{hho}\\
&  =-i\underset{k,d}{\mathrm{Tr}}\left[  [V(t),\rho^{I}(t_{0})]\right]  -%
{\textstyle\int\limits_{t_{0}}^{t}}
dt_{1}~\underset{k,d}{\mathrm{Tr}}\left[  [V(t),[V(t_{1}),\rho^{I}%
(t_{0})]]\right] \nonumber\\
&  +i%
{\textstyle\iint\limits_{t_{0},t_{0}}^{t,t_{1}}}
dt_{1}dt_{2}\underset{k,d}{~\mathrm{Tr}}\left[  [V(t),[V(t_{1}),[V(t_{2}%
),\rho^{I}(t_{0})]]]\right] \nonumber\\
&  +%
{\textstyle\iiint\limits_{t_{0},t_{0},t_{0}}^{t,t_{1},t_{2}}}
dt_{1}dt_{2}dt_{3}~\underset{k,d}{\mathrm{Tr}}[[V(t),[V(t_{1}),[V(t_{2}%
),[V(t_{3}),\rho^{I}(t_{3})]]]]\nonumber
\end{align}
At this stage, conceptual difficulties as well as calculation heaviness make
the generalization of Eq.(\ref{rohPed}) nontrivial. First, a back-action of
the cavity on the mesoscopic density matrix should be taken into account. This
means that expression (\ref{rhoTot}) cannot be used to express $\rho^{I}%
(t_{0})$ and $\rho^{I}(t_{3})$ in Eq.(\ref{hho}). Hence, it will be more
difficult to introduce independently defined mesoscopic correlators in the
expression of $\partial\rho_{cav}^{I}(t)/\partial t$. Besides, the dynamics of
the system is not anymore Markovian in the general case, so that $\rho
_{cav}^{I}(t)$ does not appear naturally in the right member of Eq.(\ref{hho}%
). Finally, even in a case where a generalization of the Markovian
Eq.(\ref{rohPed}) would be possible, due to the iterative structure of
Eq.(\ref{hho}), the number of mesoscopic correlators to define would explode,
and the explicit calculation of these correlators from the mesoscopic circuit
Hamiltonian would be a lengthy task. In fact, all these difficulties stem from
the fact that the trace on the mesoscopic degrees of freedom is performed
after the time evolution of $\rho^{I}(t)$ is expressed. It is thus crucial to
use a calculation method where the electronic degrees of freedom are
integrated earlier, i.e. at the level of the device Hamiltonian. This is why
we will develop an efficient quantum path integral description of Mesoscopic
QED in the next section.

\section{General description of Mesoscopic QED with the quantum path integral
formalism\label{GENERAL}}

\begin{figure}[ptb]
\includegraphics[width=1.\linewidth]{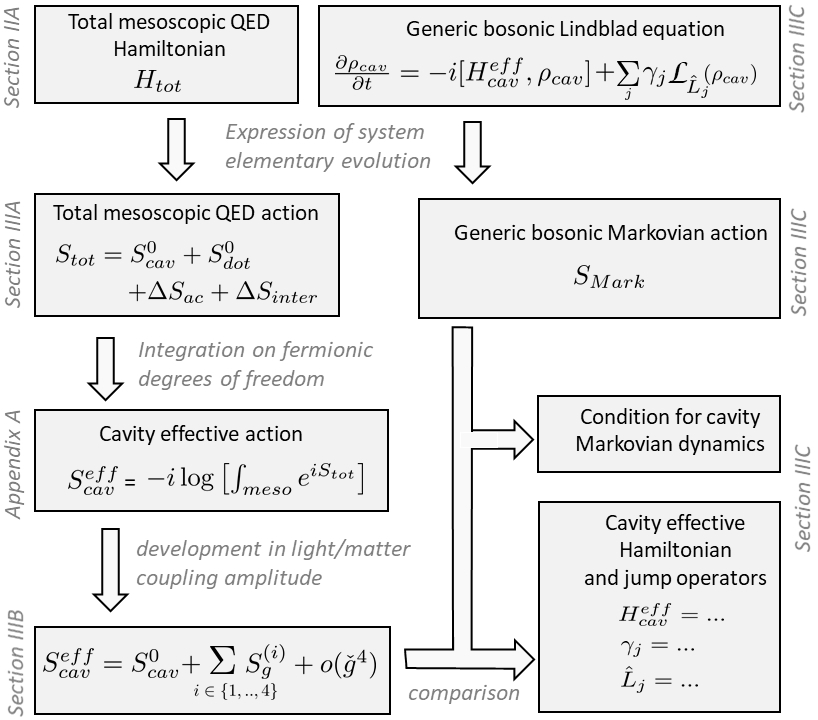}\caption{Synoptic table of
the theoretical approach introduced in section III}%
\label{FigTheo}%
\end{figure}

This section describes a general method based on the quantum path integral
description to describe the effective behavior of a microwave cavity coupled
to a mesoscopic circuit. From the Mesoscopic QED Hamiltonian of Eq.(\ref{Htot}%
), we express the global quantum action of the system (see section
\ref{GeneAct}). The fermionic degrees of freedom in this action can be
integrated out to obtain the cavity effective action (see section \ref{dev4}).
We compare this action to the action given by a generic Lindblad description
of a cavity dynamics (see section \ref{Lind}). This enables us to establish a
criterion to have a cavity Lindblad dynamics\textbf{
}at fourth order in the light/matter coupling. When this criterion is
fulfilled, we can finally write the cavity effective Lindblad equation. This
approach is summarized in the synoptic table of Figure \ref{FigTheo}%
.

\subsection{Quantum action of the whole Mesoscopic QED device\label{GeneAct}}

A generic description of Mesoscopic QED can be built by expressing the
Schwinger-Keldysh partition function of the system with a quantum path
integral along the Keldysh contour\cite{Kamenev}. To this end, we define,
along the forward and backward branches of the Keldysh contour, the fields
$\varphi_{\pm}(t)$, $\bar{\varphi}_{\pm}(t)$, $\psi_{\pm,d}(t)$ and $\bar
{\psi}_{\pm,d}(t)$, which correspond to a possible \textquotedblleft
realization\textquotedblright\ of the operators $\hat{a}$, $\hat{a}^{\dag}$,
$\hat{c}_{d}$ and $\hat{c}_{d}^{\dag}$ over time\cite{Coh}. It is convenient
to define the average and relative field components $\varphi_{cl/q}%
(t)=(\varphi_{+}(t)\pm\varphi_{-}(t))/\sqrt{2}$, $\bar{\varphi}_{cl/q}%
(t)=(\bar{\varphi}_{+}(t)\pm\bar{\varphi}_{-}(t))/\sqrt{2}$, $\psi
_{0/1,d}(t)=(\psi_{+,d}(t)\pm\psi_{-,d}(t))/\sqrt{2}$, and $\bar{\psi}%
_{0/1,d}(t)=(\bar{\psi}_{+,d}(t)\mp\bar{\psi}_{-,d}(t))/\sqrt{2}$. These
quantities can be grouped into vectorial fields $\varphi(t)=\mathstrut
^{t}\{\varphi_{cl}(t),\varphi_{q}(t)\}$, $\bar{\varphi}(t)=\{\bar{\varphi
}_{cl}(t),\bar{\varphi}_{q}(t)\}$, $\psi(t)=\mathstrut^{t}\{\psi_{0}%
(t),\psi_{1}(t)\}$ and $\bar{\psi}(t)=\{\bar{\psi}_{0}(t),\bar{\psi}_{1}%
(t)\}$. Note that in the case of a mesoscopic circuit with several discrete
orbitals, the fields $\psi_{0}(t)$ and $\psi_{1}(t)$ have an orbital structure
$\psi_{m}(t)=\mathstrut^{t}\{\psi_{m,d_{1}}(t),...,\psi_{m,d_{N}}(t)\}$ with
$m\in\{0/1\}$. In the main text of this article, all the fields have a time
argument $t$, which is omitted for brevity, except when two times $t$ and
$t^{\prime}$ are involved in an equation. The global Schwinger-Keldysh
partition function \ $Z$ of the mesoscopic QED device and the corresponding
quantum action $S_{tot}$ can be obtained directly from Hamiltonian
(\ref{Htot}) by considering the elementary evolution of the system along the
Keldysh contour\cite{Kamenev}. This gives%
\begin{equation}
Z=\int d[\bar{\varphi},\varphi,\bar{\psi},\psi]e^{iS_{tot}(\bar{\varphi
},\varphi,\bar{\psi},\psi)} \label{Z}%
\end{equation}
with%
\begin{align}
S_{tot}(\bar{\varphi},\varphi,\bar{\psi},\psi)  &  =S_{cav}^{0}(\bar{\varphi
},\varphi)+S_{meso}^{0}(\bar{\psi},\psi)\\
&  +\Delta S_{ac}(\bar{\varphi},\varphi)+\Delta S_{inter}(\bar{\varphi
},\varphi,\bar{\psi},\psi)\text{.}\nonumber
\end{align}
Above, $d[\bar{\varphi},\varphi,\bar{\psi},\psi]$ is the differential element
associated to the fields $\bar{\varphi},\varphi,\bar{\psi}$ and $\psi$. The
term%
\begin{equation}
S_{cav}^{0}(\bar{\varphi},\varphi)=\int\limits_{t}%
\begin{bmatrix}
\bar{\varphi}_{cl} & \bar{\varphi}_{q}%
\end{bmatrix}%
\begin{bmatrix}
0 & D_{t}-\frac{i\Lambda_{0}}{2}\\
D_{t}+\frac{i\Lambda_{0}}{2} & i\Lambda_{0}(1+2n_{B})
\end{bmatrix}%
\begin{bmatrix}
\varphi_{cl}\\
\varphi_{q}%
\end{bmatrix}
\label{S0}%
\end{equation}
is the bare cavity action, with $D_{t}=i\partial t-\omega_{0}$, $n_{B}%
=1/(e^{\omega_{0}/k_{B}T}-1)$ and $\Lambda_{0}$ a damping rate due to the
cavity bath treated in the Markovian approximation\cite{Torre:2013}. For
compactness, we note $\int_{-\infty}^{+\infty}dt=\int\nolimits_{t}$. The
cavity drive brings a contribution:%
\begin{equation}
\Delta S_{ac}(\bar{\varphi},\varphi)=-\sqrt{2}\int_{t}\left(  \bar{\varphi
}_{q}+\varphi_{q}\right)  \varepsilon_{ac}(t)\text{.} \label{Sac}%
\end{equation}
The bare action from the mesoscopic circuit is%
\begin{equation}
S_{meso}^{0}(\bar{\psi},\psi)=\int\limits_{t,t^{\prime}}\bar{\psi}(t)\check
{G}^{-1}(t,t^{\prime})\psi(t^{\prime}) \label{Sdd}%
\end{equation}
with $\check{G}$ the mesoscopic circuit Green's function in the absence of
light/matter coupling. The contribution from the light/matter coupling is%
\begin{equation}
\Delta S_{inter}(\bar{\varphi},\varphi,\bar{\psi},\psi)=-\int%
\limits_{t,t^{\prime}}\bar{\psi}(t)(\check{v}(\bar{\varphi},\varphi
,t)\delta(t-t^{\prime}))\psi(t^{\prime}) \label{Sinter}%
\end{equation}
with $\iint_{-\infty}^{+\infty}dt~dt^{\prime}=\int_{t,t^{\prime}}$ and
$\check{v}$ a light/matter coupling function. Both $\check{G}$ and $\check{v}$
are defined below.

The unperturbed mesoscopic circuit Green's function which appears in
Eq.(\ref{Sdd}) has the structure $\check{G}(t,t^{\prime})=\int%
\limits_{^{\omega}}\check{G}(\omega)e^{i\omega(t^{\prime}-t)}$
with\cite{Kamenev}
\begin{equation}
\check{G}(\omega)=%
\begin{bmatrix}
\tilde{G}_{r}(\omega) & \tilde{G}_{K}(\omega)\\
\tilde{0} & \tilde{G}_{a}(\omega)
\end{bmatrix}
\label{Gdef}%
\end{equation}
the $2\times2$ mesoscopic Keldysh space. Above, $\tilde{0}$ is a matrix full
of zeros in the $N\times N$ the mesoscopic orbitals space. The retarded,
advanced and Keldysh components $\tilde{G}_{r/a/K}(\omega)$ of $\check{G}$
also have a $N\times N$ structure in the mesoscopic orbitals space. In the
absence of superconducting correlations in a circuit, the elements of
$\tilde{G}_{r}$, $\tilde{G}_{a}$ and $\tilde{G}_{K}$ in the line $d$ and
column $d^{\prime}$ can be defined as
\begin{equation}
G_{r}^{d,d^{\prime}}(t,t^{\prime})=-i\theta(t)\left\langle \{\hat{c}%
_{d}(t),\hat{c}_{d^{\prime}}^{\dag}(t^{\prime})\}\right\rangle \text{,}
\label{AA}%
\end{equation}%
\begin{equation}
G_{a}^{d,d^{\prime}}(t,t^{\prime})=i\theta(-t)\left\langle \{\hat{c}%
_{d}(t),\hat{c}_{d^{\prime}}^{\dag}(t^{\prime})\}\right\rangle \label{BB}%
\end{equation}
and
\begin{equation}
G_{K}^{d,d^{\prime}}(t,t^{\prime})=-i\left\langle [\hat{c}_{d}(t),\hat
{c}_{d^{\prime}}^{\dag}(t^{\prime})]\right\rangle \label{CC}%
\end{equation}
respectively. We also use the stationary relations $\tilde{G}_{r/a/K}%
(t,t^{\prime})=\int\limits_{^{\omega}}\tilde{G}_{r/a/K}(\omega)e^{i\omega
(t^{\prime}-t)}$. Importantly, the index $d\in\lbrack1,N]$ in the
above Green's functions runs only on the set of confined discrete orbitals of
the mesoscopic circuit (like for instance quantum dot orbitals) which remain
after the leads' orbital continua have been integrated out. The leads
contribute to $\check{G}$ through self energy terms which depend on the tunnel
rates between the mesoscopic orbitals $d$ and the leads. At this
stage, we do not give a more explicit expression for $\check{G}$ because we
consider a generic mesoscopic circuit. An example of expression for $\check
{G}$ will be given in section \ref{DOUBLEDOT} for a non-interacting double dot
(see Eqs. (\ref{Gmeso})-(\ref{mkdot})).

The light matter coupling occurs in Eq.(\ref{Sinter}) through the term
\begin{equation}
\check{v}(\bar{\varphi},\varphi,t)=\check{g}\frac{\left(  \bar{\varphi}%
_{cl}(t)+\varphi_{cl}(t)\right)  \check{\sigma}_{0}+\left(  \bar{\varphi}%
_{q}(t)+\varphi_{q}(t)\right)  \check{\sigma}_{1}}{\sqrt{2}} \label{vchap}%
\end{equation}
Above, we use matrices $\check{\sigma}_{0[1]}=\mathring{\sigma}_{0[1]}%
\otimes\widetilde{1}$, where $\mathring{\sigma}_{0}$ and $\mathring{\sigma
}_{1}$ correspond to the identity and the first Pauli matrix in the Keldysh
subspace of the mesoscopic circuit (index $0/1$) and $\widetilde{1}$ is the
identity in the mesoscopic orbitals subspace. We also note $\check
{g}=\mathring{\sigma}_{0}\otimes\tilde{g}$ with $\tilde{g}=diag[g_{1}%
,...,g_{N}]$ a diagonal matrix in the mesoscopic orbitals subspace. More
generally, the superscripts $\circ$ and $\sim$ decorate a matrix in the
$2\times2$ mesoscopic Keldysh subspace and the $N\times N$ mesoscopic orbital
subspace, respectively. The superscript $\vee$ decorates a matrix in the
tensor product of these two spaces. The notation $g$ used previously
corresponds to $g=\max_{d}[g_{d}]$.

\subsection{\label{dev4}Effective cavity action to fourth order in\textbf{
}$g$}

In order to obtain an effective description of the cavity dynamics solely, one
must integrate out the electronic degrees of freedom in Eq.(\ref{Z}). For
simplicity, we will disregard Coulomb interactions in the mesoscopic circuit.
In this case, the mesoscopic QED action is quadratic with respect the
electronic fields $\psi$ and $\bar{\psi}$, and one can thus perform a
straightforward Gaussian integration of Eq.(\ref{Z}) on these fields (in the
interacting case, it is possible to use more elaborate integration
procedures\cite{Kamenev}). The resulting effective cavity action
$S_{cav}^{eff}(\bar{\varphi},\varphi)$ can be simplified after a systematic
expansion with respect to the light/matter coupling matrix $\check{g}$ (see
Appendix A for details). We work to fourth order in\textbf{ }$g$ in order to
capture essential non-linear electron/photon interaction effects. In order to
simplify the final expression of $S_{cav}^{eff}$, we assume that the dressed
cavity linewidth is much smaller than $\omega_{0}$ and the width of the
mesoscopic resonances linewidth. This criterion is largely satisfied in
experiments as well as for the parameters used in this manuscript. We finally
obtain the expression
\begin{equation}
S_{cav}^{eff}(\bar{\varphi},\varphi)=S_{cav}^{0}(\bar{\varphi},\varphi)+%
{\textstyle\sum\limits_{i\in\{2,3,4\}}}
\Delta S_{g}^{(i)}(\bar{\varphi},\varphi)+o(\check{g}^{4})\text{.} \label{S}%
\end{equation}
Above, $\Delta S_{g}^{(i)}$ is the mesoscopic circuit contribution to
$S_{cav}^{eff}$ to $i^{th}$ order in\textbf{ }$g$. The first order
contribution in\textbf{ }$g$ can be disregarded because it is not resonant
with the cavity (see Eqs.(\ref{Seff1}) and (\ref{Heff1}) of
Appendix A).

The second order contribution
\begin{equation}
\Delta S_{g}^{(2)}(\bar{\varphi},\varphi)=-\int_{t}%
\begin{bmatrix}
\bar{\varphi}_{cl} & \bar{\varphi}_{q}%
\end{bmatrix}
.%
\begin{bmatrix}
0 & \chi_{2}^{\ast}\\
\chi_{2} & \lambda_{2}%
\end{bmatrix}
.%
\begin{bmatrix}
\varphi_{cl}\\
\varphi_{q}%
\end{bmatrix}
\label{S2}%
\end{equation}
involves the semiclassical charge susceptibility
\begin{equation}
\chi_{2}=-\frac{i}{2}\int\limits_{\omega}\underset{d}{\mathrm{Tr}}\left[
\tilde{G}_{K}(\omega)\tilde{g}\left(  \tilde{G}_{a}(\omega-\omega_{0}%
)+\tilde{G}_{r}(\omega+\omega_{0})\right)  \tilde{g}\right]  \label{Chi2}%
\end{equation}
of the mesoscopic circuit at frequency $\omega_{0}$ and the correlation
function%
\begin{align}
\lambda_{2}  &  =-\frac{i}{2}\int\limits_{\omega}\underset{d}{\mathrm{Tr}%
}[\tilde{G}_{K}(\omega)\tilde{g}\tilde{G}_{K}(\omega+\omega_{0})\tilde
{g}\label{landa}\\
&  +\tilde{G}_{a}(\omega)\tilde{g}\tilde{G}_{r}(\omega+\omega_{0})\tilde
{g}+\tilde{G}_{r}(\omega)\tilde{g}\tilde{G}_{a}(\omega+\omega_{0})\tilde
{g}]\text{.}\nonumber
\end{align}
We note $\int\limits_{\omega}=\int\limits_{-\infty}^{+\infty}\frac{d\omega
}{2\pi}$, and $\underset{d}{\mathrm{Tr}}$ the trace operator on the mesoscopic
orbital index $d$. Note that $\chi_{2}$ has already been introduced in other
works\cite{Cottet:2011,Bruhat:2016,Viennot:2014,Schiro:2014,Dmytruk:2015,Dartiailh:2017,Dmytruk:2016}%
, essentially for studying the semiclassical behavior of a Mesoscopic QED
device to second order in\textbf{ }$g$. A cavity frequency shift is caused by
$\operatorname{Re}[\chi_{2}]$ whereas $\operatorname{Im}[\chi_{2}]$
renormalizes the bare cavity linewidth $\Lambda_{0}$ of Eq.(\ref{S0}). The
parameter $\lambda_{2}$ is necessary to describe the quantum regime of
Mesoscopic QED, but it has been disregarded so far. From Eq.(\ref{landa}) with
$\tilde{G}_{K}(\omega)=-\tilde{G}_{K}(\omega)^{\dag}$ and $\tilde{G}%
_{a}(\omega)=\tilde{G}_{r}(\omega)^{\dag}$, one can check that $\lambda_{2}$
is purely imaginary.

For $\varepsilon_{p}\neq0$, we obtain a third order term $S_{g}^{(3)}(t)$
in\textbf{ }$g$ which can be expressed as
\begin{align}
\Delta S_{g}^{(3)}(\bar{\varphi},\varphi)  &  =-i\int_{t}e^{-2i\omega_{0}t}%
\begin{bmatrix}
\bar{\varphi}_{cl} & \bar{\varphi}_{q}%
\end{bmatrix}
.%
\begin{bmatrix}
0 & U_{cl}/2\\
U_{cl}/2 & U_{q}%
\end{bmatrix}
.%
\begin{bmatrix}
\bar{\varphi}_{cl}\\
\bar{\varphi}_{q}%
\end{bmatrix}
\nonumber\\
&  -i\int_{t}e^{2i\omega_{0}t}%
\begin{bmatrix}
\varphi_{cl} & \varphi_{q}%
\end{bmatrix}
.%
\begin{bmatrix}
0 & -U_{cl}^{\ast}/2\\
-U_{cl}^{\ast}/2 & U_{q}^{\ast}%
\end{bmatrix}
.%
\begin{bmatrix}
\varphi_{cl}\\
\varphi_{q}%
\end{bmatrix}
\label{S3}%
\end{align}
with%
\begin{align}
U_{cl}  &  =-\frac{\beta_{p}}{2}\int\limits_{\omega}\underset{k,d}{\mathrm{Tr}%
}[\check{\sigma}_{1}\check{g}\check{G}(\omega)\check{g}\check{G}(\omega
+\omega_{0})\check{g}\check{G}(\omega-\omega_{0})]\nonumber\\
&  +\underset{k,d}{\mathrm{Tr}}[\check{G}(\omega)\check{\sigma}_{1}\check
{g}\check{G}(\omega+\omega_{0})\check{g}\check{G}(\omega-\omega_{0})\check
{g}])\text{,} \label{Uclq}%
\end{align}%
\begin{equation}
U_{q}=-\frac{\beta_{p}}{2}\int\limits_{\omega}\underset{k,d}{\mathrm{Tr}%
}[\check{\sigma}_{1}\check{g}\check{G}(\omega)\check{\sigma}_{1}\check
{g}\check{G}(\omega+\omega_{0})\check{g}\check{G}(\omega-\omega_{0})]
\label{Uqq}%
\end{equation}
and
\begin{equation}
\beta_{p}=\varepsilon_{p}t_{0}/2\text{.} \label{epred}%
\end{equation}
Above, we note $\underset{k,d}{\mathrm{Tr}}$ the trace operator on both the
mesoscopic orbital index $d$ and the Keldysh index $k$. The prefactor%
\begin{equation}
t_{0}=\mathcal{G}_{0}^{r}(2\omega_{0})+\mathcal{G}_{0}^{a}(-2\omega_{0})
\label{tzero}%
\end{equation}
takes into account how the mesoscopic circuit feels the ac drive through the
cavity, with
\begin{equation}
\mathcal{G}_{0}^{r/a}(\omega)=(\omega-\omega_{0}\pm i\frac{\Lambda_{0}}%
{2})^{-1} \label{gbare}%
\end{equation}
the bare cavity retarded/advanced Green's function [see Eq.(\ref{Eac}) for a
semiclassical picture of this effect]. Subsequently, the reaction of the
mesoscopic circuit to the ac drive affects the cavity effective behavior, as
described by the terms in $U_{q}$ and $U_{cl}$. Importantly, these terms can
be significant because the smallness of $t_{0}$ can be compensated by the use
of a large enough drive amplitude $\beta_{p}$. Interestingly, the coefficient
$U_{cl}$ corresponds to the semiclassical joint response of the mesoscopic
charge to the cavity field in $\hat{a}$ and to the drive in $\beta_{p}$ (see
Appendix B1, Eq.(\ref{NT})).

Finally, we find a fourth order contribution in\textbf{ }$g$, which occurs
even for $\beta_{p}=0$, i.e.%
\begin{equation}
\Delta S_{g}^{(4)}(\bar{\varphi},\varphi)=-\int\limits_{t}%
\begin{bmatrix}
\bar{\varphi}_{cl}\bar{\varphi}_{cl} & \bar{\varphi}_{cl}\bar{\varphi}_{q} &
\bar{\varphi}_{q}\bar{\varphi}_{q}%
\end{bmatrix}
.\mathcal{A}.%
\begin{bmatrix}
\varphi_{cl}\varphi_{cl}\\
\varphi_{cl}\varphi_{q}\\
\varphi_{q}\varphi_{q}%
\end{bmatrix}
\label{S4}%
\end{equation}
with%
\begin{equation}
\mathcal{A=}%
\begin{bmatrix}
0 & \chi_{4}^{\ast} & -U_{4}^{\ast}\\
\chi_{4} & \lambda_{4} & V_{4}^{\ast}\\
U_{4} & V_{4} & W_{4}%
\end{bmatrix}
\text{,} \label{AAA}%
\end{equation}%
\begin{equation}
\chi_{4}=i(\mathcal{N}_{q,cl,cl,cl}+\mathcal{N}_{cl,q,cl,cl})\text{,}%
\end{equation}%
\begin{equation}
\lambda_{4}=i(\mathcal{N}_{cl,q,cl,q}+\mathcal{N}_{cl,q,q,cl}+\mathcal{N}%
_{q,cl,cl,q}+\mathcal{N}_{q,cl,q,cl})\text{,}%
\end{equation}%
\begin{equation}
V_{4}=i(\mathcal{N}_{q,q,cl,q}+\mathcal{N}_{q,q,q,cl})\text{,} \label{taa}%
\end{equation}%
\begin{equation}
U_{4}(\omega_{0})=i\mathcal{N}_{q,q,cl,cl}\text{,}%
\end{equation}%
\begin{equation}
W_{4}(\omega_{0})=i\mathcal{N}_{q,q,q,q}%
\end{equation}%
\begin{align}
\mathcal{N}_{f,f^{\prime},l,l^{\prime}}  &  =-\int\limits_{\omega
}\underset{k,d}{\mathrm{Tr}}\left[  \frac{1}{8}\check{G}(\omega)\hat{\sigma
}_{f}\check{g}\check{G}_{+}\hat{\sigma}_{l}\check{g}\check{G}(\omega
)\hat{\sigma}_{f^{\prime}}\check{g}\check{G}_{+}\hat{\sigma}_{l^{\prime}%
}\check{g}\right. \nonumber\\
&  +\left.  \frac{1}{4}\check{G}(\omega)\hat{\sigma}_{f}\check{g}\check{G}%
_{+}\hat{\sigma}_{f^{\prime}}\check{g}\check{G}(\omega+2\omega_{0})\hat
{\sigma}_{l}\check{g}\check{G}_{+}\hat{\sigma}_{l^{\prime}}\check{g}\right]
\label{NNN}%
\end{align}
and $\check{G}_{+}=\check{G}(\omega+\omega_{0})$. Note that $\lambda_{4}$ and
$W_{4}$ are purely imaginary due to $\tilde{G}_{K}(\omega)=-\tilde{G}%
_{K}(\omega)^{\dag}$ and $\tilde{G}_{a}(\omega)=\tilde{G}_{r}(\omega)^{\dag}$.
The coefficient $\chi_{4}$ corresponds to the second order semiclassical
response function of the quantum dot to the cavity electric field (see
Appendix B1, Eq.(\ref{NT})). The other coefficients $\lambda_{4}$, $U_{4}$,
$V_{4}$ and $W_{4}$ are necessary to describe quantum fluctuations of the
cavity field. In summary, Eqs. (\ref{S}) - (\ref{NNN}) describe the effective
action of a microwave cavity in a generic Mesoscopic QED\ device to fourth
order in the light/matter coupling. This requires to introduce new types of
quantum dot correlators than the known $\chi_{2}$. We will discuss the
physical effect of the new correlators $\lambda_{2}$, $U_{cl}$, $U_{q}$,
$\chi_{4}$, $\lambda_{4}$, $U_{4}$, $V_{4}$ and $W_{4}$ in the next sections.
Importantly, one has to choose an appropriate technique to obtain an explicit
description of the cavity dynamics out of the cavity effective action. In the
following we will consider situations such that an effective Lindblad equation
on the cavity density matrix can be used.

\subsection{Correspondence between the cavity effective action and a photonic
Lindblad equation\label{Lind}}

The most popular description of Circuit QED is the Lindblad equation which
describes the evolution of the cavity density matrix. Below, we come back to
this description, already illustrated by our Eq.(\ref{rohPed}), to clarify the
physical meaning of the different terms in the cavity action of section
\ref{dev4}.

\subsubsection{Cavity effective Lindblad equation up to third order in\textbf{
}$g$}

In the limit of low couplings $g_{d}$ and limited cavity drive $\beta_{p}$,
the cavity field remains small so that one can truncate the cavity effective
action to third order in\textbf{ }$g$. In this case, we show below that it is
always possible to establish a Lindblad equation on the cavity density matrix.
Thereby, we clarify the physical meaning of the terms in $U_{cl}$ and $U_{q}$.

When a cavity follows a Lindblad description, the time derivative of its
density matrix $\rho_{cav}(t)$ can be expressed as\cite{Haroche: 2006}:%
\begin{equation}
\frac{\partial\rho_{cav}(t)}{\partial t}=-i[H_{cav}^{eff},\rho_{cav}(t)]+%
{\textstyle\sum\limits_{j}}
\gamma_{j}\mathcal{D}_{\hat{L}_{j}}(\rho_{cav}(t)) \label{ME}%
\end{equation}
with $H_{cav}^{eff}$ the effective cavity Hamiltonian, $\gamma_{j}$ the rate
of a dissipative process corresponding to the jump operator $\hat{L}_{j}$ and
$\mathcal{D}_{\hat{L}_{j}}(\rho_{cav})$ defined in Eq.(\ref{Lj}). Let us
assume that the effective Hamiltonian has the generic form%
\begin{equation}
H_{cav}^{eff}=(\omega_{0}+\Delta\omega_{0})\hat{a}^{\dag}\hat{a}+i\rho
_{p}e^{-i2\omega_{0}t}\hat{a}^{\dag2}-i\rho_{p}^{\ast}e^{i2\omega_{0}t}\hat
{a}^{2} \label{Hefff}%
\end{equation}
and the dissipative processes are characterized by $(\gamma_{j},\hat{L}%
_{j})\in\mathcal{P}$ with%
\begin{equation}
\mathcal{P}=\{(\gamma_{loss},\hat{a}),(\gamma_{gain},\hat{a}^{\dag}%
),(\gamma_{p},\hat{a}+e^{i\varphi_{p}}e^{-i2\omega_{0}t}\hat{a}^{\dag
})\}\text{.} \label{P2}%
\end{equation}
The above real parameters $\Delta\omega_{0}$, $\rho_{p}$,
$\gamma_{loss}$, $\gamma_{gain}$, $\gamma_{p}$ and $\varphi_{p}$ are
unspecified for the moment. The action corresponding to the
master equation (\ref{ME}) can be expressed as (see details in Appendix C)
\begin{align}
S_{Mark}(t)  &  =\int\limits_{t}%
\begin{bmatrix}
\bar{\varphi}_{cl} & \bar{\varphi}_{q}%
\end{bmatrix}
.%
\begin{bmatrix}
0 & F_{t}-i\frac{\gamma_{-}}{2}\\
F_{t}+i\frac{\gamma_{-}}{2} & i\gamma_{+}%
\end{bmatrix}
.%
\begin{bmatrix}
\varphi_{cl}\\
\varphi_{q}%
\end{bmatrix}
\nonumber\\
&  +\int\limits_{t}e^{-i2\omega_{0}t}%
\begin{bmatrix}
\bar{\varphi}_{cl} & \bar{\varphi}_{q}%
\end{bmatrix}
.%
\begin{bmatrix}
0 & -i\rho_{p}\\
-i\rho_{p} & i\gamma_{p}e^{i\varphi_{p}}%
\end{bmatrix}
.%
\begin{bmatrix}
\bar{\varphi}_{cl}\\
\bar{\varphi}_{q}%
\end{bmatrix}
\nonumber\\
&  +\int\limits_{t}e^{i2\omega_{0}t}%
\begin{bmatrix}
\varphi_{cl} & \varphi_{q}%
\end{bmatrix}
.%
\begin{bmatrix}
0 & i\rho_{p}^{\ast}\\
i\rho_{p}^{\ast} & i\gamma_{p}e^{-i\varphi_{p}}%
\end{bmatrix}
.%
\begin{bmatrix}
\varphi_{cl}\\
\varphi_{q}%
\end{bmatrix}
\label{SM}%
\end{align}
with
\begin{equation}
\gamma_{-}=\gamma_{loss}-\gamma_{gain}\text{,} \label{gmoins}%
\end{equation}%
\begin{equation}
\gamma_{+}=\gamma_{loss}+\gamma_{gain}+2\gamma_{p} \label{gplus}%
\end{equation}
and $F_{t}=i\partial t-\omega_{0}-\Delta\omega_{0}$. It is possible
to perform an exact identification between the action of Eq.(\ref{SM}) and the
cavity effective action to third order in\textbf{ }$g$ (i.e. Eqs.(\ref{S2}%
)+(\ref{S3})) by using parameters $\Delta\omega_{0}$, $\rho_{p}$,
$\gamma_{loss}$, $\gamma_{gain}$, $\gamma_{p}$ and $\varphi_{p}$ given by the
relations:
\begin{equation}
\Delta\omega_{0}=\operatorname{Re}[\chi_{2}]\text{,} \label{domega0}%
\end{equation}%
\begin{equation}
\rho_{p}=U_{cl}/2\text{,} \label{p1}%
\end{equation}%
\begin{equation}
\gamma_{p}e^{i\varphi_{p}}=-U_{q}\text{,} \label{ggp}%
\end{equation}%
\begin{equation}
\gamma_{loss}=\gamma_{loss}^{0}-\gamma_{p}\text{,} \label{gg1}%
\end{equation}
and%
\begin{equation}
\gamma_{gain}=\gamma_{gain}^{0}-\gamma_{p} \label{gg2}%
\end{equation}
with%
\begin{equation}
\gamma_{loss}^{0}=\Lambda_{0}(1+n_{B})-\operatorname{Im}[\chi_{2}%
+\frac{\lambda_{2}}{2}]\text{,} \label{pp5}%
\end{equation}%
\begin{equation}
\gamma_{gain}^{0}=\Lambda_{0}n_{B}+\operatorname{Im}[\chi_{2}-\frac
{\lambda_{2}}{2}] \label{p5}%
\end{equation}
and $\gamma_{p}>0$ by definition.

We now comment on the physical effect of the components (\ref{domega0}%
)-(\ref{p5}). As found
previously\cite{Cottet:2011,Bruhat:2016,Viennot:2014,Schiro:2014,Dmytruk:2015,Dartiailh:2017,Dmytruk:2016}%
, the cavity frequency shift $\Delta\omega_{0}$ is directly set by the real
part of $\chi_{2}$. A comparison between Eqs.(\ref{S0}) and (\ref{SM})
indicates that the cavity intrinsic linewidth $\Lambda_{0}$ is also shifted by
$\Delta\Lambda_{0}=-2\operatorname{Im}[\chi_{2}]$. The dissipative processes
with rates $\gamma_{loss}$ and $\gamma_{gain}$ correspond to standard
single-photon emission and absorption which are widely considered in circuit
QED. One can see from Eqs.(\ref{gg1})-(\ref{p5}) that $Im[\chi_{2}]$
contributes to the asymmetry between the photon loss and gain rates
$\gamma_{loss}$ and $\gamma_{gain}$ whereas $Im[\lambda_{2}]$ contributes
equally to $\gamma_{loss}$ and $\gamma_{gain}$. The coefficients $\rho_{p}$
and $\gamma_{p}$ account for the effect of the ac drive since they are nonzero
only for $\beta_{p}\neq0$. From Eq.(\ref{p1}), $U_{cl}$ generates the
two-photon coherent drive in $\rho_{p}$ of Eq.(\ref{Hefff}). Such a
term can be obtained with a degenerate parametric amplifier (see for instance
section 5.1.1 of Ref.\cite{Millburn}). It was also obtained in
Ref.\cite{Leghtas:2015} by using a complex configuration with two microwave
cavities coupled nonlinearly and subject to two off resonant drives. Finally,
the dissipative process with a rate $\gamma_{p}$ generated by $U_{q}$
is less usual. Its jump operator $L_{p}=\hat
{a}+e^{i\varphi_{p}}e^{-i2\omega_{0}t}\hat{a}^{\dag}$ corresponds to a
time-dependent coherent superposition of photon absorption and emission
operators. From Eqs. (\ref{gg1}) and (\ref{gg2}), one could believe that
$\gamma_{p}$ decreases the single-photon loss and gain rates, but this is not
effective because the rates $\gamma_{+}$ and $\gamma_{-}$ through which
$\gamma_{loss}$ and $\gamma_{gain}$ occur in the cavity action do not depend
on $\gamma_{p}$. Indeed, from Eqs.(\ref{gmoins}), (\ref{gplus}), (\ref{gg1})
and (\ref{gg2}), one has $\gamma_{-}=\gamma_{loss}^{0}-\gamma_{gain}^{0}$ and
$\gamma_{+}=\gamma_{loss}^{0}+\gamma_{gain}^{0}$. There remains a term in
$\gamma_{p}$ which occurs through the second and third lines of Eq.(\ref{SM})
on the same footing as $\rho_{p}$. We will illustrate the effect of this
peculiar term in section \ref{SW} for the case of a double quantum dot and
check that it corresponds to a \textquotedblleft squeezing
dissipation\textquotedblright. In fact, such an effect can also be
obtained by using a broadband squeezed bath input\cite{Lu:2015} or a cavity
damping modulation\cite{Didier:2014}. It leads to the relaxation of the cavity
to a squeezed state. In these Refs., squeezing superoperators are used to
describe this effect, instead of the jump operator $L_{p}$, but one can check
that there is a formal equivalence between the two descriptions\cite{AF}%
. Importantly, in our work, we have used a range of $\gamma_{p}$
such that one has $\gamma_{loss}>0$ and $\gamma_{gain}>0$, as required by the
definition of the Lindblad equation (\ref{ME}). When the drive amplitude
$\beta_{p}$ becomes so large that $\gamma_{loss}<0$ and/or $\gamma_{gain}<0$,
we expect that higher order terms in $\beta_{p}$ become relevant, which
introduces new terms in the cavity action which are not necessarily Markovian.
In this case, the Lindblad Eq.(\ref{ME}) is not relevant anymore. This limit
is beyond the scope of this article.

\subsubsection{Cavity effective Lindblad equation to fourth order in\textbf{
}$g$\label{g4res}}

We now investigate the possibility to identify the path integral approach of
section \ref{GENERAL} with a Lindblad description up to fourth order
in\textbf{ }$g$. We expect an extra contribution%
\begin{equation}
H_{cav}^{eff,4}=K\hat{a}^{\dag2}\hat{a}^{2} \label{H4}%
\end{equation}
to the effective Hamiltonian (\ref{Hefff}), which corresponds to a Kerr
photonic interaction. We also expect dissipative processes with rates and jump
operators $(\gamma_{j},\hat{L}_{j})\in\mathcal{P}_{4}$ with%
\begin{equation}
\mathcal{P}_{4}=\{(K_{loss},\hat{a}^{2}),(K_{gain},\hat{a}^{\dag2}),(D,\hat
{a}^{\dag}\hat{a})\}\text{.} \label{P4}%
\end{equation}
The three processes in the above ensemble correspond respectively to
two-photon loss, two-photon gain and pure dephasing. This leads to an action
contribution (see Appendix A)%
\begin{equation}
S_{Mark}^{(4)}=-\int\limits_{t}%
\begin{bmatrix}
\bar{\varphi}_{cl}\bar{\varphi}_{cl} & \bar{\varphi}_{cl}\bar{\varphi}_{q} &
\bar{\varphi}_{q}\bar{\varphi}_{q}%
\end{bmatrix}
.\mathcal{A}_{M}.%
\begin{bmatrix}
\varphi_{cl}\varphi_{cl}\\
\varphi_{cl}\varphi_{q}\\
\varphi_{q}\varphi_{q}%
\end{bmatrix}
\label{Smark4}%
\end{equation}
with
\begin{equation}
\mathcal{A}_{M}=\left[
\begin{tabular}
[c]{rrr}%
$0$ & $i\frac{K_{-}}{2}+K$ & $-\frac{iD}{2}$\\
$-i\frac{K_{-}}{2}+K$ & $-i\left(  D+2K_{+}\right)  $ & $-i\frac{K_{-}}{2}%
+K$\\
$-\frac{iD}{2}$ & $i\frac{K_{-}}{2}+K$ & $0$%
\end{tabular}
\ \ \ \ \ \ \ \ \ \ \right]  \label{AM}%
\end{equation}
and $K_{-}=K_{loss}-K_{gain}$, $K_{+}=K_{loss}+K_{gain}$. To establish a
mapping with the path integral description, we now have to compare the above
matrix $\mathcal{A}_{M}$ with the matrix $\mathcal{A}$ of Eq. (\ref{AAA})
which occurs in the effective action of the Mesoscopic QED device to fourth
order in\textbf{ }$g$. Strikingly, $\mathcal{A}_{M}$ and $\mathcal{A}$ cannot
be mapped in all situations. This is possible when the condition%
\begin{equation}
\mathcal{C}_{Ldb}=(W_{4}=0)\&(\operatorname{Re}[U_{4}]=0)\&(V_{4}=\chi
_{4}^{\ast}) \label{Cmark}%
\end{equation}
is fulfilled. Equation (\ref{Cmark}) represents\textbf{ }a
sufficient condition to have a description of the cavity dynamics in terms of
a Lindblad equation to fourth order in\textbf{ }$g$. For a given
mesoscopic circuit, one can test this condition by evaluating numerically the
different fourth order mesoscopic correlators. When condition (\ref{Cmark}) is
valid, one has%
\begin{equation}
K=\operatorname{Re}[\chi_{4}]\text{,} \label{m1}%
\end{equation}%
\begin{equation}
K_{loss/gain}=\mp\operatorname{Im}[\chi_{4}]+\frac{\operatorname{Im}[U_{4}%
]}{2}-\frac{\operatorname{Im}[\lambda_{4}]}{4} \label{mm5}%
\end{equation}
and%
\begin{equation}
D=-2\operatorname{Im}[U_{4}]\text{.} \label{m4}%
\end{equation}
Hence, $\operatorname{Re}[\chi_{4}]$ generates the effective Kerr interaction
(\ref{H4}). Remarkably, there exists an analogy between the expressions of the
rates for the single and two-photon stochastic processes, Eqs. (\ref{mm5}) and
Eqs. (\ref{pp5})-(\ref{p5}). Indeed, $\operatorname{Im}[\chi_{4}]$ provides an
opposite contribution to two-photon loss and gain, like $\operatorname{Im}%
[\chi_{2}]$ does for single-photon processes. In contrast, $\operatorname{Im}%
[\lambda_{4}]-2\operatorname{Im}[U_{4}]$ provides the same contribution to
two-photon loss and gain, like $\operatorname{Im}[\lambda_{2}]$ does for
single-photon processes. The term in $\operatorname{Im}[U_{4}]$ also
contributes to photonic dephasing (term in $D$). This last effect does not
have any analogue to second order in\textbf{ }$g$.

We could not find other contributions to  the Hamiltonian
(\ref{H4}) and  the jump operator ensemble $\mathcal{P}_{4}$ of
Eq. (\ref{P4}) to extend the mapping between the path integral approach and
the Lindblad description beyond the regime of validity of Eq.(\ref{Cmark}).
It would be interesting to find a systematic method to derive a
cavity evolution equation from the cavity action, in order to establish the
necessary conditions for having the Lindblad
description\textbf{.} Importantly, to fourth order in\textbf{
}$g$, a systematic mapping cannot be expected since the dynamics of the cavity
is not necessarily Markovian. For instance, there can be \textquotedblleft
memory\textquotedblright\ effects due to a coherent exchange of energy between
the cavity and the mesoscopic circuit. This will be illustrated in the case of
a non-interacting double quantum dot in section \ref{cat}.

\subsubsection{Summary: total photonic Lindblad equation up to fourth order
in\textbf{ }$g$ in the interaction picture}

In practice, it is convenient to study the cavity dynamics in an interaction
picture by considering the time evolution of the cavity density operator
$\rho_{cav}^{I}(t)=e^{i\omega_{0}\hat{a}^{\dag}\hat{a}t}\rho_{cav}%
(t)e^{-i\omega_{0}\hat{a}^{\dag}\hat{a}t}$. In this picture, Eqs. (\ref{ME}),
(\ref{Hefff}), (\ref{P2}), (\ref{H4}) and (\ref{P4}) lead to%
\begin{equation}
\frac{\partial\rho_{cav}^{I}(t)}{\partial t}=-i[H_{cav}^{eff,I},\rho_{cav}%
^{I}]+%
{\textstyle\sum\limits_{j}}
\gamma_{j}\mathcal{D}_{\hat{L}_{j}}(\rho_{cav}^{I}) \label{MElin}%
\end{equation}
with%
\begin{equation}
H_{cav}^{eff,I}=\Delta\omega_{0}\hat{a}_{I}^{\dag}\hat{a}_{I}+i\rho_{p}\hat
{a}_{I}^{\dag2}-i\rho_{p}^{\ast}\hat{a}_{I}^{2}+K\hat{a}_{I}^{\dag2}\hat
{a}_{I}^{2} \label{H3}%
\end{equation}
and dissipative processes $(\gamma_{j},\hat{L}_{j})\in\mathcal{P}_{I}$ with
\begin{align}
\mathcal{P}_{I}  &  \mathcal{=}\{(\gamma_{loss},\hat{a}_{I}),(\gamma
_{gain},\hat{a}_{I}^{\dag}),(\gamma_{p},\hat{a}_{I}+e^{i\varphi_{p}}\hat
{a}_{I}^{\dag}),\nonumber\\
&  (K_{loss},\hat{a}_{I}^{2}),(K_{gain},\hat{a}_{I}^{\dag2}),(D,\hat{a}%
_{I}^{\dag}\hat{a}_{I})\} \label{P3}%
\end{align}
with $\hat{a}_{I}=e^{-i\omega_{0}t}\hat{a}$.

Interestingly, Eq.(\ref{MElin}) appears as a generalization to fourth order
in\textbf{ }$g$ of Eq.(\ref{rohPed}) obtained with the direct density matrix
approach. Indeed, one can check that these two Eqs. agree to second order
in\textbf{ }$g$, provided the assumption $\Lambda_{0}=0$ of section
\ref{direct} is used. For this purpose, one must use the equalities%
\begin{equation}
\chi_{2}=\chi_{B}(\omega_{0})-\chi_{A}(\omega_{0}) \label{lalla1}%
\end{equation}
and%
\begin{equation}
\left.  \lambda_{2}\right\vert _{\omega_{0}\neq0}=2i\operatorname{Im}\left[
\chi_{A}(\omega_{0})+\chi_{B}(\omega_{0})\right]  \label{lalla2}%
\end{equation}
which are derived in Appendix D.

\section{The case of a double quantum dot in a cavity\label{DOUBLEDOT}}

\subsection{Circuit description}

We now apply the results of section \ref{GENERAL} to the case of a
spin-degenerate double quantum dot coupled to a microwave cavity, represented
schematically in Fig.\ref{Fig0}, panels (a) and (b). This circuit encloses two
quantum dots $L$ and $R$ with a tunnel coupling $t_{LR}$ such that $\hat
{H}_{meso}$ includes a term $t_{LR}\hat{c}_{L}^{\dag}\hat{c}_{R}+t_{LR}^{\ast
}\hat{c}_{R}^{\dag}\hat{c}_{L}$. The dot $L(R)$ is contacted to a normal metal
reservoir with a tunnel rate $\Gamma_{L(R)}$. Equation (\ref{Htot}) gives
$\Gamma_{d}=2\pi\Sigma_{k\in C}\left\vert t_{k,d}\right\vert ^{2}$ for $d\in
L(R)$. The rate $\Gamma_{d}$ can be considered as energy-independent in the
framework of a wide band approximation for the reservoirs with $\left\vert
t_{k,d}\right\vert ^{2}$ independent of $k$. In the following we consider the
case $\Gamma_{L}=\Gamma_{R}=\Gamma$. A bias voltage $V$ is applied between the
two normal metal contacts. The orbital energy $\omega_{L(R)}$ of dot $L(R)$
can be finely tuned with an electrostatic gate. In principle, $\omega_{L(R)}$
can also be shifted by a fraction of $eV$ which depends on the ratio of the
junctions capacitances. Here we will assume that this shift is
negligible\cite{NoteBias}. We will also disregard Coulomb interactions in the
double dot. This basic case presents essential ingredients of mesoscopic QED:
the cavity electric field can couple to both the internal transition between
the $L$ and $R$ orbitals of the double dot, and to tunnel transitions between
the dots and the continuum of states of the normal metal reservoirs.

\subsection{Unperturbed mesoscopic Green's function of the double
dot\label{DQD}}

The unperturbed mesoscopic circuit Green's function $\check{G}$ of the double
dot, whose inverse appears in Eq. (\ref{Sdd}), must be calculated in the
absence of light/matter coupling (i.e. $g_{L}=0$ and $g_{R}=0$). It can be
obtained by performing the inversion
\begin{equation}
\check{G}(\omega)=%
\begin{bmatrix}
\tilde{G}_{r}^{-1}(\omega) & \tilde{M}_{K}\\
\tilde{0} & \tilde{G}_{a}^{-1}(\omega)
\end{bmatrix}
^{-1} \label{Gmeso}%
\end{equation}
with\cite{Jauho:1994,Talbo:2018}
\begin{equation}
\tilde{G}_{r(a)}^{-1}(\omega)=%
\begin{bmatrix}
\omega-\omega_{L}\pm i\frac{\Gamma}{2} & -t_{LR}\\
-t_{LR}^{\ast} & \omega-\omega_{R}\pm i\frac{\Gamma}{2}%
\end{bmatrix}
\label{Grdot}%
\end{equation}
and%
\begin{equation}
\tilde{M}_{K}=%
\begin{bmatrix}
i\Gamma(1-2n_{F,L}(\omega)) & 0\\
0 & i\Gamma(1-2n_{F,R}(\omega))
\end{bmatrix}
\text{.} \label{mkdot}%
\end{equation}
Equations (\ref{Grdot}) and (\ref{mkdot}) stem from the explicit definitions
(\ref{AA})-(\ref{CC}) of the Green's functions $\tilde{G}_{r/a/K}(\omega)$ in
terms of fermionic operators and the expression of the double dot circuit
Hamiltonian (see Eq. (\ref{HHmeso}) with $g_{L(R)}=0$). Since we consider a
spin degenerate situation with non-interacting quantum dots, the spin degree
of freedom is omitted in the above orbital subspace structure. We will restore
it later in numerical evaluations by taking into account an implicit
multiplication by a factor 2 in the traces operator over the orbital index
$d$. The Fermi occupation function $n_{F,L(R)}(\omega)=(1+\exp[(\omega
\mp(eV_{b}/2))/k_{B}T])^{-1}$ of the $L(R)$ contact is affected by the bias
voltage $V_{b}$. For later use, we also define the lesser self energy of the
double dot\cite{Jauho:1994}%
\begin{equation}
\tilde{\Sigma}^{<}(\omega)=%
\begin{bmatrix}
i\Gamma n_{F,L}(\omega) & 0\\
0 & i\Gamma n_{F,R}(\omega)
\end{bmatrix}
\end{equation}
and the light/matter coupling matrix
\begin{equation}
\check{g}=diag[g_{L},g_{R},g_{L},g_{R}]\text{.}%
\end{equation}

\subsection{Choice of parameters}

For simplicity, we will use a nonzero $g_{L}$ and $g_{R}=0$, which corresponds
to DQD experiments realized so far, where a very asymmetric microwave coupling
to the two dots is engineered. In experiments realized with standard coplanar
microwave resonators, the light matter coupling is typically $g_{L}%
\sim0.001\omega_{0}$\cite{Cottet:2017}. In a more recent design based on high
kinetic inductance superconducting nanowire resonators, $g_{L}\sim
0.03\omega_{0}$ was reached\cite{Samkharadze:2017}. However, since the rms
voltage of these resonators is \cite{Samkharadze:2016} $V_{rms}=20~\mu
V\simeq4.9~\mathrm{GHz}$ for $\omega_{0}\sim4~\mathrm{GHz}$, one can reach
$g_{L}\sim\omega_{0}$, in principle, by using a galvanic coupling between one
of the dots and the cavity. In this work, we consider the regime
$\Lambda_{0}\ll\Gamma$ explored experimentally, with $\Gamma\geq
0.005\omega_{0}$. We also use $g_{L}/\omega_{0}\leqslant0.125$ and $\beta
_{p}g_{L}^{3}/\omega_{0}^{3}\leqslant0.001$

Since we develop the cavity action with respect to $g_{L}$ and $\beta_{p}$,
the amplitude of these two parameters must not be too large. Besides, having
$\Gamma\neq0$ is crucial for ensuring the validity of our perturbation scheme.
Indeed, in the absence of dissipation, the correlators $\chi_{2}$ and
$\chi_{4}$ are expected to diverge at $\omega_{DQD}=\omega_{0}$ and/or
$\omega_{DQD}=2\omega_{0}$\cite{perturbation}. However, giving a simple
analytic criterion for the regime of validity of our development is very
complex because of the many parameters involved in the problem and because
these parameters occur in the system description through complicated
functional dependences (see the expressions of $\chi_{2}$, $\lambda_{2}$,
$U_{cl}$, $U_{q}$, $\chi_{4}$, $\lambda_{4}$, $U_{4}$, $V_{4}$ and $W_{4}$).
Alternatively, one can check that the next-orders mesoscopic correlators in
$g^{6}$ and $g^{8}$ are negligible. This is discussed in details in Appendix
H. We have checked that we remain on the safe side with the parameters used in
the present work. 

\subsection{The low coupling limit: squeezed photonic vacuum induced by a
double quantum dot\label{SQ}}

\subsubsection{Evaluation of the Lindblad equation coefficients to third order
in $g_{L}$}

\begin{figure}[ptb]
\includegraphics[width=1\linewidth]{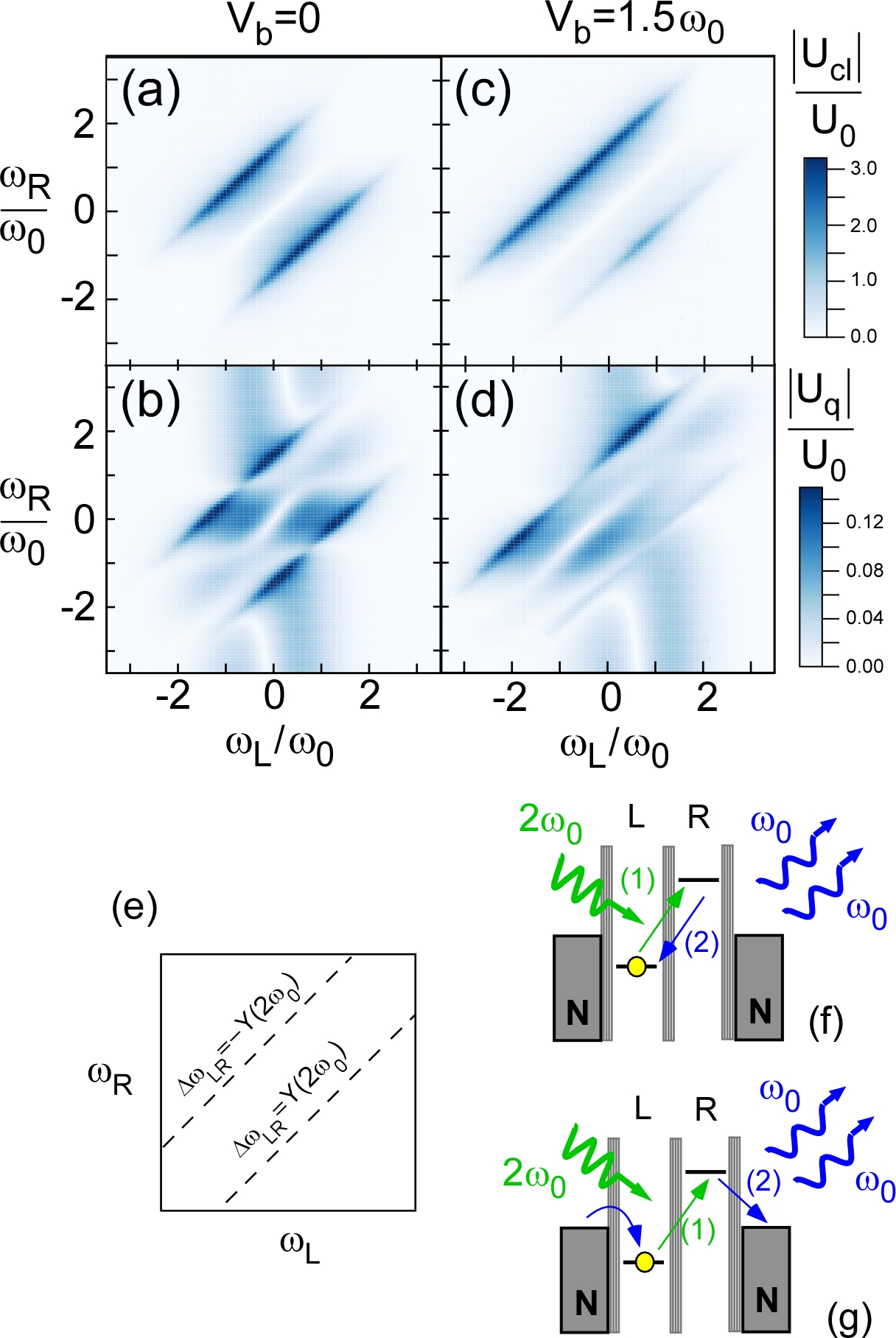}\newline\caption{{}Panels (a),
(b), (c) and (d): Absolute values of the coefficients $U_{cl}$ and $U_{q}$
which account for the effect of the $2\omega_{0}$ drive of the cavity at order
3 in the photon/dot coupling $g_{L}$, versus the dot orbital energies
$\omega_{L}$ and $\omega_{R}$. Panels (a) and (b) correspond to a bias voltage
$V_{b}=0$ and panels (c) and (d) to $eV_{b}=1.5\omega_{0}$. The other
parameters are $\Gamma=0.1\omega_{0}$, $t_{LR}=0.7\omega_{0}$, $k_{B}%
T=0.275\omega_{0}$, $g_{R}=0$, and $\Lambda_{0}=5.10^{-5}\omega_{0}$. We use a
normalization factor $U_{0}=g_{L}^{3}\beta_{p}/\omega_{0}^{2}$. Panel (e)
indicates the positions of the two-photon resonances $\omega_{DQD}=2\omega
_{0}$ between the dot internal degree of freedom and the cavity, which are
obtained for $\Delta\omega_{LR}\simeq\pm R(2\omega_{0})$. Panels (f) and (g):
examples of coherent and dissipative processes in $g_{L}^{3}$ involving the
$2\omega_{0}$ drive, for $\Delta\omega_{LR}\simeq\pm R(2\omega_{0})$. When the
internal transition of the double dot matches $2\omega_{0}$, it can absorb a
$2\omega_{0}$ photon. This enables the emission of two $\omega_{0}$ photons
upon electronic transitions which are internal to the dot (panel (f)) or
involve the normal metal contacts (panel (g)).}%
\label{Fig1}%
\end{figure}We have seen above that $U_{cl}$ corresponds to a coherent
two-photon drive whereas $U_{q}$ corresponds to an unusual form of squeezing
dissipation. In this section, we evaluate these coefficients in the double dot
case. Figure \ref{Fig1} shows $\left\vert U_{cl}\right\vert $ and $\left\vert
U_{q}\right\vert $ versus the dot orbital energies $\omega_{L}$ and
$\omega_{R}$, for moderate tunnel rates $\Gamma=0.1\omega_{0}$ and a moderate
interdot hopping $t_{LR}=0.1\omega_{0}$. We use a zero bias voltage in panels
(a) and (b) and a nonzero bias voltage $V_{b}=1.5\omega_{0}$ in panels (c) and
(d). Both $U_{cl}$ and $U_{q}$ show strong resonances which appear as diagonal
lines in Fig.\ref{Fig1}. These lines correspond to resonances of the cavity
with the double dot internal degree of freedom (see panel (e)). More
precisely, the bonding and antibonding states of the double dot, which result
from the tunnel coupling between the left and right orbitals, have energies
$\omega_{\mp}=(\omega_{L}+\omega_{R}\mp\sqrt{\Delta\omega_{LR}^{2}+4t_{LR}%
^{2}})/2$ with $\Delta\omega_{LR}=\omega_{L}-\omega_{R}$ the dots orbital
detuning. In principle, single-photon resonances $\omega_{DQD}=\omega_{0}$,
with $\omega_{DQD}=\omega_{+}-\omega_{-}$ by definition, are expected for
$\Delta\omega_{LR}=\pm R(\omega_{0})$ with $R(\omega_{0})=\sqrt{\omega_{0}%
^{2}-4t_{LR}^{2}}$, and two-photon resonances $\omega_{DQD}=2\omega_{0}$ are
expected for $\Delta\omega_{LR}=\pm R(2\omega_{0})$. In Fig.\ref{Fig1}, only
the two-photon resonances are visible because we use $2t_{LR}>\omega_{0}$ and
therefore the condition $\Delta\omega_{LR}=\pm R(\omega_{0})$ can never be
satisfied. Panels (f) and (g) show some examples of two-photon processes which
are expected to contribute to the resonances at $\Delta\omega_{LR}=\pm
R(2\omega_{0})$. A photon with frequency $2\omega_{0}$ can be converted into
two-photons with frequency $\omega_{0}$, in tunneling sequences which can be
either purely coherent (panel (f)) or dissipative (panel (g)). Interestingly,
the gate voltage area where the two-photon resonances appear is modified when
a nonzero bias voltage is used (panels (c) and (d)). This is because the third
order processes such as the one of panels (f) and (g) require that the double
dot bonding and antibonding states are occupied and empty respectively, and
the transport processes induced by a nonzero $V_{b}$ modify the occupation of
these states. Therefore using a nonzero bias voltage can be useful to trigger
two-photon processes, especially in case of weak tunability of $\omega_{L(R)}%
$, which can happen for some types of quantum dots. Interestingly, $\left\vert
U_{q}\right\vert $ also shows broad vertical resonances (for $\omega_{L}$
constant) outside of the gap between the $\Delta\omega_{LR}=R(2\omega_{0})$
and $\Delta\omega_{LR}=-R(2\omega_{0})$ resonances (see panels (b) and (d)).
These resonances are due to tunneling between the left dot and the left
reservoir, due to the conditions $g_{L}\neq0$ and $\Gamma\neq0$. As expected,
these resonances shift with $V_{b}$ (compare panels (b) and (d)) and get
thinner when $\Gamma$ decreases (not shown). The transition between the right
reservoir and the right dot is not directly coupled to the cavity since
$g_{R}=0$, but a broad horizontal resonance also appears in Fig.\ref{Fig1}b
between the lines $\Delta\omega_{LR}=R(2\omega_{0})$ and $\Delta\omega
_{LR}=-R(2\omega_{0})$ because the hybridization between the left and right
orbitals enables tunneling to the right reservoir. Note that the horizontal
and vertical resonances induced by the presence of the normal metal reservoirs
are visible in $\left\vert U_{q}\right\vert $ but not in $\left\vert
U_{cl}\right\vert $. This can be explained by the fact that tunneling to the
normal metal reservoirs is a stochastic effect which impacts more directly the
dissipative processes in $\gamma_{p}$ (or $U_{q}$) than the coherent drive in
$\rho_{p}$ generated by $U_{cl}$.

\subsubsection{Stationary Wigner function of the cavity to third order in
$g_{L}$\label{SW}}

To characterize the effects of the terms in $U_{cl}$ and $U_{q}$, we now
calculate analytically the stationary cavity Wigner function which follows
from Eq.(\ref{MElin}) to third order in\textbf{ }$g$, i.e. assuming that the
terms in $K$, $K_{loss}$, $K_{gain}$ and $D$ are negligible. The cavity Wigner
function can be defined quite generally as
\begin{equation}
W(\alpha,\alpha^{\ast},t)=\frac{1}{\pi^{2}}\int d^{2}\beta e^{(\beta^{\ast
}\alpha-\alpha^{\ast}\beta)}\left\langle e^{\beta\hat{a}_{I}^{\dag}-\beta
_{I}^{\ast}\hat{a}_{I}}\right\rangle _{t}\text{.}\label{Wdef}%
\end{equation}
Following the method of Ref.\cite{Millburn}, one can show that Eq.
(\ref{MElin}) leads to the evolution equation%
\begin{align}
\frac{\partial}{\partial t}W &  =\left(  -i\Delta\omega_{0}\left[
\frac{\partial}{\partial\alpha^{\ast}}\alpha^{\ast}-\frac{\partial}%
{\partial\alpha}\alpha\right]  \right)  W\label{Wevol}\\
&  +\left(  \frac{\gamma_{+}}{2}\frac{\partial}{\partial\alpha}\frac{\partial
}{\partial\alpha^{\ast}}+\frac{\gamma_{-}}{2}\left(  \frac{\partial}%
{\partial\alpha}\alpha+\frac{\partial}{\partial\alpha^{\ast}}\alpha^{\ast
}\right)  \right)  W\nonumber\\
&  -\left(  2\rho_{p}\frac{\partial}{\partial\alpha}\alpha^{\ast}+2\rho
_{p}^{\ast}\frac{\partial}{\partial\alpha^{\ast}}\alpha\right)  W\nonumber\\
&  -\gamma_{p}\left(  \frac{e^{-i\varphi_{p}}}{2}\frac{\partial^{2}}%
{\partial\alpha^{\ast2}}+\frac{e^{i\varphi_{p}}}{2}\frac{\partial^{2}%
}{\partial\alpha^{2}}\right)  W\nonumber
\end{align}
(see details in Appendix E). The term in $\gamma_{p}$ in
Eq.(\ref{Wevol}) describes a squeezing dissipation similar to
Refs.\cite{Didier:2014,Lu:2015}.  In the stationary regime, the
solution of this equation is:%
\begin{equation}
W(\alpha,\alpha^{\ast},t\rightarrow+\infty)=\frac{1}{\pi\sqrt{A^{2}%
-4\left\vert B\right\vert ^{2}}}\exp\left(  \frac{P}{A^{2}-4\left\vert
B\right\vert ^{2}}\right)  \label{Wfinal}%
\end{equation}
with%
\begin{equation}
P=A\left\vert \alpha\right\vert ^{2}+B^{\ast}\alpha^{2}+B\alpha^{\ast
2}\label{PP}%
\end{equation}
and, to third order in\textbf{ }$g$ and first order in $\varepsilon_{p}$,%
\begin{equation}
A=-\gamma_{+}/2\gamma_{-}\label{Afull}%
\end{equation}
and%
\begin{equation}
B=\left(  \rho_{p}\frac{\gamma_{+}}{\gamma_{-}}-\gamma_{p}\frac{e^{i\varphi
_{p}}}{2}\right)  /\left(  \gamma_{-}+2i\Delta\omega_{0}\right)
\text{.}\label{Bfull}%
\end{equation}
Equation (\ref{Wfinal}) describes a squeezed cavity vacuum. The major axis of
the squeezed Gaussian is tilted by an angle $\theta=\arg[B]/2$ from the
$\operatorname{Re}[\alpha]$ axis. The fields quadratures along the $\theta$
and $\theta+\pi/2$ angles have the variances $\Delta X_{\pm}=\sqrt
{-(A/2)\pm\left\vert B\right\vert }$. Strinkingly, from Eq.(\ref{Bfull}), the
coherent drive in $\rho_{p}$ and the dissipation processes in $\gamma_{p}$ can
both contribute to cavity squeezing and interfere constructively or
destructively depending on the value of the phase $\varphi_{p}$.
Note that expression (\ref{Wfinal}) is valid for any type of
mesoscopic circuit with internal degrees of freedom coupled to the cavity
electric field, as long as (\ref{MElin}) can be treated to third order
in\textbf{ }$g$. In Appendix F, we study in more details the
influence of the double dot parameters on the photonic squeezing.
Note that squeezing has already been found in various mesoscopic
QED configurations\cite{Mendes:2015,Mendes:2016,Grimsmo:2016}.

\subsection{Photonic Schr\"{o}dinger cat states produced by a double quantum
dot\label{cat}}

Obtaining Schr\"{o}dinger cat states is useful to study the quantum behavior
of a device on a fundamental level as well as to develop quantum computers. To
obtain such states with our device, we need to invoke the fourth order terms
in $g_{L}$ of Eqs. (\ref{H4}) or (\ref{P4}), which will generate
multistability in the cavity behavior. For simplicity, we will perform the
study of this situation in the particular case where the system dynamics
can be described by a Lindblad equation. This limit
presents the advantage of remaining formally simple while demonstrating
interesting potentialities of Mesoscopic QED.

\subsubsection{Double dot correlation functions to fourth order in $g_{L}%
$\label{Correl4}}

\begin{figure}[ptb]
\includegraphics[width=1\linewidth]{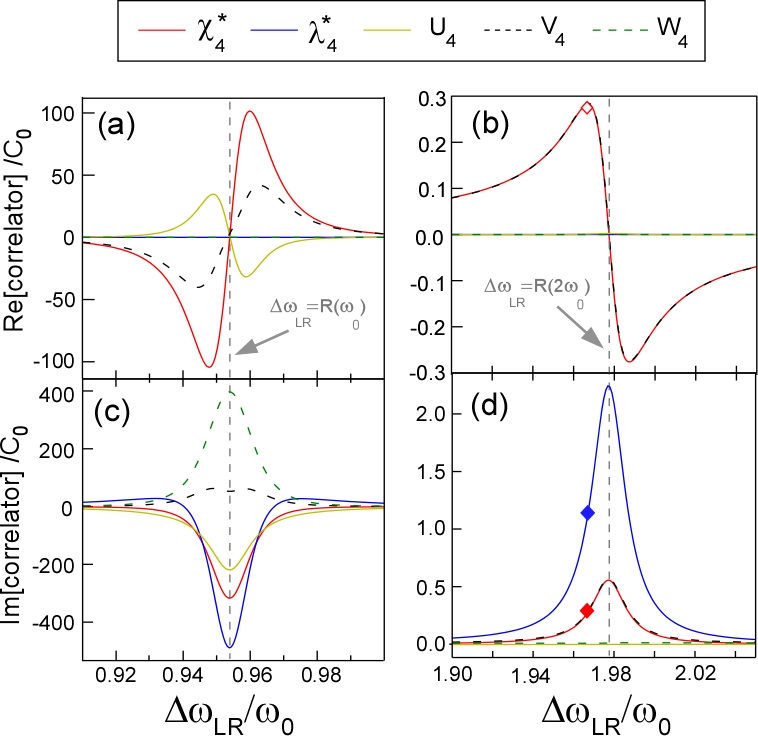}\newline\caption{Fourth order
electronic correlation functions versus $\Delta\omega_{LR}$ calculated for
$\omega_{av}=(\omega_{L}+\omega_{R})/2=0.989\omega_{0}$, $\Gamma
=0.01\omega_{0}$, $t_{LR}=0.15\omega_{0}$, $k_{B}T=0.3\omega_{0}$, $eV_{b}=0$,
$g_{R}=0$, $\beta_{p}=0.35$, and $\Lambda_{0}=10^{-4}\omega_{0}$. Panels (a)
and (b) show the real parts of the correlators and panels (c) and (d) the
imaginary parts. The left panels show the area $\Delta\omega_{LR}\sim
R(\omega_{0})$ (which implies $\omega_{DQD}\sim\omega_{0}$) whereas the right
panels show $\Delta\omega_{LR}\sim R(2\omega_{0})$ (which implies
$\omega_{DQD}\sim2\omega_{0}$). All correlation functions are normalized by
$C_{0}=g_{L}^{4}/\omega_{0}^{3}$. The full and empty diamonds correspond to
reference points for a comparison with Fig.\ref{Fig6}. The mapping condition
(\ref{Cmark}) is satisfied when the red full lines and black dashed lines
coincide in the top and bottom panels ($\chi_{4}^{\ast}=V_{4}$), the green
dashed line is close to $0$ in both panels ($W_{4}=0$) and the yellow line is
close to zero in the top panel ($\operatorname{Re}[U_{4}]=0$). This is true
for panels (b) and (d).}%
\label{Fig5}%
\end{figure}

\begin{figure}[ptb]
\includegraphics[width=1\linewidth]{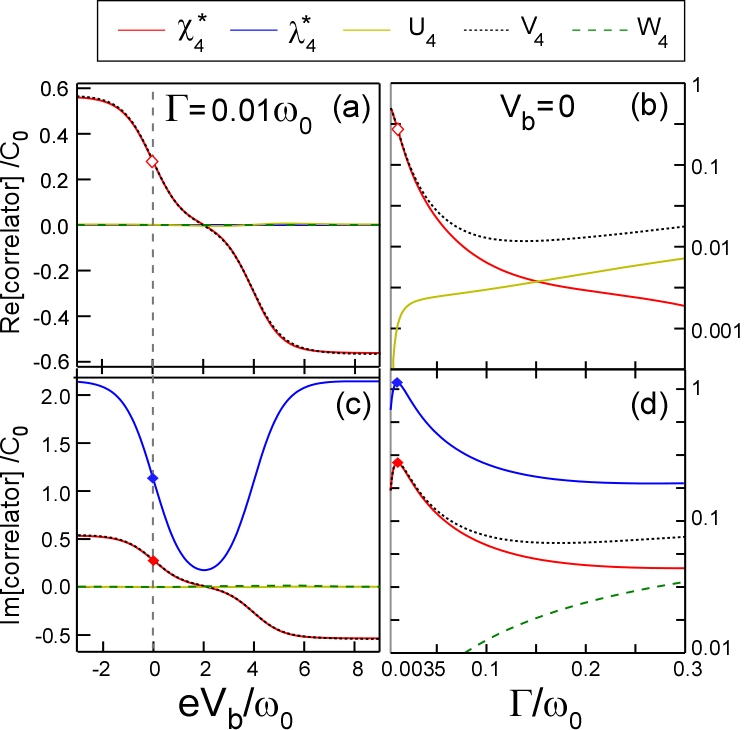}\newline\caption{Fourth order
electronic correlation functions versus $V_{b}$ for $\Gamma=0.01\omega_{0}$
[panels (a) and (c)] and versus $\Gamma$ for $V_{b}=0$ [panels (b) and (d)].
We use $\omega_{av}=0.989\omega_{0}$ and $\Delta\omega_{LR}=1.967\omega_{0}$.
The other parameters are the same as in Fig.\ref{Fig5}. The diamonds
correspond to reference points identical to those of Fig.\ref{Fig5}.}%
\label{Fig6}%
\end{figure}

\begin{figure}[ptb]
\includegraphics[width=1\linewidth]{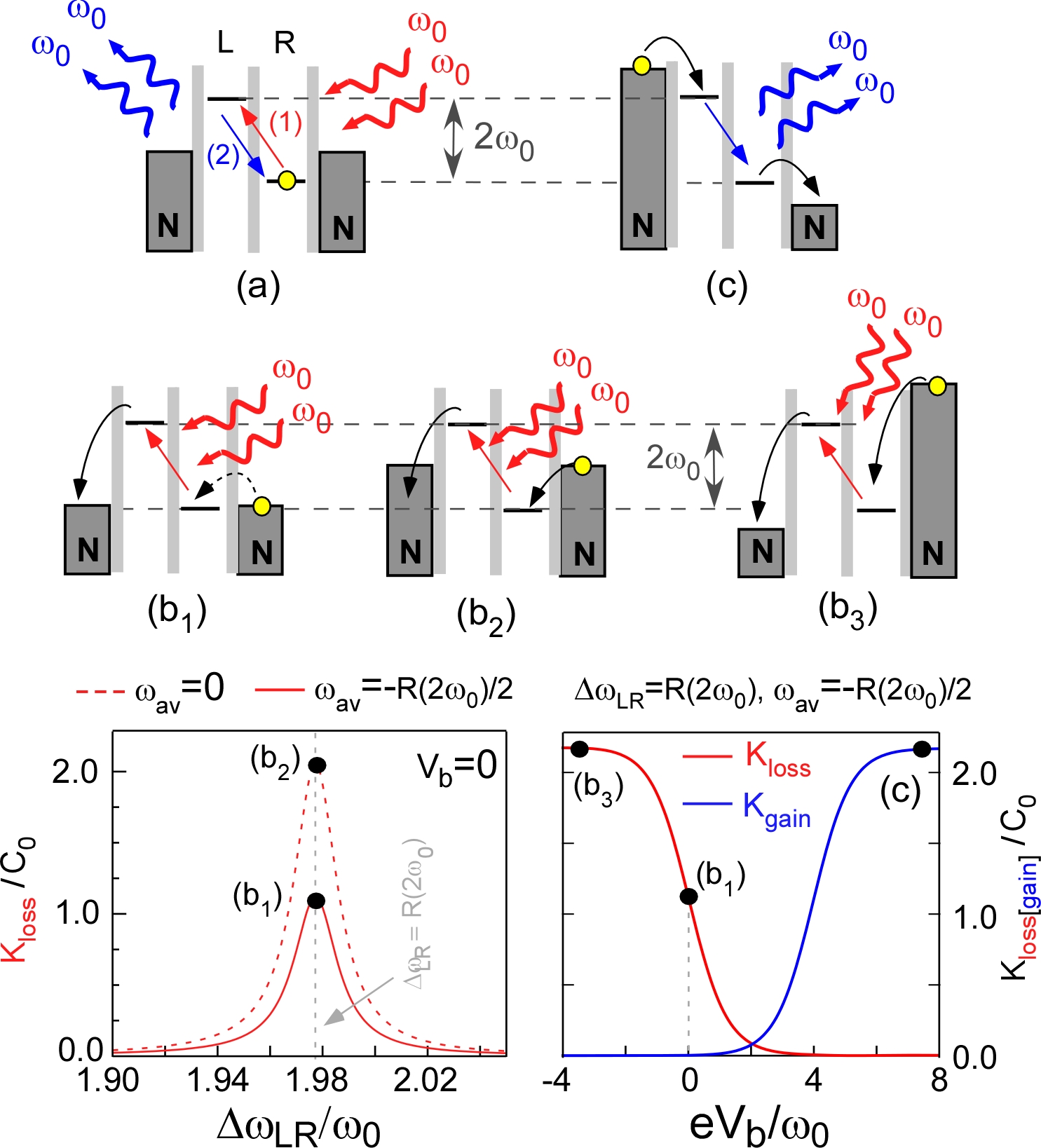}\newline\caption{Example of
processes at fourth order in the light/matter coupling $g_{L}$. Panel (a)
shows a fully coherent process which involves only the internal transition of
the double dot and can contribute to the term in $K$. Panels (b$_{\mathrm{1}}%
$), (b$_{\mathrm{2}}$), (b$_{\mathrm{3}}$) show processes which involve
irreversible tunneling to the normal metal reservoirs and contribute to
$K_{loss}$ for different configurations of dot orbital energies and bias
voltage. Panel (c) shows a process which contributes to $K_{gain}$ in the
presence of a finite bias voltage. The left bottom plot shows $K_{loss}$
versus $\omega_{LR}$ for two difference values of $\omega_{av}$, i.e.
$\omega_{av}=R(2\omega_{0})/2$ (full red line) and $\omega_{av}=0$ (dashed red
line). The right bottom plot shows $K_{loss}$ (red full line) and $K_{gain}$
(blue full line) versus $V_{b}$ for $\omega_{av}=R(2\omega_{0})/2$. The same
parameters as in Figs.\ref{Fig5} and \ref{Fig6} are used.}%
\label{Fig10}%
\end{figure}

In the double dot case, can the  Lindblad description 
\ hold to fourth order in $g_{L}$, or equivalently, can the condition
$\mathcal{C}_{Ldb}$ of Eq.(\ref{Cmark}) be satisfied? To answer this question,
we show in Fig.\ref{Fig5} the dependence of the coefficients $\chi_{4}$,
$\lambda_{4}$, $U_{4}$, $V_{4}$ and $W_{4}$ on $\Delta\omega_{LR}$, for a zero
bias voltage ($V_{b}=0$) and low tunnel rates ($\Gamma=0.01\omega_{0}$).
Figures \ref{Fig5}a and \ref{Fig5}c show that $\mathcal{C}_{Ldb}$ is not true
when the double dot is resonant with the cavity ($\omega_{DQD}=\omega_{0}$
i.e. $\Delta\omega_{LR}\sim R(\omega_{0})$). This is not surprising, because,
in this case, real energy exchanges between the double dot and the cavity are
possible, leading to vacuum Rabi oscillations in the case of low $\Gamma$ and
$\Lambda_{0}$. Hence, for the cavity, the mesoscopic circuit represents a
\textquotedblleft bath with memory\textquotedblright, which is incompatible
with an effective Markovian dynamics. Another interesting regime is
$\omega_{DQD}=2\omega_{0}$ i.e. $\Delta\omega_{LR}\sim\pm R(2\omega_{0})$,
because the electronic correlation functions in $g_{L}^{4}$ present resonances
in this area, as already seen for $U_{cl}$ and $U_{q}$ to third order in
$g_{L}$. The Lindblad  condition (\ref{Cmark}) is satisfied for
$\Delta\omega_{LR}\sim\pm R(2\omega_{0})$ with small values of $\Gamma$ and
$t_{LR}$, and $V_{b}=0$ (see Figs.\ref{Fig5}b and d) as well as a nonzero
$V_{b}$ (see Figs.\ref{Fig6}a and \ref{Fig6}b). More generally, 
the Lindblad condition $\mathcal{C}_{Ldb}$ is satisfied when the cavity and
double dot are off-resonant (for single photon exchange) and the dot-lead and
dot-dot couplings weak enough \textbf{(}$\Gamma,t\ll\omega_{0}-\omega_{DQD}%
$\textbf{). \ } \ One may attribute this result to the fact that, in this regime,
there can only be virtual energy exchanges between the cavity and the double
dot, which occur on a timescale which is very short in comparison with the
typical timescale for the evolution of the cavity. The condition
$\mathcal{C}_{Ldb}$ is not valid anymore for higher tunnel rates
$\Gamma>0.1\omega_{0}$ (see Figs.\ref{Fig6}b and \ref{Fig6}d). Indeed, in this
case the resonances at $\Delta\omega_{LR}\sim R(\omega_{0})$ and $\Delta
\omega_{LR}\sim R(2\omega_{0})$ start overlapping and the distinction between
real and virtual energy exchanges between the cavity and the double dot
becomes less clear. The  condition $\mathcal{C}_{Ldb}$
is not valid either for $\Delta\omega_{LR}\sim R(2\omega_{0})$
and $t_{LR}$ large ($t_{LR}>0.3\omega_{0}$) (not shown). This is why, in the
rest of this section, we will focus on the Lindbladian dynamics of the cavity
for $\Delta\omega_{LR}\sim R(2\omega_{0})$, $t_{LR}\lesssim0.15\omega_{0}$ and
$\Gamma\lesssim0.1\omega_{0}$. Note that for $\Gamma\rightarrow0$, the
imaginary part of the correlators vanishes (see the very left of
Fig.\ref{Fig6}d for the onset of this effect). Since we are interested in the
effect of a genuinely dissipative mesoscopic circuit, we will only consider
the case $\Gamma\geq0.005\omega_{0}$ in the following. In particular, we will
consider the working point $\Gamma\simeq0.01\omega_{0}$ where $\left\vert
\operatorname{Im}[\chi_{4}]\right\vert $ and $\left\vert \operatorname{Im}%
[\lambda_{4}]\right\vert $ have a local maximum (see very left of
Fig.\ref{Fig6}d). Figure \ref{Fig10} represents some possible photonic
processes at fourth order in $g_{L}$ in this limit (see panels (a),
(b$_{\mathrm{1}}$), (b$_{\mathrm{2}}$), (b$_{\mathrm{3}}$) and (c)), for
different configurations of dot orbital energies and bias voltage. It also
shows $K_{loss}$ and $K_{gain}$ versus $\Delta\omega_{LR}$ and $V_{b}$ for the
parameters of Fig.\ref{Fig5} and Fig.\ref{Fig6} with $\Gamma=0.01\omega_{0}$
and $\Delta\omega_{LR}=R(2\omega_{0})$. In these conditions, one can check
that for $V_{b}=0$, the two-photon stochastic dissipation rate $K_{loss}$ is
the dominant stochastic rate in Eq.(\ref{P3}), i.e. $K_{gain}$, $D$,
$\gamma_{loss}$ and $\gamma_{gain}$ are much weaker. The rate $K_{loss}$
corresponds to the type of processes represented in Fig.\ref{Fig10}, panels
(b$_{\text{1}}$) and (b$_{\mathrm{2}}$), where two-photons can be absorbed
simultaneously by the double dot circuit because the double dot is resonant
with $2\omega_{0}$, and this absorption is made irreversible by the presence
of the normal metal reservoirs. The working point $\omega_{av}=0$ and
$\Delta\omega_{LR}=R(2\omega_{0})$ corresponds to a maximal $K_{loss}$ for
$V_{b}=0$ (see point (b$_{\mathrm{2}}$)). For comparison, in the configuration
of (b$_{\text{1}}$), $K_{loss}$ is weaker because the filling of the lower dot
level is less efficient. Remarkably, a nonzero $V_{b}$ can be used to obtain a
nonzero $K_{gain}$ and change the relative values of $K_{loss}$ and $K_{gain}$
(see bottom right panel of Fig.\ref{Fig10}). For $V_{b}<0$, $K_{loss}$
increases because the filling of the lower dot level and/or the emptying of
the upper dot level by the normal metal reservoirs becomes more efficient and
this enhances the \textquotedblleft reset\textquotedblright\ of the double dot
between two-photon pair absorption processes (Fig.\ref{Fig10}b$_{\text{3}}$).
For $V_{b}>0$ and sufficiently large, the filling of the upper dot level and
emptying of the lower dot level are favored, which causes photon pair emission
processes (see Fig.\ref{Fig10}c) while $K_{loss}$ vanishes. In this limit, the
emission of photon pairs is obtained without any need for an ac cavity
excitation ($\varepsilon_{p}=0$) because the mesoscopic bias in $V_{b}$
provides the energy for this process. The Kerr interaction $K$, which
corresponds to the processes of Fig.\ref{Fig10}a, varies like
$\operatorname{Re}[\chi_{4}]$ which is represented in Figs.\ref{Fig5} and
\ref{Fig6}. Strinkingly, for $V_{b}=0$, $K$ cancels at $\Delta\omega
_{LR}=R(2\omega_{0})$ where $K_{loss}$ is maximal (see Figs.\ref{Fig5}b and
\ref{Fig5}b). Importantly, in all these plots, the order of magnitude of
$K_{loss}$, $K_{gain}$ and $K$ is given by the constant $C_{0}=g_{L}%
^{4}/\omega_{0}^{3}$. Using the typical value $\omega_{0}=2\pi\times
5~\mathrm{GHz}$ and the ratio $g_{L}=0.125$ which is strong but experimentally
feasible, in principle (see section IV.C), one finds $C_{0}=2\pi
\times1.2~\mathrm{MHz}$. We will see in next sections that this is sufficient
to obtain sizeable non-linear signatures in the cavity response.

\subsubsection{Average photon number\label{avN}}

Before studying the full quantum behavior of the cavity through the Wigner
function $W$, it is useful to study the mean value of $\left\langle \hat
{a}\right\rangle $ which can be expressed analytically. This can reveal a
multistable behavior which is expected for driven nonlinear systems
\cite{Millburn} and which will be useful to obtain photonic Schr\"{o}dinger
cats. From the Lindblad equation (\ref{ME}) with the fourth order terms
(\ref{H4}) and (\ref{P4}) included and $\left\langle \hat{a}\right\rangle
=\alpha_{av}e^{-i\omega_{0}t}$, one gets
\begin{equation}
U_{cl}\alpha_{av}^{\ast}-\left(  \frac{\Lambda_{0}+\Delta\Lambda_{0,4}}%
{2}+i\chi_{2}+2i\chi_{4}\left\vert \alpha_{av}\right\vert ^{2}\right)
\alpha_{av}=0 \label{aav}%
\end{equation}
with
\begin{equation}
\Delta\Lambda_{0,4}=\operatorname{Im}[\lambda_{4}-4(\chi_{4}+U_{4})]
\end{equation}
the renormalization of the cavity linewidth to fourth order in $g_{L}$.
Equation (\ref{aav}) bears similarities with the result given by semiclassical
approaches (see Appendix B), but the term $\Delta\Lambda_{0,4}$ is specific to
a full quantum-mechanical treatment. The\ solution $\alpha_{av}=0$ is obvious.
However, in principle, Eq.(\ref{aav}) can also give nonzero values of
$\alpha_{av}=\alpha_{av}^{\pm}$ given by\begin{figure}[ptb]
\includegraphics[width=1.\linewidth]{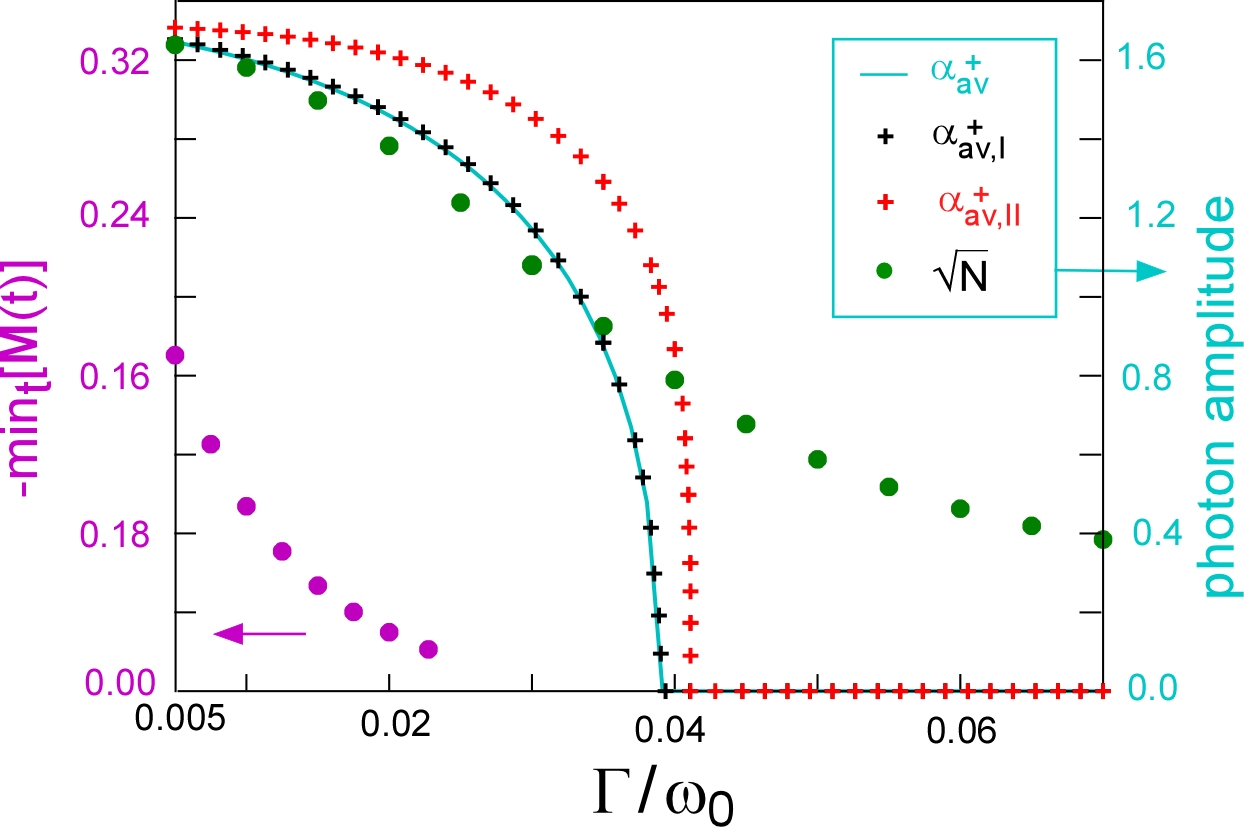}\newline\caption{Various
characteristics of the cavity response versus the tunnel rate to the normal
metal reservoirs $\Gamma$ in the presence of the cavity drive in
$\varepsilon_{p}$ treated at fourth order in $g_{L}$. We use $\omega
_{av}=0.989\omega_{0}$, $g_{L}=0.125\omega_{0}$ and $\Delta\omega
_{LR}=R(2\omega_{0})\simeq1.978\omega_{0}$. The other parameters are the same
as in Fig.\ref{Fig5}. The full cyan line, the black crosses and the red
crosses show the semiclassical photon amplitudes $\alpha_{av}^{+}$ and its
approximations $\alpha_{av,I}^{+}$ and $\alpha_{av,II}^{+}$ of
Eqs.(\ref{alphasc}), (\ref{alphPapp}) and (\ref{alphaCrude}), respectively.
The green dots show the square root of the average photon\ number $N$ in the
cavity in stationary conditions, obtained from Eq.(\ref{MElin}). The magenta
dots show the maximum negativity of the Wigner function over time $t$ and the
quadratures $\alpha,\alpha^{\ast}$ for the protocol discussed in section
\ref{Wnum} where the cavity drive in switched on suddenly.}%
\label{Fig9}%
\end{figure}%
\begin{equation}
\alpha_{av}^{\pm}=\frac{1}{\left\vert \chi_{4}\right\vert }\sqrt
{\frac{-\operatorname{Re}[\chi_{2}^{ren}\chi_{4}^{\ast}]\pm\sqrt{\Delta}}{2}}
\label{alphasc}%
\end{equation}
with
\begin{equation}
\Delta=\left\vert \chi_{4}\right\vert ^{2}\left\vert U_{cl}\right\vert
^{2}-\operatorname{Im}\left[  \chi_{2}^{ren}\chi_{4}^{\ast}\right]  ^{2}%
\end{equation}
and $\chi_{2}^{ren}=\chi_{2}-i(\Lambda_{0}+\Delta\Lambda_{0,4})/2$.
Importantly, $\alpha_{av}$ must be real. Hence, from Eq.(\ref{alphasc}) for
low amplitudes of $\beta_{p}$, the only possible solution is $\alpha_{av}=0$
since $\Delta<0$. For a stronger drive ($\left\vert U_{cl}\right\vert
>\left\vert \operatorname{Im}\left[  \chi_{2}^{ren}\chi_{4}^{\ast}\right]
/\chi_{4}\right\vert $), $\Delta$ becomes positive. Then, the comparison
between $\sqrt{\Delta}$ and $\pm\operatorname{Re}[\chi_{4}^{\ast}\chi
_{2}^{ren}]$ sets whether there are 0, 1 or 2 values of $\alpha_{av}$ allowed
by Eq.(\ref{alphasc}). Finally, two values for $\alpha_{av}$ are possible for
each value of $\varphi_{av}$, i.e.%
\begin{equation}
\varphi_{av}^{\pm}=-\frac{1}{2}\arg\left[  \frac{i\chi_{2}^{ren}+2i\chi
_{4}\alpha_{av}^{\pm~2}}{U_{cl}}\right]  +n\pi
\end{equation}
with $n\in\{0,1\}$. In some cases, we find that $\alpha_{av}^{+}$ and
$\alpha_{av}^{-}$ can be both solution to Eq. (\ref{aav}). However, for
simplicity, we focus below on the situation of moderate interdot hopping
($t_{LR}=0.15\omega_{0}$), moderate tunnel rates ($0.005\omega_{0}\leq
\Gamma\leq0.1\omega_{0}$) and a zero bias voltage ($V_{b}=0$), where one has
typically a single nonzero solution $\alpha_{av}^{+}$. In particular, for the
parameters considered in Fig. \ref{Fig9}, one has $\left\vert U_{cl}%
\right\vert \gg\left\vert U_{q}\right\vert $, $K=\operatorname{Re}[\chi
_{4}]\ll-\operatorname{Im}[\chi_{4}]$ and $\operatorname{Im}[\chi_{4}]<0$.
Therefore, one has $\alpha_{av}^{+}\simeq\alpha_{av,I}^{+}$ with%
\begin{equation}
\alpha_{av,I}^{+}=\sqrt{\frac{\operatorname{Im}[\chi_{2}^{ren}]+\sqrt
{\left\vert U_{cl}\right\vert ^{2}-\operatorname{Re}\left[  \chi_{2}%
^{ren}\right]  ^{2}}}{-2\operatorname{Im}[\chi_{4}]}} \label{alphPapp}%
\end{equation}
This quantity is represented with black crosses in Fig.\ref{Fig9}, and is in
excellent agreement with the exact $\alpha_{av}^{+}$ represented with a cyan
line. Equation (\ref{alphPapp}) shows the crucial role of the two-photon
dissipation provided by the term in $\operatorname{Im}[\chi_{4}]$ for the
creation of nonzero photon states (if one had $\left\vert \chi_{4}\right\vert
\rightarrow0$, $\alpha_{av}^{+}$ would diverge and thus become physically
irrelevant). Of course, it is also necessary to have a high enough $U_{cl}$. A
crudest approximation is obtained by using $\chi_{2}^{ren}=0$, which yields
\begin{equation}
\alpha_{av,II}^{+}=\sqrt{\frac{2\left\vert \rho_{p}\right\vert }%
{K_{loss}-K_{gain}}} \label{alphaCrude}%
\end{equation}
(see red crosses in Fig.\ref{Fig9}). This expression shows well that the
nonzero $\alpha_{av}^{+}$ results from a balance between two-photon coherent
injection and two-photon dissipation. In contrast, the effect of the
Hamiltonian Kerr term $K$ is negligible in Fig.\ref{Fig9}. The comparison
between $\alpha_{av,I}^{+}$ and $\alpha_{av,II}^{+}$ shows that the
single-photon processes described by $\chi_{2}^{ren}$ slightly decrease the
amplitude of $\alpha_{av}^{+}$ and the range of $\ \Gamma$ for which cavity
bistability is obtained. Note that in principle, one has to study the
stability of the $\alpha_{av}^{\pm}$ solutions to determine their relevance.
We will omit such a study because the cavity Wigner function calculated in
section \ref{Wnum} can provide this information for the regime we are
interested in.

\begin{figure}[ptb]
\includegraphics[width=1\linewidth]{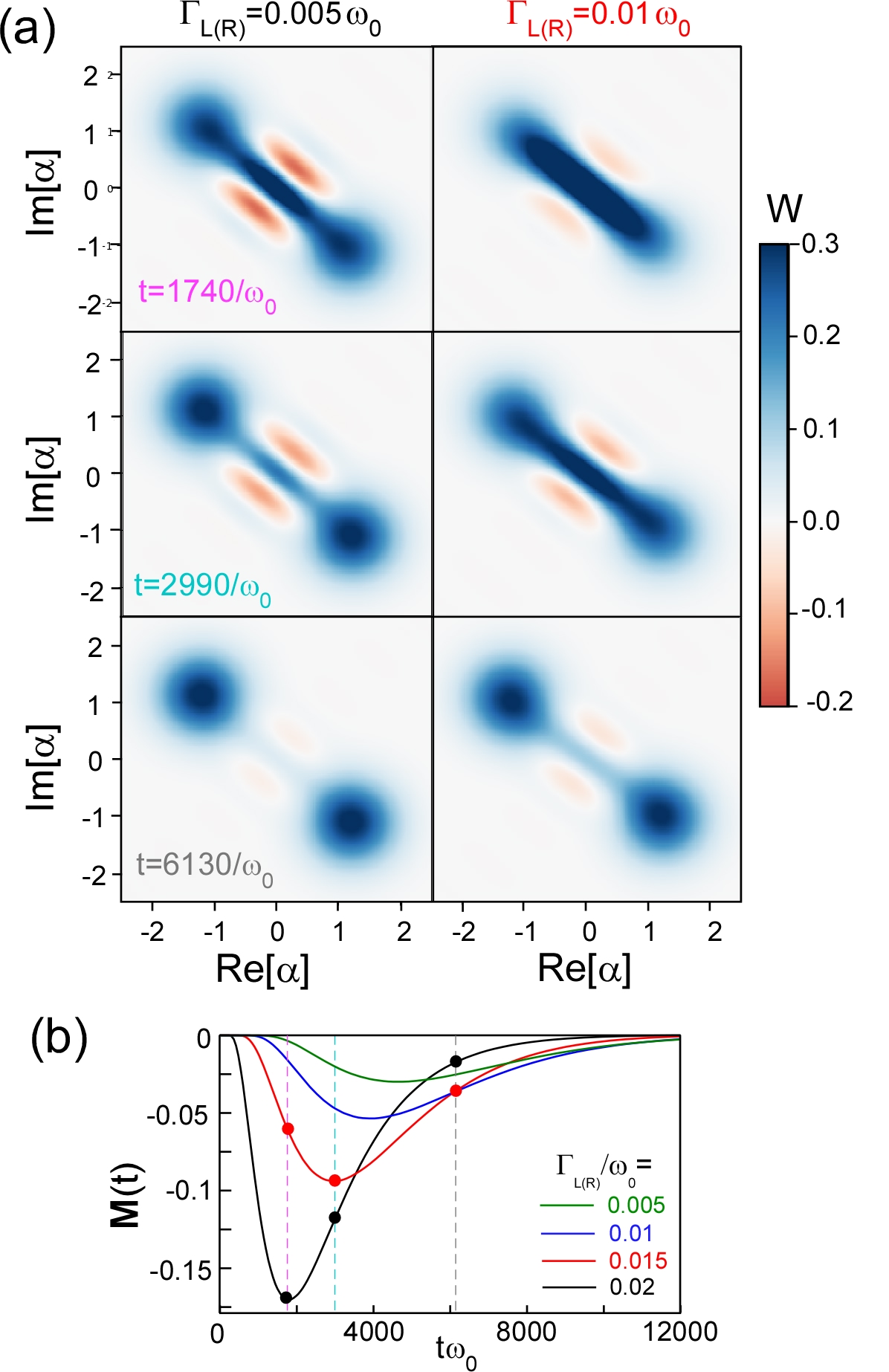}\newline\caption{{}Panel (a) and
(b): Wigner function $W$ of the cavity for tunnel rates $\Gamma=0.005\omega
_{0}$ and $\Gamma=0.01\omega_{0}$ and different times $t$ after switching on
the cavity drive in $\varepsilon_{p}$ ($t\omega_{0}=1740$, $2990$, $6130$ from
top to bottom). The other parameters are the same as in Fig.\ref{Fig9}.
\ Panel (c): Minimum $\mathbf{M}(t)$ of the Wigner function $W$ over the field
quadratures, versus $t$, for th same protocole as in panels (a) and (b), and
different tunnel rates. The black and red curves correspond to panels (a) and
(b) respectively.  Panel (d): Relaxation of $\mathbf{M}(t)$ versus
time, starting from the initial state shown in panel (a) at $t=1740/\omega
_{0}$, for different values of $\Gamma$ .}%
\label{Fig8}%
\end{figure}

\subsubsection{Cavity Wigner function to fourth order in $g_{L}$ in
non-stationary conditions\label{Wnum}}

So far, we have studied the cavity Wigner function $W$ in stationary
conditions. We now assume that the cavity is initially in the stationary
vacuum state obtained in the absence of the microwave drive ($\beta_{p}=0$).
We want to study the time evolution of $W$ when we switch on $\beta_{p}$ at
$t=0$. However, since we have derived the terms in $\beta_{p}$ in
Eq.(\ref{MElin}) in stationary conditions (see Eq.(\ref{eac}) and Appendix A),
one has to be careful about the validity of this equation which could be
jeopardized by the sudden rise of $\beta_{p}$. In fact, Eq. (\ref{MElin}) will
still be valid in the transient regime if we impose two constraints on the
rise time of $\beta_{p}$. On the one hand, we will assume that this rise time
is much longer that the correlation time $\sim1/\Gamma$ associated to
tunneling to the mesoscopic reservoirs, so that the terms $U_{cl}$ and $U_{q}$
in the cavity effective action can still be defined at any time from
Eqs.(\ref{Uclq}) and (\ref{Uqq}) with a prefactor $\beta_{p}$ which depends on
$t$. On the other hand, we will assume that the rise time of $\beta_{p}$ is
much faster than the cavity characteristic evolution time (visible in
Fig.\ref{Fig8}c). In these conditions, it is sufficient to use the Lindblad
equation (\ref{MElin}) with terms (\ref{H4}) and (\ref{P4}) which depend on
$\beta_{p}(t)=\beta_{p}\theta(t)$ with $\theta(t)$ the Heavidside function.

We compute $W(t)$ numerically by using the function \textquotedblleft
mesolve\textquotedblright\ from the qutip~package to solve Eq.(\ref{MElin}%
)\cite{qutip:2012}. For moderate tunnel rates, the cavity evolves towards a
coherent superposition of two coherent states (see Fig.\ref{Fig8}a a and b).
The nonclassicality of $W(t)$ is revealed by the red areas where $W(t)<0$. At
large times, there remains only two positive spots in the Wigner function,
which are approximately centered on the average values $\alpha_{av}%
^{+}e^{i\varphi_{av}^{+}}$ and $-\alpha_{av}^{+}e^{i\varphi_{av}^{+}}$
determined in section \ref{avN}. Therefore, these two solutions represent
cavity stable states in stationary conditions. Accordingly, we have checked
that the square root $\sqrt{N}$ of the average number $N=\left\langle \hat
{a}^{\dag}\hat{a}\right\rangle $ of photons in the cavity calculated
numerically for $t\rightarrow+\infty$ matches $\alpha_{av}^{+}$ when the
tunnel rate $\Gamma$ is small (see green dots in Fig.\ref{Fig9}). For higher
tunnel rates this is not the case anymore because $\alpha_{av}^{+}=0$ whereas
$W(t)$ corresponds to a squeezed vacuum.  From the case
$\Gamma/\omega_{0}=0.005$ (panel (a)) to the case $\Gamma/\omega_{0}=0.01$
(panel (b)), the semiclassical minima of the Wigner function $W(t)$ keep
approximately the same position in the quadratures space, because $\left\vert
\rho_{p}\right\vert $ and $K_{loss}$ are both approximately divided by two,
and because from Eq.(\ref{alphaCrude}) it is the ratio between $K_{loss}$ and
$\left\vert \rho_{p}\right\vert $ which roughly determines the value of
$\left\vert \alpha_{av}\right\vert $ (one has $\left\vert \rho_{p}\right\vert
=7.7~10^{-4}\omega_{0}$ and $K_{loss}=5.4~10^{-4}\omega_{0}$ in the black
case, and $\left\vert \rho_{p}\right\vert =3.7~10^{-4}\omega_{0}$,
$K_{loss}=2.7~10^{-4}\omega_{0}$ in the red case, and $K_{gain}$ is negligible
in both cases).

Figure \ref{Fig8}c represents the time evolution of the minimum negativity
$\mathbf{M}(t)=\min_{\alpha,\alpha^{\ast}}[W(t)]$ of the Wigner function
$W(t)$ over the fields quadratures $(\alpha,\alpha^{\ast})$. When
$\Gamma$ increases, the minimum of $\mathbf{M}(t)$ over time is reached later.
This is because, from the black curve ($\Gamma/\omega_{0}=0.005$) to the red
curve ($\Gamma/\omega_{0}=0.1$), the rate $K_{loss}$ which determines the
speed at which the system is attracted to its two semiclassical
minima\cite{Leghtas:2015}, is divided by two. Looking at panel (c), one could
believe that the Wigner function negativity relaxes faster in the black case,
but this is an impression which is due to the fact that the minimum of
$\mathbf{M}(t)$ is reached earlier. For a more rigorous comparison of
relaxation in the different cases, we have plotted, in panel (d),
$\mathbf{M}(t)$ for different values of $\Gamma$ when the cavity is
initialized in all cases in the state corresponding to panel (a), upper graph,
marked with a pink star, at time $t=1740/\omega_{0}$. One can see that the
relaxation of $\mathbf{M}(t)$ is similar in all cases, because this relaxation
is set by the value of $\gamma_{loss}^{0}$ (see Ref.\cite{Leghtas:2015}),
which varies only weakly from one case to the other (for instance, one has
$\gamma_{loss}^{0}=1.10~10^{-4}\omega_{0}$ and $\gamma_{loss}^{0}%
=1.16~10^{-4}\omega_{0}$ in the black and red cases, respectively). The facts
mentioned above that $\rho_{p}$ and $K_{loss}$ vary a lot from the black to
the red case whereas $\gamma_{loss}^{0}$ is approximately unchanged deserves
an explanation. The parameters $\rho_{p}$ and $K_{loss}$ describe two-photon
resonance effects and the system is tuned at the two-photon resonant point
$\omega_{DQD}=2\omega_{0}$ where $\rho_{p}$ and $K_{loss}$ present strong
resonances. They are thus very sensitive to the value of $\Gamma$ at this
working point. This is not the case for the parameter $\gamma_{loss}^{0}$
because it describes a single-photon effect which presents a resonance only at
$\omega_{DQD}=\omega_{0}$. 

Figure \ref{Fig9} shows with magenta dots the minimum negativity $\min
_{t}\left[  \mathbf{M}(t)\right]  $ of $W(t)$ over $\alpha,\alpha^{\ast}$ and
the time $t$, as a function of $\Gamma$. This quantity decreases more quickly
with $\Gamma$ than the amplitude of the semiclassical solution $\alpha
_{av}^{+}$. However, it is striking that a genuinely dissipative circuit such
as a double quantum dot circuit is able to induce non classical cavity states
thanks to the two-photon irreversible tunneling processes represented by
$K_{loss}$. Note that in Ref.\cite{Leghtas:2015}, a two-photon dissipation
term similar to $\rho_{p}$ and a two-photon drive term similar to $K_{loss}$
were obtained artificially by using an auxiliary cavity and two microwave
tones. Photonic Schr\"{o}dinger cats were obtained experimentally due to these
effects. In our case, a single drive at $2\omega_{0}$ and the inclusion of a
double dot in a single cavity are used to obtain these effects. For a typical
cavity frequency $\omega_{0}\sim2\pi\times5~\mathrm{GHz}$, the required tunnel
rates $\Gamma\sim0.01\omega_{0}$ correspond to $0.2~\mathrm{\mu eV}$, a value
which can be reached in practice\cite{Gustavsson:2006,Bruhat:2016b}. With the
simple protocol considered in this section, the photonic quantum superposition
survives for a duration of the order of $8000/\omega_{0}\simeq0.25~\mathrm{\mu
s}$ which is much longer than the time scale $1/\Gamma=100/\omega_{0}%
\simeq3~\mathrm{ns}$ associated to dissipative tunneling between the dots and
the normal reservoirs.

\section{Discussion}

In this section, we discuss our results in the light of various recent
References. Interestingly, Ref.\cite{Muller:2017} has proposed a
method to combine the Lindblad description of a cavity coupled to a DQD and
the Keldysh description of the dissipation provided by a bath of phonons
coupled to the DQD. The aim of this Ref. is to study photon emission in the
off resonant regime $\omega_{DQD}\neq\omega_{0}$ with the cavity driven at a
frequency $\omega_{d}\sim\omega_{0}$ and the DQD dc voltage biased. Two-photon
processes in $K$, $K_{loss}$, $K_{gain}$ and $D$ are disregarded. Although
this situation is physically different from the one we consider in section IV,
it is interesting to draw a technical comparison with our approach. In
Ref.\cite{Muller:2017}, the Keldysh framework is used to perform a
diagrammatic calculation of the phonon-induced rates in the effective Lindblad
equation of the cavity and DQD. This calculation is perturbative with respect
to both the cavity-DQD coupling and the DQD-phonon bath coupling. Besides, the
intrinsic cavity damping and the damping due to the fermionic leads of the DQD
are implicitly assumed to be very small, so that they do not enter in the
Keldysh diagrammatics and are added in the final Lindblad equation, as
independent terms. Consequently, some of the phonon-induced rates have a
denominator in $(\omega_{DQD}\pm\omega_{0})^{2}$, or $(\omega_{DQD}^{2}%
-\omega_{0}^{2})^{2}$, and thus diverge at $\omega_{DQD}=\omega_{0}$ (see
Eq.(22) and Fig.4 of Ref.\cite{Muller:2017}). A regularization of these
divergences by the system baths is missing and would require higher orders
perturbation series. In contrast, our approach is perturbative only in the
cavity-double dot coupling. We do not have divergences in our effective rates
at $\omega_{DQD}=\omega_{0}$ or $\omega_{DQD}=2\omega_{0}$ because these are
naturally regularized by the tunneling rate $\Gamma$ to the fermionic leads,
which appears in the mesoscopic Green's functions (\ref{Gmeso})-(\ref{mkdot}).
This is essential to depict situations such as the one considered in our
section IV.

In section \ref{Wnum}, we have considered a system working point $\Delta
\omega_{LR}=R(2\omega_{0})$ such that the effective Kerr nonlinearity of the
cavity $K$ vanishes, and the effective two-photon dissipation $K_{loss}$ and
two-photon drive in $\rho_{p}$ generate Schr\"{o}dinger cats in a transient
regime. Similarly, it has been shown experimentally with Josephson circuits
that the combination of $K_{loss}$ and $\rho_{p}$ enables the autonomous
preparation of Schr\"{o}dinger cat states\cite{Leghtas:2015,Touzard:2018}, but
also the protection of these cats again certain types of
decoherence\cite{Lescanne:2019}. This represents an important research
direction in the context of the development of quantum computing schemes which
require to fight calculation errors caused by decoherence. A cavity coupled to
a double quantum dot could represent an alternative way to implement this
"$K_{loss}$\&$\rho_{p}$" qubit scheme. Interestingly, the preparation and
protection of Schr\"{o}dinger cat states can also be obtained in Josephson
circuits by combing a Kerr nonlinearity $K$ with $\rho_{p}$%
\cite{Puri:2017,Grimm:2019,Puri:2019}. In our device, the required $K$ can be
obtained simultaneously with the two-photon loss $K_{loss}$ ($K\neq0$ and
$K_{loss}\neq0$), or almost separately ($K\neq0$, $K_{loss}\rightarrow0$) by
working slightly away from the $\Delta\omega_{LR}=R(2\omega_{0})$ resonance
(see panel (b) and (d) of Fig.\ref{Fig5} and Eqs.(\ref{m1}) and (\ref{mm5})).
Therefore it would also be interesting to investigate the potentialities of
the cavity+ double-dot device to implement the "$K$\&$\rho_{p}$" qubit scheme
or even a hybrid "$K_{loss}$\&$K$\&$\rho_{p}$" qubit scheme.

Note that our formalism is suitable for describing experiments which involve
quantum conductors with internal degrees of freedom coupled to the cavity
electric field. For the particular case of quantum conductors with no internal
degrees of freedom such as tunnel junctions or quantum point contacts, see for
instance Refs.\cite{Mendes:2015,Grimsmo:2016} for second order effects in the
light/matter coupling, and Ref. \cite{Mendes:2016} for higher orders. In these
References, the coupling of the source or drain of the conductors to the
cavity electric field is favored by a galvanic coupling scheme (i.e. the
source or drain of the device is directly connected to the cavity central
conductor). We do not consider such a\ coupling but rather an electrostatic
coupling of the mesoscopic circuit internal degrees of freedom to the cavity
electric field because this is favored by most designs used in mesoscopic QED
experiments where ac gates are placed between the circuit internal sites and
the cavity central conductor. Our approach nevertheless takes into account
photo-assisted dot-lead tunneling. For instance, in Ref.\cite{Bruhat:2016},
the coefficient $\chi_{2}$ reveals signatures of photo-assisted tunneling
between a quantum dot and a superconducting contact.

\section{Conclusion}

In this work, we have developed a quantum nonlinear description of mesoscopic
QED experiments. More precisely, we have used a quantum path integral approach
to express the effective action of a microwave cavity with bare frequency
$\omega_{0}$, coupled to a generic mesoscopic circuit, and excited by a
microwave drive at frequency $2\omega_{0}$. We have developed this action to
fourth order in the cavity/circuit coupling. This development reveals
photon/photon interactions mediated by the mesoscopic circuit. We have
investigated the possibility to establish a Lindblad description of the cavity
dynamics from the cavity action. This is always possible to third order in the
light matter coupling. In this limit, the cavity is subject to a coherent
photon pair drive\cite{Millburn} and a squeezing
dissipation\cite{Didier:2014,Lu:2015} mediated by the mesoscopic circuit.
To fourth order in the light/matter coupling, we identify
sufficient conditions in which a Markovian Lindblad description of the cavity
dynamics is still possible. This condition has to be tested for a given
circuit configuration by evaluating numerically different mesoscopic
correlators. In the Lindblad framework, the mesoscopic circuit
enables Kerr photon/photon interactions and two-photon loss/gain stochastic processes.

We have shown an example of application of our formalism to the case of a
resonator coupled to a double quantum dot with normal metal contacts.
The Lindblad condition is satisfied when the cavity and double dot
are off-resonant (for single photon exchange) and the dot-lead and dot-dot
couplings weak enough\textbf{ (}$\Gamma,t\ll\omega_{0}-\omega_{DQD}$%
\textbf{)}\textbf{.} We have studied how nonlinear effects such
as cavity squeezing, and photonic Schr\"{o}dinger cat states can occur, with a
non-trivial influence of dissipative mesoscopic transport. In particular,
quantum superpositions of photonic states can occur thanks to two-photon
dissipation caused by tunneling processes inside the double dot circuit.\ The
cavity squeezing effect also depends non-trivially on the dissipative tunnel
rates between the dots and normal reservoirs (see Appendix F).

We anticipate that the quantum regime of Mesoscopic QED\ conceals many more
surprises which our approach can reveal. Indeed, our method can be extended
straightforwardly to more complex circuit geometries with multiple quantum
dots and ferromagnetic or superconducting reservoirs. The effect of Coulomb
interactions inside the quantum dots also represents a rich field of
investigation\cite{Desjardins:2017}. For simplicity, we have studied
Lindbladian  situations, which are Markovian by
definition. However, our cavity action fully includes non-Markovian effects
and it could be exploited in the non-Markovian regime by using a more general
technical framework\cite{Kamenev}. Therefore, our work should be instrumental
to develop Mesoscopic QED in the quantum nonlinear regime. Interestingly, the
description of the effective dynamics of microwave cavities coupled to
dissipative Josephson circuits is also an important topic which lacks of
systematic approaches beyond the second order in the light/matter
interaction\cite{Azouit:2017,Forni:2018}. Our path integral approach could be
used to tackle this problem.

\textit{Acknowledgements: ZL acknowledges support from ANR project ENDURANCE,
and EMERGENCES grant ENDURANCE of Ville de Paris. TK acknowledges support from
Quantera project SuperTop.}

\section*{Appendix A: Details on the derivation of the cavity effective
action}

Here, we give more details on the derivation of Eqs.(\ref{S}) - (\ref{NNN}).
The drive at frequency $2\omega_{0}$ is not resonant with the cavity, and will
affect the photonic dynamics only indirectly thanks to the nonlinearity of the
mesoscopic circuit. To emphasize this fact and simplify the calculation of the
cavity effective action, it is convenient to make a displacement of the cavity
fields%
\begin{equation}%
\begin{bmatrix}
\phi_{cl}(t)\\
\phi_{q}(t)
\end{bmatrix}
=%
\begin{bmatrix}
\varphi_{cl}(t)\\
\varphi_{q}(t)
\end{bmatrix}
+%
\begin{bmatrix}
\int_{\omega}\sqrt{2}\varepsilon_{ac}^{\ast}(\omega)\mathcal{G}_{0}^{A}%
(\omega)e^{i\omega t}\\
0
\end{bmatrix}
\end{equation}
with the cavity drive $\varepsilon_{ac}$ defined temporally in Eq.(\ref{eac})
and $\mathcal{G}_{0}^{A}$ the bare cavity green's function defined in
Eq.(\ref{gbare}). In this framework, the action of the system becomes%
\begin{equation}
Z=\int d[\bar{\phi},\phi]e^{iS_{cav}^{0}(\bar{\phi},\phi)}\int d[\bar{\psi
},\psi]e^{iS_{meso}(\bar{\varphi},\varphi,\bar{\psi},\psi)} \label{Zchang}%
\end{equation}
with $S_{cav}^{0}$ defined in Eq.(\ref{S0}),%
\begin{equation}
S_{meso}(\bar{\phi},\phi,\bar{\psi},\psi)=\int\limits_{t,t^{\prime}}\bar{\psi
}(t)(\check{G}^{-1}(t,t^{\prime})-\check{v}_{\Sigma}^{\bar{\phi},\phi
}(t,t^{\prime}))\psi(t^{\prime})\text{,} \label{Sd}%
\end{equation}%
\begin{equation}
\check{v}_{\Sigma}^{\bar{\phi},\phi}(t,t^{\prime})=\left(  \check{v}(\bar
{\phi},\phi,t)+\check{v}_{ac,1}(t)+\check{v}_{ac,1}^{\dag}(t)\right)
\delta(t-t^{\prime}) \label{vvv}%
\end{equation}
and%
\begin{equation}
\check{v}_{ac,1}(t)=\frac{\check{g}}{2}\left(  \varepsilon_{p}\mathcal{G}%
_{0}^{R}(2\omega_{0})e^{-i2\omega_{0}t}+\varepsilon_{p}^{\ast}\mathcal{G}%
_{0}^{R}(-2\omega_{0})e^{i2\omega_{0}t}\right)  \text{.} \label{vv}%
\end{equation}
In Eqs.(\ref{Sd})-(\ref{vv}), the ac drive now modifies directly the potential
seen by the electrons of the mesoscopic circuit. The coefficients in
$\mathcal{G}_{0}^{R}$ in Eq. (\ref{vv}) express how the ac drive is seen by
electrons after a transduction by the cavity. They lead to the occurrence of
the factor $t_{0}$ in Eq.(\ref{epred}).

To eliminate the electronic degrees of freedom from Eq.(\ref{Zchang}), we
perform a Gaussian integration of (\ref{Zchang}) with respect to the
$\bar{\psi}$ and $\psi$ fields. This Gaussian integration is possible because,
in the absence of Coulomb interactions, the system action is quadratic with
respect to the electronic fields. This gives%
\begin{equation}
Z=\int d[\bar{\phi},\phi]e^{iS_{cav}^{0}(\bar{\phi},\phi)}\Xi(\bar{\phi},\phi)
\end{equation}
with%
\begin{equation}
\Xi(\bar{\phi},\phi)=\det_{t,k,d}[\check{1}-\check{m}] \label{detdet}%
\end{equation}
and%
\begin{equation}
\check{m}=\check{G}\circ\check{v}_{\Sigma}^{\bar{\phi},\phi} \label{mjaime}%
\end{equation}
Above, $\circ$ denotes a convolution on the time variables and a matrix
product on the mesoscopic orbital degrees of freedom, and $\det_{t,k,d}$ is a
generalized determinant on the time, Keldysh and orbital spaces which is
defined such that\cite{ZJ}
\begin{align}
&  Log[\Xi(\bar{\phi},\phi)]\label{dev}\\
&  =-\int\limits_{t}\underset{k,d}{\mathrm{Tr}}\left[  \check{m}%
(t,t)+\frac{\left.  \check{m}\circ\check{m}\right\vert _{t,t}}{2}%
+\frac{\left.  \check{m}\circ\check{m}\circ\check{m}\right\vert _{t,t}}%
{3}+...\right]  \text{.}\nonumber
\end{align}
In this work, we build a perturbation theory where the development
parameter is the matricial function $\check{m}$ which appears in Eq.(\ref{dev}).

The next step is to express Eq. (\ref{dev}) in terms of dot Green's functions.
This can generate many terms with a complex structure, but significant
simplifications can be performed in the limit where the dressed cavity has a
sufficient finesse. For brevity we only discuss the development of
the first and second order terms%
\begin{equation}
C_{1}=-\int\limits_{t}\underset{k,d}{\mathrm{Tr}}\check{m}(t,t)_{\varepsilon
_{p}=0}%
\end{equation}
and%
\begin{equation}
C_{2}=-\int\limits_{t}\underset{k,d}{\mathrm{Tr}}\left[  \left.  \check
{m}\circ\check{m}\right\vert _{t,t}/2\right]  _{\varepsilon_{p}=0}%
\end{equation}
in Eq.(\ref{dev}), in the absence of the $2\omega_{0}$ drive ($\varepsilon
_{p}=0$).

Let us first calculate $C_{1\text{. }}$From the definitions of
$\check{m}$ and $\check{v}_{\Sigma}^{\bar{\phi},\phi}$, one has:%
\begin{equation}
C_{1}=-\int\limits_{t,t^{\prime}}\underset{k,d}{\mathrm{Tr}}[\check
{G}(t,t^{\prime})\check{v}(\bar{\phi},\phi,t^{\prime})]\text{.}%
\end{equation}
Then, using the definition (\ref{vchap}) of $\check{v}$ in terms of fermionic
fields and the expression (\ref{Gdef}) of $\check{G}$ in terms of Keldysh
components, we obtain
\begin{multline}
-\sqrt{2}C_{1}\\
=%
{\textstyle\sum\limits_{d}}
g_{d}(\tilde{G}_{r}^{d,d}(t=0)+\tilde{G}_{a}^{d,d}(t=0))\int\limits_{t}\left(
\bar{\varphi}_{cl}(t)+\varphi_{cl}(t)\right) \\
+%
{\textstyle\sum\limits_{d}}
g_{d}\tilde{G}_{K}^{d,d}(t=0)\int\limits_{t}\left(  \bar{\varphi}%
_{q}(t)+\varphi_{q}(t)\right) \nonumber
\end{multline}
Using the general relation\cite{rel} $\tilde{G}_{r}^{d,d}(t=0)+\tilde{G}%
_{a}^{d,d}(t=0)=0$ and the definition (\ref{CC}) of $\tilde{G}_{K}^{d,d}$, one
can check%
\begin{equation}
C_{1}=-\frac{i}{\sqrt{2}}%
{\textstyle\sum\limits_{d}}
g_{d}\left(  2n_{d,0}-1\right)  \int\limits_{t}\left(  \bar{\varphi}%
_{q}(t)+\varphi_{q}(t)\right)  \label{Seff1}%
\end{equation}
where $n_{d,0}=\left\langle \hat{c}_{d}^{\dag}\hat{c}_{d}\right\rangle
_{\check{g}=0}$ is the average occupation of level $d$ in the absence of
light/matter coupling. A comparison of this term with Eq.(\ref{Sac}) shows
that $C_{1}$ corresponds to a cavity dc drive
\begin{equation}
H_{cav,1}^{eff}=%
{\textstyle\sum\limits_{d}}
g_{d}\left(  n_{d,0}-\frac{1}{2}\right)  \left(  \hat{a}^{\dag}+\hat
{a}\right)  \label{Heff1}%
\end{equation}
which can be disregarded in our study due to its non-resonant nature.

We now calculate $C_{2\text{. }}$From the definitions of $\check{m}$ and
$\check{v}_{\Sigma}^{\bar{\phi},\phi}$, one has:%
\begin{equation}
C_{2}=-\int\limits_{t,t^{\prime}}\underset{k,d}{\mathrm{Tr}}[\check
{G}(t,t^{\prime})\check{v}(\bar{\phi},\phi,t^{\prime})\check{G}(t^{\prime
},t)\check{v}(\bar{\phi},\phi,t)]/4\text{.}%
\end{equation}
Using the definition (\ref{vchap}) of $\check{v}$ in terms of fermionic fields
and introducing Fourier transforms, one gets%
\begin{align}
C_{2}  &  =-\iint\limits_{\omega_{1},\omega_{2}}\underset{k,d}{\mathrm{Tr}%
}[\check{G}(\omega_{1})\check{g}\bar{\phi}_{\Sigma}(\omega_{3}-\omega
_{1})\check{G}(\omega_{3})\check{g}\bar{\phi}_{\Sigma}(\omega_{1}-\omega
_{3})]/4\nonumber\\
&  -\iint\limits_{\omega_{1},\omega_{2}}\underset{k,d}{\mathrm{Tr}}[\check
{G}(\omega_{1})\check{g}\bar{\phi}_{\Sigma}(\omega_{3}-\omega_{1})\check
{G}(\omega_{3})\check{g}\phi_{\Sigma}(\omega_{3}-\omega_{1})]/4\nonumber\\
&  -\iint\limits_{\omega_{1},\omega_{2}}\underset{k,d}{\mathrm{Tr}}[\check
{G}(\omega_{1})\check{g}\phi_{\Sigma}(\omega_{1}-\omega_{3})\check{G}%
(\omega_{3})\check{g}\bar{\phi}_{\Sigma}(\omega_{1}-\omega_{3})]/4\nonumber\\
&  -\iint\limits_{\omega_{1},\omega_{2}}\underset{k,d}{\mathrm{Tr}}[\check
{G}(\omega_{1})\check{g}\phi_{\Sigma}(\omega_{1}-\omega_{3})\check{G}%
(\omega_{3})\check{g}\phi_{\Sigma}(\omega_{3}-\omega_{1})]/4
\end{align}
with%
\begin{equation}
\phi_{\Sigma}(\omega_{3}-\omega_{1})=\phi_{cl}(\omega_{3}-\omega_{1}%
)\check{\sigma}_{0}+\phi_{q}(\omega_{3}-\omega_{1})\check{\sigma}_{1}%
\end{equation}
and%
\begin{equation}
\bar{\phi}_{\Sigma}(\omega_{3}-\omega_{1})=\bar{\phi}_{cl}(\omega_{3}%
-\omega_{1})\check{\sigma}_{0}+\bar{\phi}_{q}(\omega_{3}-\omega_{1}%
)\check{\sigma}_{1}\text{.}%
\end{equation}
Assuming that the dressed cavity has a good quality factor ($\Lambda_{0}%
^{app}=\Lambda_{0}+\Delta\Lambda_{0}\ll\omega_{0}$ has to be checked a
posteriori), the terms $\phi_{\Sigma}(\omega_{1}-\omega_{3})$ and
$\phi_{\Sigma}(\omega_{3}-\omega_{1})$ have a weak overlap and therefore the
first and fourth line of the above expression, which contains products
$\bar{\phi}_{cl(q)}\bar{\phi}_{cl[q]}$ or $\phi_{cl(q)}\phi_{cl(q)}$, are
negligible. A change of frequency variables in the remaining terms (which
contain contributions in $\bar{\phi}_{cl(q)}\phi_{cl[q]}$ only) gives%
\begin{align}
C_{2}  &  =-\iint\limits_{\omega_{1},\omega}\underset{k,d}{\mathrm{Tr}}%
[\check{G}(\omega_{1})\check{g}\bar{\phi}_{\Sigma}(\omega)\check{G}%
(\omega+\omega_{1})\check{g}\phi_{\Sigma}(\omega)]/4\nonumber\\
&  -\iint\limits_{\omega_{1},\omega}\underset{k,d}{\mathrm{Tr}}[\check
{G}(\omega_{1})\check{g}\phi_{\Sigma}(\omega)\check{G}(\omega_{1}%
-\omega)\check{g}\bar{\phi}_{\Sigma}(\omega)]/4\text{.}%
\end{align}
Then, we assume that the dressed cavity linewidth is much smaller than the
mesoscopic resonances linewidth ($\Lambda_{0}+\Delta\Lambda_{0}\ll\Gamma$ has
to be checked a posteriori, with $\Gamma$ the order of magnitude of the tunnel
rates to the mesoscopic reservoirs). In this case, the terms in $\check{G}$ in
the above integral vary very slowly in the frequency area $\omega_{0}%
-\Lambda_{0}^{app}\lesssim\omega\lesssim\omega_{0}+\Lambda_{0}^{app}$ where
$\phi_{\Sigma}(\omega)$ and $\bar{\phi}_{\Sigma}(\omega)$ contribute
significantly to the cavity action, and one can thus use $\omega\simeq
\omega_{0}$ in these terms. This gives%
\begin{align}
C_{2}  &  =-\iint\limits_{\omega_{1},\omega}\underset{k,d}{\mathrm{Tr}}%
[\check{G}(\omega)\check{g}\bar{\phi}_{\Sigma}(\omega_{1})\check{G}(\omega
_{0}+\omega)\check{g}\phi_{\Sigma}(\omega_{1})]/4\nonumber\\
&  -\iint\limits_{\omega_{1},\omega}\underset{k,d}{\mathrm{Tr}}[\check
{G}(\omega)\check{g}\phi_{\Sigma}(\omega_{1})\check{G}(\omega-\omega
_{0})\check{g}\bar{\phi}_{\Sigma}(\omega_{1})]/4\text{.}%
\end{align}
Finally we can come back to the time representation for the cavity fields%
\begin{align}
C_{2}  &  =-\iint\limits_{\omega,t}\underset{k,d}{\mathrm{Tr}}[\check
{G}(\omega)\check{g}\bar{\phi}_{\Sigma}(t)\check{G}(\omega_{0}+\omega
)\check{g}\phi_{\Sigma}(t)]/4\nonumber\\
&  -\iint\limits_{\omega,\omega}\underset{k,d}{\mathrm{Tr}}[\check{G}%
(\omega)\check{g}\phi_{\Sigma}(t)\check{G}(\omega-\omega_{0})\check{g}%
\bar{\phi}_{\Sigma}(t)]/4\text{.} \label{resres}%
\end{align}
A rearrangement of these terms leads to an action contribution similar to that
of Eq.(\ref{S2}), with fields $\bar{\varphi},~\varphi$ replaced by $\bar{\phi
}$,$~\phi$. A similar treatment can be performed for higher order terms of
Eq.(\ref{dev}) and terms which depend on $\varepsilon_{p}$. For instance, the
contribution in $g^{4}$ corresponds to 6 terms similar to those of
Eq.(\ref{resres}). We finally obtain, after some algebra and term regrouping,
a cavity effective Schwinger-Keldysh partition function $Z=\int d[\bar{\phi
},\phi]e^{iS_{cav}^{eff}(\bar{\phi},\phi)}$ with $S_{cav}^{eff}$ defined in
Eq.(\ref{S}). The final step is to come back to an expression of the cavity
action with the fields $\bar{\varphi},\varphi$. We disregard terms of order
$g^{4}\varepsilon_{p}$, since we assume that both $g^{4}$ and $\varepsilon
_{p}$ are small. In this case, one obtains $Z=\int d[\bar{\varphi}%
,\varphi]e^{i(S_{cav}^{eff}(\bar{\varphi},\varphi)+\Delta\tilde{S}_{ac}%
(\bar{\varphi},\varphi))}$ where $\Delta\tilde{S}_{ac}(\bar{\varphi},\varphi)$
is a drive term similar to the term $\Delta S_{ac}(\bar{\varphi},\varphi)$ of
Eq.(\ref{Sac}), but with an amplitude $\varepsilon_{p}$ which has a
renormalization in $g^{2}\varepsilon_{p}$. However, since this ac drive is non
resonant with the cavity, one can disregard $\Delta\tilde{S}_{ac}$. Therefore,
one can use $Z\simeq\int d[\bar{\varphi},\varphi]\exp[iS_{cav}^{eff}%
(\bar{\varphi},\varphi)]$. In particular, one gets the expression
\begin{equation}
\mathcal{A=}i\left[
\begin{tabular}
[c]{rrr}%
$\mathcal{N}_{cl,cl,cl,cl}$ & $\mathcal{N}_{cl,cl,cl,q}$ & $\mathcal{N}%
_{cl,cl,q,q}$\\
$\mathcal{N}_{cl,q,cl,cl}$ & $\mathcal{N}_{cl,q,cl,q}$ & $\mathcal{N}%
_{cl,q,q,q}$\\
$\mathcal{N}_{q,q,cl,cl}$ & $\mathcal{N}_{q,q,cl,q}$ & $\mathcal{N}_{q,q,q,q}$%
\end{tabular}
\ \ \ \ \ \ \right]  \label{Ainit}%
\end{equation}
for the matrix which occurs in the expression (\ref{S4}), with coefficients
$\mathcal{N}_{f,f^{\prime},l,l^{\prime}}$ defined in Eq.(\ref{NNN}). Using the
cyclic property of the trace in Eq.(\ref{NNN}) and the properties $\tilde
{G}_{K}(\omega)=-\tilde{G}_{K}(\omega)^{\dag}$ and $\tilde{G}_{a}%
(\omega)=\tilde{G}_{r}(\omega)^{\dag}$, one can check that there exists
relations between the different components of $\mathcal{A}$ in Eq.(\ref{Ainit}%
) so that one finally gets expression (\ref{AAA}).

\section*{Appendix B: Semiclassical description of Mesoscopic QED}

\subsection*{B1. Direct semiclassical description of Mesoscopic QED}

It is useful to reconsider the problem of Mesoscopic QED with a direct
semiclassical approach (without the path integral formulation) in order to
gain more physical insight into the new coefficients $\chi_{4}$ and $U_{cl}$
which appear in Eqs.(\ref{S3}) and (\ref{S4}). Equation (\ref{Htot}) gives the
photonic equation of motion in the Heisenberg picture:%
\begin{equation}
\frac{d}{dt}\hat{a}(t)=-i\omega_{0}\hat{a}(t)-\frac{i}{\hbar}\sum
\limits_{d}g_{d}\hat{n}_{d}(t)-\frac{\Lambda_{0}}{2}\hat{a}(t)-i\varepsilon
_{ac}(t)\text{.} \label{em}%
\end{equation}
In a semiclassical picture, the operator $\hat{a}(t)$ in the above equation
can be treated as a classical quantity $a(t)=\hat{a}(t)=\left\langle \hat
{a}(t)\right\rangle $. In this case, the average electron number operator
$\left\langle \hat{n}_{d}(t)\right\rangle =\left\langle \hat{c}_{d}^{\dag
}(t)\hat{c}_{d}(t)\right\rangle $ in orbital $d$ can be calculated as the
response to the \textquotedblleft classical\textquotedblright\ excitations
$g_{d^{\prime}}(a^{\dag}(t)+a(t))$ , with $d^{\prime}\in\lbrack1,N]$, which we
will write in a matrix form as
\begin{equation}
\tilde{E}_{ac}(t)=\tilde{g}(a^{\dag}(t)+a(t))\text{.} \label{ac}%
\end{equation}
At this stage, although $a(t)$ is expected to have a dominant contribution in
$e^{-i\omega_{0}t}$, it is essential to take into account weak components in
$e^{\pm i2\omega_{0}t}$ to describe the effect of the drive in $\beta_{p}$ on
$\left\langle \hat{n}_{d}(t)\right\rangle $. It is sufficient to estimate
these components from Eq.(\ref{em}) treated to order $0$ in $g$, because this
is enough to obtain a $\beta_{p}g^{3}$ contribution to the photonic field, as
we will see below. Hence, we use%
\begin{equation}
\tilde{E}_{ac}(t)=\tilde{g}\left(  \alpha e^{-i\omega_{0}t}+\alpha^{\ast
}e^{i\omega_{0}t}+\operatorname{Re}[t_{0}\varepsilon_{p}e^{-i2\omega_{0}%
t}/2]\right)  \label{Eac}%
\end{equation}
with $t_{0}$ defined by Eq.(\ref{tzero}). The amplitude $\alpha$ is not
specified since it must be determined self-consistently from Eq.(\ref{em}) and
the response of the average dot charges to $\tilde{E}_{ac}(t)$. From the
Keldysh description of mesoscopic transport\cite{Jauho:1994}, this response is
given by%
\begin{equation}
\sum\limits_{d}g_{d}\left\langle \hat{n}_{d}(t)\right\rangle =-iTr_{d}%
[\tilde{g}\mathbf{\tilde{G}}_{<}(t,t)] \label{gn}%
\end{equation}
where the lesser Green's function of the dots $\mathbf{\tilde{G}}_{<}$ in the
presence of $\tilde{E}_{ac}(t)$ can be expressed as%
\begin{align}
&  \mathbf{\tilde{G}}_{<}(t,t)\label{hh}\\
&  =%
{\textstyle\iiint}
\frac{d\omega}{2\pi}dt_{1}dt_{2}e^{-i\omega(t_{1}-t_{2})}\mathbf{\tilde{G}%
}_{r}(t,t_{1})\tilde{\Sigma}^{<}(\omega)\mathbf{\tilde{G}}_{a}(t_{2}%
,t)\text{.}\nonumber
\end{align}
Above, $\tilde{\Sigma}^{<}(\omega)$ is the lesser self energy of the dots
illustrated in section \ref{DQD} for the double dot case. The mesoscopic
retarded and advanced Green's functions $\mathbf{\tilde{G}}_{r(a)}$ in the
presence of $\tilde{E}_{ac}(t)$ can be calculated in terms of the unperturbed
mesoscopic Green's functions $\tilde{G}_{r(a)}$ defined in section
\ref{GeneAct} by using the Dyson equation%
\begin{equation}
\mathbf{\tilde{G}}_{J}(t,t^{\prime})=\tilde{G}_{J}(t,t^{\prime})+%
{\textstyle\int}
\frac{dt_{1}}{\hbar}\tilde{G}_{J}(t,t_{1})\tilde{E}_{ac}(t_{1})\mathbf{\tilde
{G}}_{J}(t_{1},t^{\prime}) \label{Dyson}%
\end{equation}
with $J\in r(a)$.

The combination of Eqs.(\ref{gn}), (\ref{hh}) and (\ref{Dyson}) gives, by
keeping only resonant contributions in $e^{-i\omega_{0}t}$,%

\begin{equation}
\sum\limits_{d}g_{d}\left\langle \hat{n}_{d}\right\rangle \simeq\left(
\alpha\chi_{2}+2\alpha\left\vert \alpha\right\vert ^{2}\chi_{4}+i\alpha^{\ast
}U_{cl}\right)  e^{-i\omega_{0}t}\text{.} \label{NT}%
\end{equation}
Tedious algebra is necessary to identify the coefficients which appear in
Eq.(\ref{NT}) with the correlation functions $U_{cl}$ and $\chi_{4}$ defined
in the main text, especially in the multi-orbital case $N>1$. Equation
(\ref{NT}) shows that $\chi_{2}$ is the linear response function of the dots
charge to the excitation in $\alpha e^{-i\omega_{0}t}$, and $\chi_{4}$ is the
second order response function to the same excitation, whereas $U_{cl}$
appears as a transduction coefficient for the field component in $\alpha
^{\ast}e^{i\omega_{0}t}$ into a resonant term in $e^{-i\omega_{0}t}$ thanks to
the energy provided by the drive in $\varepsilon_{p}$. One can finally inject
Eq.(\ref{NT}) into the statistical average of Eq.(\ref{em}) to obtain%
\begin{equation}
0=\alpha^{\ast}U_{cl}-\left(  i\chi_{2}+\frac{\Lambda_{0}}{2}+2i\left\vert
\alpha\right\vert ^{2}\chi_{4}\right)  \alpha\text{.} \label{res1}%
\end{equation}
For this last step, we have used the resonant approximation $a(t)\simeq\alpha
e^{-i\omega_{0}t}$ in Eq.(\ref{em}) and disregarded the term in $\varepsilon
_{ac}(t)$ in the right member of (\ref{em}) because it is not resonant with
the cavity. One can see along this calculation that $\varepsilon_{p}$ plays a
crucial role in intermediary steps of the calculation for the description of
two-photon processes, but its direct contribution to (\ref{res1}) can be
disregarded. A\ similar fact happens with the path integral approach where
$\varepsilon_{p}$ produces indirectly the $S_{g}^{(3)}(t)$ term whereas its
direct contribution $\Delta S_{ac}(t)$ can be disregarded from the effective
action $S_{cav}^{eff}(t)$ in the resonant approximation. Note that
Eq.(\ref{res1}) is in full agreement with the result given by a direct
calculation of the semiclassical cavity steady states with the path integral
description (see Appendix B2).\begin{figure}[ptb]
\includegraphics[width=1\linewidth]{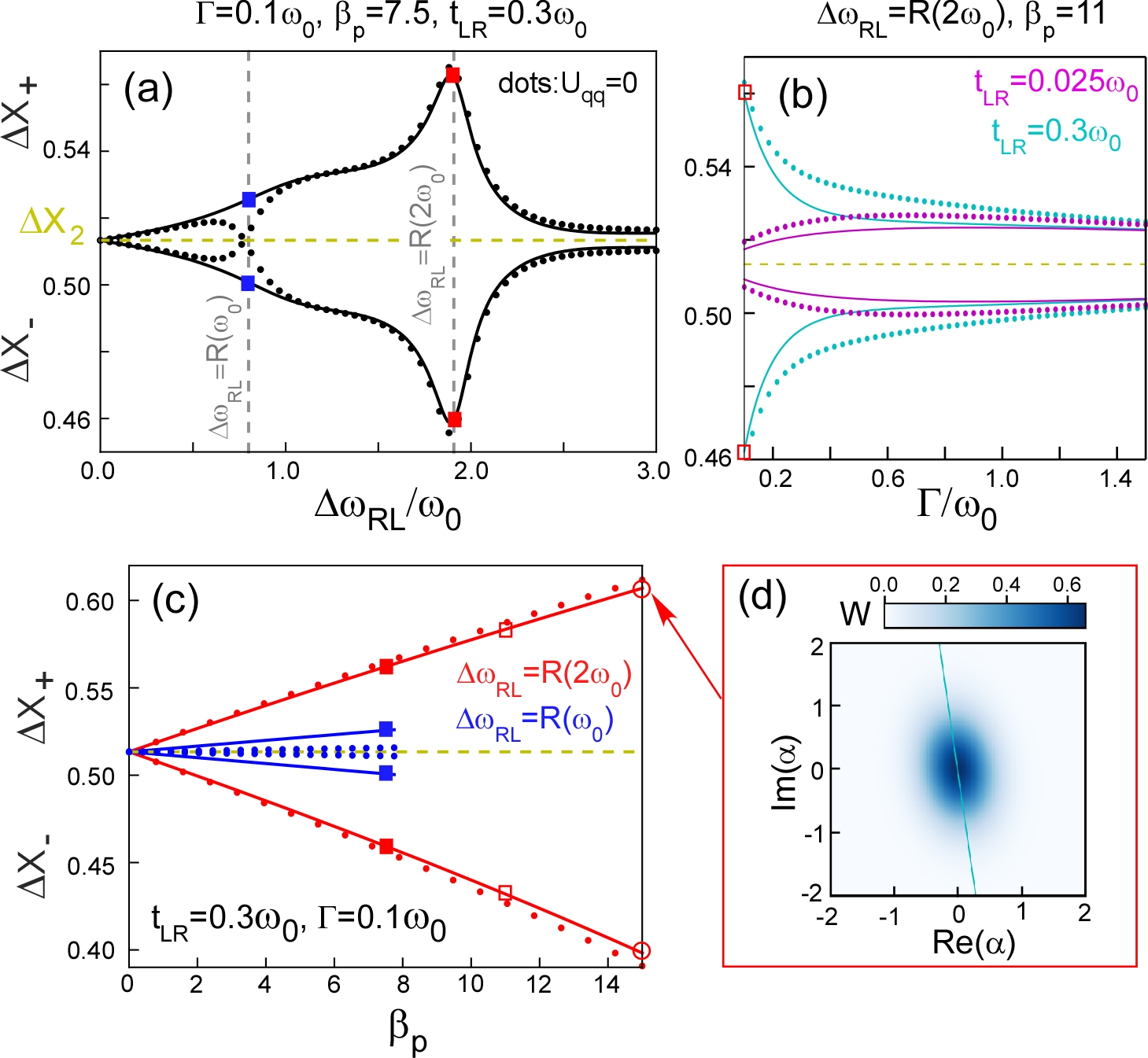}\newline\caption{{}Panels (a),
(b) and (c): Cavity field quadratures $\Delta X_{\pm}$ versus $\Delta
\omega_{RL}$, $\Gamma$ and $\beta_{p}$ respectively. In panel (a), we use
$t_{LR}=0.3\omega_{0}$, $\Gamma=0.1\omega_{0}$ and $\beta_{p}=7.5$. In panel
(b) we use $t_{LR}=0.3\omega_{0}$ (cyan lines) or $t_{LR}=0.025\omega_{0}$
(magenta lines), $\Delta\omega_{RL}=R(2\omega_{0})$ and $\beta_{p}=11$. In
panel (c) we use $t_{LR}=0.3$, $\Gamma=0.1\omega_{0}$ and $\Delta\omega
_{RL}=R(2\omega_{0})~$\ (red lines) or $\Delta\omega_{RL}=R(\omega_{0})$ (blue
lines). The other parameters are the same as in Fig. \ref{Fig1} with
$\omega_{av}=0$ and $g_{L}=0.01\omega_{0}$. The full lines correspond to the
result given by the full expressions (\ref{Afull}) and (\ref{Bfull}) of $A$
and $B$ whereas the dotted lines omit the contribution of $\gamma_{p}$ (or
equivalently $U_{qq}$). For reference, the second order variance $\Delta
X_{2}$ for an empty cavity (corresponding to the case for $g_{L}=0$) is also
shown as a dashed yellow line. The vertical dashed gray lines in panel (a)
indicate the resonances $\Delta\omega_{RL}=R(\omega_{0})$ and $\Delta
\omega_{RL}=R(2\omega_{0})$. The blue and red squares indicate working points
which are common to panels (a), (b) and (c). In panel (c), the plots are
restricted to the range where $\gamma_{loss}>0$ and $\gamma_{gain}>0$, which
is narrower in the case $\Delta\omega_{RL}=R(\omega_{0})$ (blue curves). Panel
d: Squeezed cavity Wigner function for the working point corresponding to the
empty red circles in panel (c). The major axis of the Wigner function is shown
as a blue line.}%
\label{Fig3}%
\end{figure}

\subsection*{B2. Semiclassical photonic amplitudes given by the path integral
description}

The possible semiclassical photonic amplitudes of the cavity in stationary
conditions can also be obtained by looking for the saddle points of the cavity
effective action\cite{Kamenev}. Since the action (\ref{S}) vanishes for
$\varphi_{cl}=0$, $\bar{\varphi}_{cl}=0$, a semiclassical solution for the
cavity field can be found at $\varphi_{q}=0$, $\bar{\varphi}_{q}=0$ and values
of $\varphi_{cl}$ and $\bar{\varphi}_{cl}$ such that $\left.  \partial
(S)/\partial\bar{\varphi}_{q}(t)\right\vert _{\varphi_{q}=0,\bar{\varphi}%
_{q}=0}=0$. This gives%
\begin{align}
-\sqrt{2}\varepsilon_{ac}(t)  &  =(i\partial_{t}-\omega_{0}+\frac{i\Lambda
_{0}}{2})\varphi_{cl}-\chi_{2}\varphi_{cl}\label{sccc}\\
&  -ie^{-2i\omega_{0}t}U_{cl}\bar{\varphi}_{cl}-\chi_{4}\bar{\varphi}%
_{cl}\varphi_{cl}\varphi_{cl}\text{.}\nonumber
\end{align}
One can disregarded $\varepsilon_{ac}(t)$ from the left member of Eq.
(\ref{sccc}) because it is not directly resonant with the cavity. Hence, one
can expect a semiclassical solution $\varphi_{sc}=\sqrt{2}\alpha
_{sc}e^{i(\varphi_{sc}-\omega_{0}t)}$ such that%
\begin{equation}
\left(  U_{cl}e^{-2i\varphi_{sc}}-\frac{\Lambda_{0}}{2}-i\chi_{2}-2i\chi
_{4}\left\vert \alpha_{sc}\right\vert ^{2}\right)  \alpha_{sc}=0 \label{EqSC}%
\end{equation}
with $\alpha_{sc}$ the semiclassical value of $\hat{a}$. Equation (\ref{EqSC})
is in full agreement with the semiclassical Eq. (\ref{res1}) if $\alpha
=\alpha_{sc}$ is used. This equation is also similar to the equation
(\ref{aav}) on the average photons amplitude $\alpha_{av}$ obtained from the
Lindblad description of the cavity dynamics, up to the term in $\Delta
\Lambda_{0,4}$ which is not present in Eq.(\ref{EqSC}). This discrepancy is
due to the fact that the equation on $\alpha_{sc}$ is obtained by disregarding
quantum fluctuations of the cavity occupation.

\section*{Appendix C:Action associated to a Lindblad equation}

Following Ref.\cite{Sieberer:2016}, the action corresponding to a Lindblad
equation with the form (\ref{ME}) can be expressed as%

\begin{equation}
S=\int\limits_{t}\left(  \bar{\varphi}_{+}(t)i\partial t\varphi_{+}%
(t)-\bar{\varphi}_{-}(t)i\partial t\varphi_{-}(t)-i\mathcal{L}(t\right)
\end{equation}
with $\varphi_{\pm}=\frac{1}{\sqrt{2}}(\varphi_{cl}\pm\varphi_{q})$,
$\bar{\varphi}_{\pm}=\frac{1}{\sqrt{2}}(\bar{\varphi}_{cl}\pm\bar{\varphi}%
_{q})$ and%
\begin{align}
-i\mathcal{L}(t)  &  =-H_{cav}^{eff}[\bar{\varphi}_{+}(t),\varphi
_{+}(t)]+H_{cav}^{eff}[\bar{\varphi}_{-}(t),\varphi_{-}(t)]\nonumber\\
&  -i\sum_{j}\gamma_{j}\hat{L}_{j}[\bar{\varphi}_{+},\varphi_{+}]\hat{L}%
_{j}^{\dag}[\bar{\varphi}_{-},\varphi_{-}]\nonumber\\
&  +\frac{i}{2}\sum_{j,s\in\{+,-\}}\gamma_{j}\hat{L}_{j}^{\dag}[\bar{\varphi
}_{s},\varphi_{s}]\hat{L}_{j}[\bar{\varphi}_{s},\varphi_{s}]\text{.}
\label{iL}%
\end{align}
This leads to Eqs.(\ref{SM}), (\ref{Smark4}), and (\ref{AM}) of the main text.
Note that this result is valid even when the dissipative rates $\gamma_{j}$
and the Hamiltonian $H_{cav}^{eff}$ are time-dependent

\section*{Appendix D: Link between the direct density matrix approach and the
path integral approach to second order in\textbf{ }$g$}

To show that the Lindblad Eqs.(\ref{rohPed}) and (\ref{MElin}) obtained with
the direct density matrix approach and the path integral approach,
respectively, agree to second order in\textbf{ }$g$, one must establish the
relation between the parameters $\chi_{A}$, $\chi_{B}$ and $\chi_{2}$,
$\lambda_{2}$ which occur in these Eqs. Note that $\chi_{2}$ and $\lambda_{2}$
have a frequency dependence which is omitted in the main text where we use
$\chi_{2}=\chi_{2}(\omega_{0})$, and $\lambda_{2}=\lambda_{2}(\omega_{0})$.
For our present purpose, it is convenient to use the inverse Fourier transform
of these quantities, defined generally as $f(t)=\int_{-\infty}^{+\infty}%
\frac{d\omega_{0}}{2\pi}f(\omega_{0})e^{-i\omega_{0}t}$. One can use the
general relation
\begin{equation}
\int_{-\infty}^{+\infty}\frac{d\omega}{2\pi}a(\omega+\omega_{0})b(\omega
)=\int_{-\infty}^{+\infty}dt~a(t)b(-t)e^{i\omega_{0}t}%
\end{equation}
where $a$ and $b$ are two generic functions, to reexpress Eqs.(\ref{Chi2}) and
(\ref{landa}) as
\begin{equation}
\chi_{2}(t)=-\frac{i}{2}\underset{d}{\mathrm{Tr}}\left[  \tilde{G}%
_{K}(t)\tilde{g}\tilde{G}_{a}(-t)\tilde{g}+\tilde{G}_{K}(-t)\tilde{g}\tilde
{G}_{r}(t)\tilde{g}\right]  \text{,} \label{bi1}%
\end{equation}%
\begin{align}
\lambda_{2}(t)  &  =-\frac{i}{2}\underset{d}{\mathrm{Tr}}\left[  \tilde{G}%
_{K}(-t)\tilde{g}\tilde{G}_{K}(t)\tilde{g}\right. \label{bi2}\\
&  +\left.  \tilde{G}_{a}(-t)\tilde{g}\tilde{G}_{r}(t)\tilde{g}+\tilde{G}%
_{r}(-t)\tilde{g}\tilde{G}_{a}(t)\tilde{g}\right]  \text{.}\nonumber
\end{align}
At this stage, it is convenient to define the lesser and greater fermionic
Green's functions
\begin{equation}
G_{<}^{d,d^{\prime}}(t)=i\left\langle \hat{c}_{d^{\prime}}^{\dag}(0)\hat
{c}_{d}(t)\right\rangle
\end{equation}
and%
\begin{equation}
G_{>}^{d,d^{\prime}}(t)=-i\left\langle \hat{c}_{d}(t)\hat{c}_{d^{\prime}%
}^{\dag}(0)\right\rangle
\end{equation}
to reexpress definitions (\ref{AA})-(\ref{CC}) as:%
\begin{equation}
G_{r}^{d,d^{\prime}}(t)=\theta(t)\left(  G_{>}^{d,d^{\prime}}(t)-G_{<}%
^{d,d^{\prime}}(t)\right)  \text{,} \label{r1}%
\end{equation}%
\begin{equation}
G_{a}^{d,d^{\prime}}(t)=\theta(-t)\left(  G_{<}^{d,d^{\prime}}(t)-G_{>}%
^{d,d^{\prime}}(t)\right)  \label{r2}%
\end{equation}
and%
\begin{equation}
G_{K}^{d,d^{\prime}}(t)=G_{<}^{d,d^{\prime}}(t)+G_{>}^{d,d^{\prime}%
}(t)\text{.} \label{r3}%
\end{equation}
Then, using Eqs.(\ref{r1})-(\ref{r3}), one can rewrite Eqs.(\ref{bi1}) and
(\ref{bi2}) as%

\begin{equation}
\chi_{2}(t)=i\theta(t)\underset{d}{\mathrm{Tr}}\left[  \tilde{G}_{<}%
(t)\tilde{g}\tilde{G}_{>}(-t)\tilde{g}-\tilde{G}_{>}(t)\tilde{g}\tilde{G}%
_{<}(-t)\tilde{g}\right]  \text{,} \label{w1}%
\end{equation}%
\begin{equation}
\lambda_{2}(t)=-i\underset{d}{\mathrm{Tr}}\left[  \tilde{G}_{<}(-t)\tilde
{g}\tilde{G}_{>}(t)\tilde{g}+\tilde{G}_{>}(-t)\tilde{g}\tilde{G}_{<}%
(t)\tilde{g}\right]  \text{.} \label{w2}%
\end{equation}
Since we consider a non-interacting case, one can use the Wick theorem to
reexpress the above equations in terms of charge correlators\cite{Zamoum:2016}%
. Indeed, using the operator $\hat{N}(t)$ of Eq. (\ref{NNN2}), one finds%
\begin{equation}
\left\langle \hat{N}(t)\hat{N}(0)\right\rangle =\left\langle \hat
{N}\right\rangle ^{2}+\underset{d}{\mathrm{Tr}}\left[  \tilde{G}_{<}%
(-t)\tilde{g}\tilde{G}_{>}(t)\tilde{g}\right]  \text{,}%
\end{equation}%
\begin{equation}
\left\langle \hat{N}(0)\hat{N}(t)\right\rangle =\left\langle \hat
{N}\right\rangle ^{2}+\underset{d}{\mathrm{Tr}}\left[  \tilde{G}_{<}%
(t)\tilde{g}\tilde{G}_{>}(-t)\tilde{g}\right]  \text{.}%
\end{equation}
This leads to%
\begin{equation}
\chi_{2}(t)=i\theta(t)\left(  \left\langle \hat{N}(0)\hat{N}(t)\right\rangle
-\left\langle \hat{N}(t)\hat{N}(0)\right\rangle \right)  \text{,}%
\end{equation}%
\begin{equation}
\lambda_{2}(t)=-i\left(  \left\langle \hat{N}(0)\hat{N}(t)\right\rangle
+\left\langle \hat{N}(t)\hat{N}(0)\right\rangle -2\left\langle \hat
{N}\right\rangle ^{2}\right)  \text{.}%
\end{equation}
A comparison of these equations with the definitions (\ref{xiA}) and
(\ref{xiB}) of $\chi_{A}(t)$ and $\chi_{B}(t)$ gives, in the frequency domain
\begin{equation}
\chi_{2}(\omega_{0})=\chi_{B}(\omega_{0})-\chi_{A}(\omega_{0})\text{,}%
\end{equation}%
\begin{equation}
\lambda_{2}(\omega_{0})=2i\left(  \operatorname{Im}[\chi_{A}(\omega_{0}%
)+\chi_{B}(\omega_{0})]+\left\langle \hat{N}\right\rangle ^{2}\delta
(\omega_{0})\right)  \text{.}%
\end{equation}
This proves the relations (\ref{lalla1}) and (\ref{lalla2}) of the main text
and the agreement between the Lindblad Eqs.(\ref{rohPed}) and (\ref{MElin}) at
second order in\textbf{ }$g$.

\section*{Appendix E: Analytical calculation of the Wigner function to third
order in\textbf{ }$g$}

The definition (\ref{Wdef}) of the Wigner function involves the correlation
function $\chi(t,\beta,\beta^{\ast})=\left\langle e^{\beta a_{I}^{\dag}%
-\beta^{\ast}a_{I}}\right\rangle _{t}$. From the expression of the effective
Hamiltonian $H_{cav}^{eff}$ and the jump operators $\hat{L}_{j}$, one can
check that $\chi$ follows\cite{Kamenev}%
\begin{align}
\frac{\partial}{\partial t}\chi &  =-i\Delta\omega_{0}\left(  -\beta
\partial_{\beta}+\beta^{\ast}\partial_{\beta^{\ast}}\right)  \chi-\gamma
_{+}\frac{\beta\beta^{\ast}}{2}\chi\nonumber\\
&  -\frac{\gamma_{-}}{2}\left(  \beta^{\ast}\partial_{\beta^{\ast}}%
+\beta\partial_{\beta}\right)  \chi-2\rho_{p}\beta^{\ast}\partial_{\beta}%
\chi-2\rho_{p}^{\ast}\beta\partial_{\beta^{\ast}}\chi\nonumber\\
&  -\frac{\beta^{2}}{2}\gamma_{p}e^{-i\varphi_{p}}\chi-\frac{\beta^{\ast2}}%
{2}\gamma_{p}e^{i\varphi_{p}}\chi\text{.}%
\end{align}
For compactness we note $\frac{\partial}{\partial\beta}=\partial_{\beta}$ and
$\frac{\partial}{\partial\beta^{\ast}}=\partial_{\beta^{\ast}}$. The above
equation is a first order differential equation which is more convenient to
solve than the second order differential equation (\ref{Wdef}). It is then
straightforward to Fourier transform $\chi$ to obtain $W(t)$.

\section*{Appendix F: Parametric control of the squeezing
effect\label{parametric}}

This Appendix discusses how the photonic squeezing effect of Section IV.D
depends on the double dot parameters. Figure \ref{Fig3} shows the cavity field
quadratures $\Delta X_{\pm}$ versus the orbital detuning $\Delta\omega
_{RL}=\omega_{R}-\omega_{L}$ (panel (a)), versus $\Gamma$ (panel (b)) and
versus the cavity drive amplitude $\beta_{p}$ (panel (c)) for a case where the
single and two-photon resonances at $\Delta\omega_{RL}=R(\omega_{0})$ and
$\Delta\omega_{RL}=R(2\omega_{0})$ are allowed. The results given by the full
expressions (\ref{Afull}) and (\ref{Bfull}) of $A$ and $B$ are shown with full
lines. For reference, the variance\begin{figure}[ptb]
\includegraphics[width=1\linewidth]{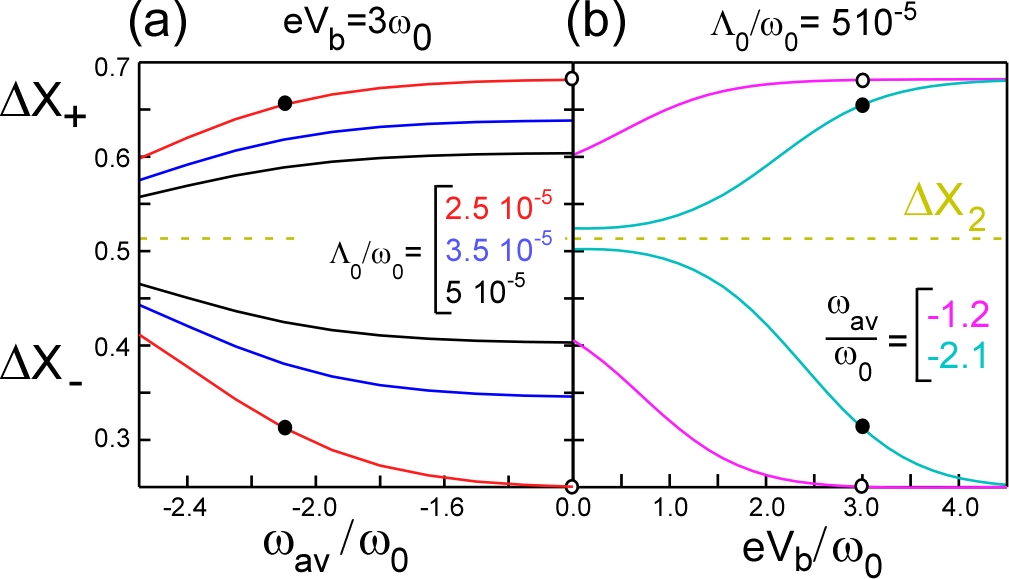}\newline\caption{{}Cavity field
quadratures $\Delta X_{\pm}$ versus $\omega_{av}=(\omega_{R}+\omega_{L})/2$
(panel (a)) \ and versus the bias voltage $V_{b}$ (panel (b)) for
$\Delta\omega_{RL}=2\omega_{0}$, $t_{LR}=0.1\omega_{0}$, $g_{L}=0.01\omega
_{0}~$and $\beta_{p}=200$. The other parameters are the same as in
Fig.\ref{Fig1}. Panel (a) considers different cavity damping rates
$\Lambda_{0}/(10^{-5}\omega_{0})=2.5$, $3.5$ and $5$ with red, blue and black
lines. Panel (b) shows results for average dot orbital energies $\omega
_{av}/\omega_{0}=-2.1$ (cyan lines) and $\omega_{av}/\omega_{0}=-1.2$ (green
lines). The circles indicate working points common to panels (a) and (b). For
reference, the second order variance $\Delta X_{2}=\sqrt{1+2n_{B}}/2$ for a
decoupled cavity ($g_{L}=0$) is also shown as a dashed yellow line. It is
independent of the value of $\Lambda_{0}$.}%
\label{Fig4}%
\end{figure}%
\begin{equation}
\Delta X_{2}=\frac{1}{2}\sqrt{\frac{\Lambda_{0}(1+2n_{B})-\operatorname{Im}%
[\lambda_{2}]}{\Lambda_{0}-2\operatorname{Im}[\chi_{2}]}}%
\end{equation}
of the cavity field to second order in $g_{L}$ is also shown as a yellow line.
One gets a squeezing effect ($\Delta X_{-}<\Delta X_{2}<\Delta X_{+}$) which
is maximal at $\Delta\omega_{RL}=R(2\omega_{0})$ (panel (a)). As visible in
panel (b), for $t_{LR}=0.3$ (cyan full line), squeezing decreases with
$\Gamma$. One could expect that higher values of $\Gamma$ are always
detrimental to squeezing. However, for a small value of $t_{LR}$ (magenta full
lines), the squeezing effect finds a local maximum for a value of $\Gamma$
which can be quite significant ($\Gamma\sim0.9\omega_{0}$ in panel (b)).

To determine the role of the parameter $U_{q}$ (or $\gamma_{p}$), we show with
dotted lines in Fig. \ref{Fig3}a, b and c, the cavity field quadratures given
by Eqs. (\ref{Afull}) and (\ref{Bfull}) with $\gamma_{p}$ omitted ($\gamma
_{p}=0$). For the moderate tunnel rate $\Gamma$ used in panel (a), the full
and dotted lines coincide around $\Delta\omega_{RL}=R(2\omega_{0})$ but not
near the single-photon resonance $\Delta\omega_{RL}=R(\omega_{0})$. For
$\Delta\omega_{RL}=R(\omega_{0})$, the dissipative\ term in $U_{q}$ is
responsible for an increase of the squeezing effect, in spite of its
dissipative nature\cite{Didier:2014,Lu:2015}. Such an effect is allowed by
Eq.(\ref{Bfull}). To see an effect of $U_{q}$ on the squeezing at the working
point $\Delta\omega_{RL}=R(2\omega_{0})$, it is necessary to increase the
value of $\Gamma$ (see panel (b)). In this case, $U_{q}$ causes a decrease of
the squeezing amplitude. To summarize, the dissipative term in $U_{q}$ can
either increase or decrease the squeezing effect, depending on the regime of
parameters. Nevertheless, to maximize the squeezing effect, it is advantageous
to use the regime $\Delta\omega_{RL}=R(2\omega_{0})$ and $\Gamma$ small, where
the effect of $U_{q}$ can be disregarded (empty red squares in Fig.\ref{Fig3}%
b). Therefore we will consider this regime in the rest of the present Appendix
and Fig.\ref{Fig4}.

The use of a double quantum dot circuit as a nonlinear element for circuit QED
can be interesting because it offers a strong tunability of the squeezing
effect, as already seen in Fig.\ref{Fig3}. Figure \ref{Fig4}a shows that the
amplitude of the squeezing effect is also strongly dependent on the average
level position $\omega_{av}=(\omega_{L}+\omega_{R})/2$. Besides, the squeezing
effect can be controlled by using a nonzero bias voltage $V_{b}$ (see
Fig.\ref{Fig4}b). This is consistent with the fact mentioned earlier that
using a nonzero $V_{b}$ modifies the orbital energy range where the drive
terms $U_{cl}$ shows strong resonances (Fig.\ref{Fig1}a and c). Note that, so
far, we have used a relatively high cavity damping rate $\Lambda_{0}$ which
limits the squeezing effect. Panels (a) and (b) of Fig. \ref{Fig4} show that
for a given set of double dot parameters, the squeezing effect increases when
$\Lambda_{0}$ decreases, as expected. Finally, Fig. \ref{Fig3}c shows an
example of cavity Wigner function corresponding to the red empty circles in
Fig.\ref{Fig3}c. Using the qutip package mesolve\cite{qutip:2012}, we have
checked that this Wigner function is in quantitative agreement with a direct
numerical treatment of Eq. (\ref{ME}). We have also checked that fourth order
corrections in $g_{L}$ are negligible for the parameters considered in the
present section. Therefore, a treatment of the master equation (\ref{ME}) to
third order in $g_{L}$ is fully justified.

Interestingly, it has also been suggested to obtain cavity squeezing by using
a single quantum dot with an ac excitation with amplitude $\varepsilon
_{p}^{\prime}$ applied directly to the dot gate\cite{Mendes:2015}. However, on
the experimental level, such a strategy is more costly since it requires to
fabricate a direct ac gate for the quantum dot. Note that
Ref.\cite{Mendes:2015} presents the cavity effective action to second order in
$g_{L}$ only. A coherent two-photon drive term in $\varepsilon_{p}^{\prime
}g_{L}^{2}$ is taken into account but the terms in $\chi_{2}$, $\lambda_{2}$
and the expected contribution in $\varepsilon_{p}^{\prime}g_{L}^{2}$ to
$U_{q}$ are disregarded. Alternatively, two-photon processes or photonic
squeezing have been found for dc voltage-biased Josephson junctions or tunnel
junctions, which have no internal degrees of
freedom\cite{Gasse:2013,Forgues:2015,Mendes:2016,Westig:2017,Grimsmo:2016}. In
our case, the dc voltage-bias is not necessary due to the presence of the dot
orbital degree of freedom.

\section*{Appendix G: Analytical expression of $\chi_{2}$ for a double quantum
dot in the sequential tunneling limit}

In the sequential tunneling limit $k_{B}T\ll\Gamma$, it is possible to obtain
a simple approximate expression of the charge susceptibility $\chi_{2}$ for
the double quantum dot of Section \ref{DOUBLEDOT}. One can use a semiclassical
framework, with a master equation description of transport through the double
quantum dot, and a resonant approximation between the double dot internal
transition and the cavity. Such an approach is described in section 4.2.1 of
Ref.\cite{Cottet:2017} and yields the expression%
\begin{equation}
\chi_{2}=\frac{2g_{t}^{2}(n_{-}-n_{+})}{\omega_{0}-\omega_{DQD}+i\Gamma
}\label{REsApprox}%
\end{equation}
Above, the factor 2 takes into account spin degeneracy. The transverse
coupling
\begin{equation}
g_{t}=(g_{R}-g_{L})\text{\textrm{sin}[}\theta\text{]}/2
\end{equation}
between the DQD internal degree of freedom and the cavity depends on the
mixing angle $\theta=\arctan[2t/(\varepsilon_{L}-\varepsilon_{R})]$ between
the left and right DQD\ orbitals. The average occupations $n_{-}$ and $n_{+}$
of the bounding and antibounding orbitals of the DQD can be expressed as%
\begin{align}
n_{+} &  =(2-f_{L}^{-}-f_{R}^{-}+(f_{L}^{-}-f_{R}^{-})\cos(\theta))\\
&  \times(f_{L}^{+}+f_{R}^{+}+(f_{L}^{+}-f_{R}^{+})\cos(\theta))/4\nonumber
\end{align}
and%
\begin{align}
n_{-} &  =(2-f_{L}^{+}-f_{R}^{+}+(f_{R}^{+}-f_{L}^{+})\cos(\theta))\\
&  \times(f_{L}^{-}+f_{R}^{-}+(f_{R}^{-}-f_{L}^{-})\cos(\theta))/4\nonumber
\end{align}
with $f_{L(R)}^{\pm}=(1+\exp[\pm(\omega_{L(R)}\mp(eV_{b}/2))/k_{B}T])^{-1}$.
\begin{figure}[ptb]
\includegraphics[width=1\linewidth]{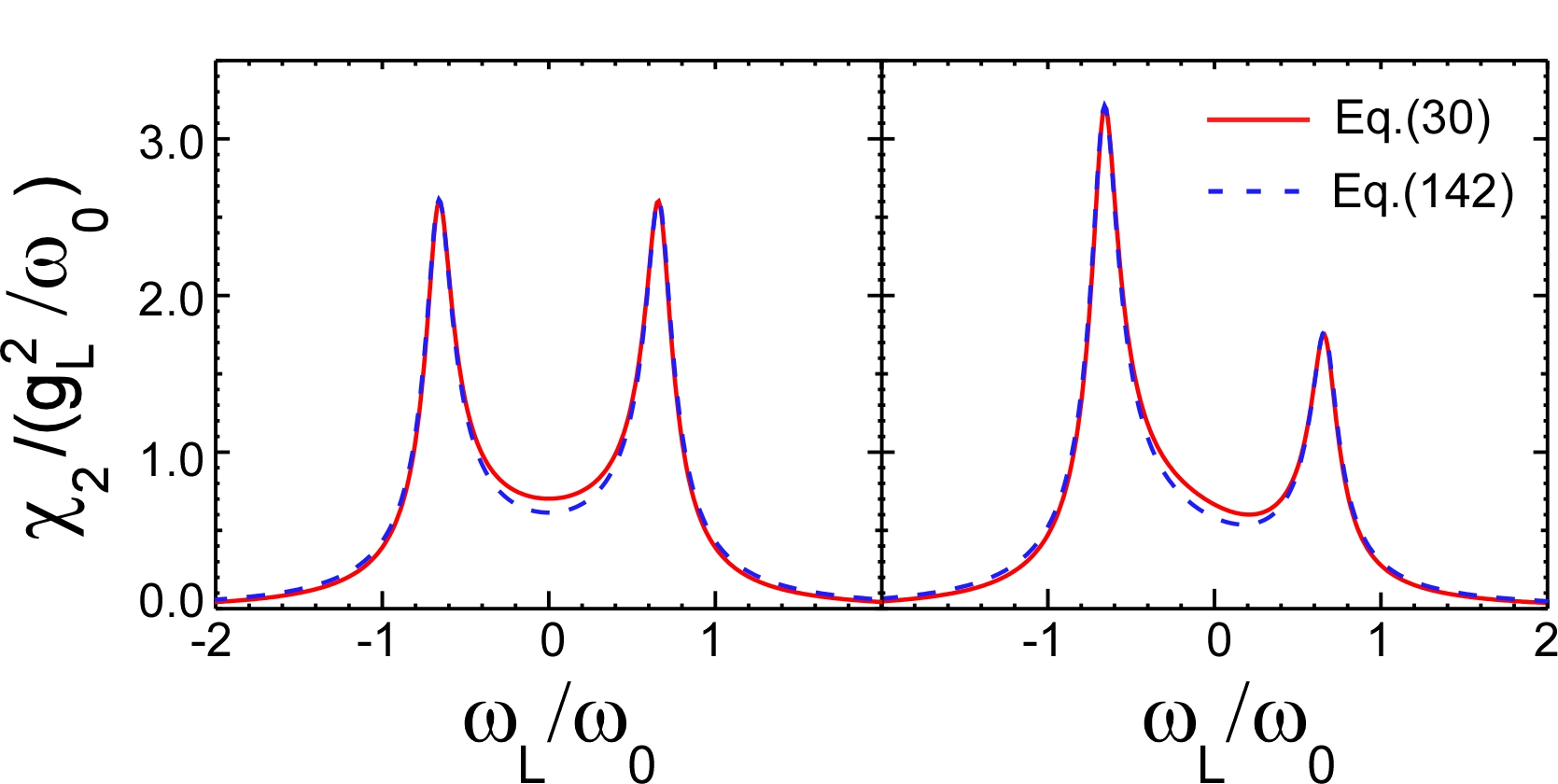}\newline%
\caption{Charge susceptibility $\chi_{2}$ of the DQD of section
\ref{DOUBLEDOT} versus $\omega_{L}$ for $V_{b}=0$ (panel (a)) and
$V_{b}=0.5\omega_{0}$(panel (b)). The full red lines and dashed blue lines
correspond to the result given by Eqs.(30) and (142), respectively. The other
parameters used are $\omega_{R}=0$, $\Gamma=0.05\omega_{0}$, $t_{LR}%
=0.375\omega_{0}$, $k_{B}T=0.5\omega_{0}$, $g_{R}=0$, and $V_{b}=0.5\omega
_{0}$.}%
\label{Chi2b}%
\end{figure}Figure \ref{Chi2b} shows a comparison between the values of
$\chi_{2}$ given by Eqs.(\ref{REsApprox}) and (\ref{Chi2}). The agreement is
very good near the resonances $\omega_{0}=\omega_{DQD}$, provided $t_{LR}%
\gg\Gamma$, because Eq.(\ref{REsApprox}) disregards photo-assisted tunneling
to the normal metal contacts, contrarily to Eq. (\ref{Chi2}). One interest of
Eq.(\ref{REsApprox}) is that it shows explicitly that divergences of $\chi
_{2}$, which should occur for $\omega_{0}=\omega_{DQD}$ in the absence of
dissipation, are regularized by the dissipative dot-lead tunneling in $\Gamma
$. More generally, in our model, dissipative tunneling to the reservoirs
prevents divergences of the cavity effective parameters, because it generates
imaginary self-energy terms in the mesoscopic Green's function of Eq.
(\ref{Gdef}). This is visible for instance in Figs.\ref{Fig1} and \ref{Fig5}
which present the numerical evaluation of these parameters versus
$\omega_{L(R)}$ or $\Delta\omega_{LR}=\omega_{L}-\omega_{R}$.

\section*{Appendix H: Validity of our perturbation scheme from the estimation
of higher order correlators in\textbf{ }$g$}

Since we develop the cavity action with respect to $\check{g}$ and $\beta_{p}%
$, the amplitude of these two parameters must not be too large. Besides,
having $\Gamma\neq0$ is crucial for ensuring the validity of our perturbation
scheme in the single or two-photon resonant regimes. Indeed, in the absence of
dissipation, the correlators $\chi_{2}$ and $\chi_{4}$ are expected to diverge
at $\omega_{DQD}=\omega_{0}$ and/or $\omega_{DQD}=2\omega_{0}$%
\cite{perturbation}.However, giving a simple analytic criterion for the regime
of validity of our description is very complex. Strictly speaking, the
development parameter in our approach is the functional matrix $\check{m}$ of
Eq.(\ref{mjaime}) which is used in the development of the cavity effective
action. It is difficult to express analytically a smallness criterion on this
quantity due to the many parameters involved through Eqs. (\ref{vchap}) and
(\ref{mjaime}) together with the light/matter coupling. This is why, out of
conciseness, we have referred to the expansion parameter as $g$ in the main
text. One can check the validity of our development a posteriori, by
estimating mesoscopic correlators which would occur in the cavity effective
action at higher orders in\textbf{ }$g$ and $\beta_{p}$ to check whether they
are negligible. Here, we present the evaluation of the generalized charge
susceptibilities $\chi_{6}$ and $\chi_{8}$ of the mesoscopic circuit at order
6 and 8 in\textbf{ }$g$ respectively. We expect the other coefficients with
the same order in\textbf{ }$g$ to have order of magnitudes similar to
$\chi_{6}$ and $\chi_{8}$ at best, similarly to what we observe at order $2$
and $4$ in\textbf{ }$g$. The parameters $\chi_{6}$ and $\chi_{8}$ can be
estimated from a semiclassical approach similar to that of Appendix B1. By
analogy with Eqs.(\ref{S2}), (\ref{S4}), (\ref{NT}), (\ref{EqSC}), one gets
\begin{equation}
S_{cav}^{eff}(\bar{\varphi},\varphi)=%
{\textstyle\sum\limits_{n\geqslant1}}
(\bar{\varphi}_{cl}\varphi_{cl})^{n}\bar{\varphi}_{q}\varphi_{cl}~\chi
_{2n}+...\label{Shigher}%
\end{equation}
Let us define%
\begin{align*}
\tilde{G}_{r,_{n}}^{j}(\omega) &  =%
{\textstyle\sum\limits_{(a_{1},a_{2},a_{n-1})\in\mathcal{S}_{n}}}
\left(  \mathcal{G}^{r}(\omega+j\omega_{0})\tilde{g}~\mathcal{G}^{r}%
(\omega+a_{1}\omega_{0})\right.  \\
&  \left.  \times\tilde{g}\mathcal{G}^{r}(\omega+a_{2}\omega_{0})...\tilde
{g}\mathcal{G}^{r}(\omega+a_{n-1}\omega_{0})\tilde{g}\mathcal{G}^{r}%
(\omega)\right)
\end{align*}
where $\mathcal{S}_{n}$ is the ensemble of number sequences $(a_{1}%
,a_{2},a_{n-1})$ such that $a_{1}=j\pm1$, $a_{k}-a_{k-1}=\pm1$ for
$k\in\lbrack2,n-1]$ and $a_{n-1}=\pm1$. One can check%
\begin{align*}
\chi_{n}(\omega_{0}) &  =-\frac{i}{2^{n-2}}Tr[\tilde{g}%
{\textstyle\sum\limits_{\substack{i\in\left[  0,n-1\right]  \\k_{i}%
+k_{n-i}=-1}}}
{\textstyle\int}
\frac{d\omega}{2\pi}\tilde{G}_{r,_{i}}^{k_{i}}(\omega)\\
&  \times\tilde{\Sigma}^{<}(\omega)\tilde{G}_{a,_{n-1-i}}^{k_{n-i}}(\omega)]
\end{align*}
with
\[
\tilde{G}_{a,_{n}}^{j}(\omega)=\left(  \tilde{G}_{r,_{n}}^{-j}(\omega)\right)
^{\dag}%
\]
\begin{figure}[ptb]
\includegraphics[width=1\linewidth]{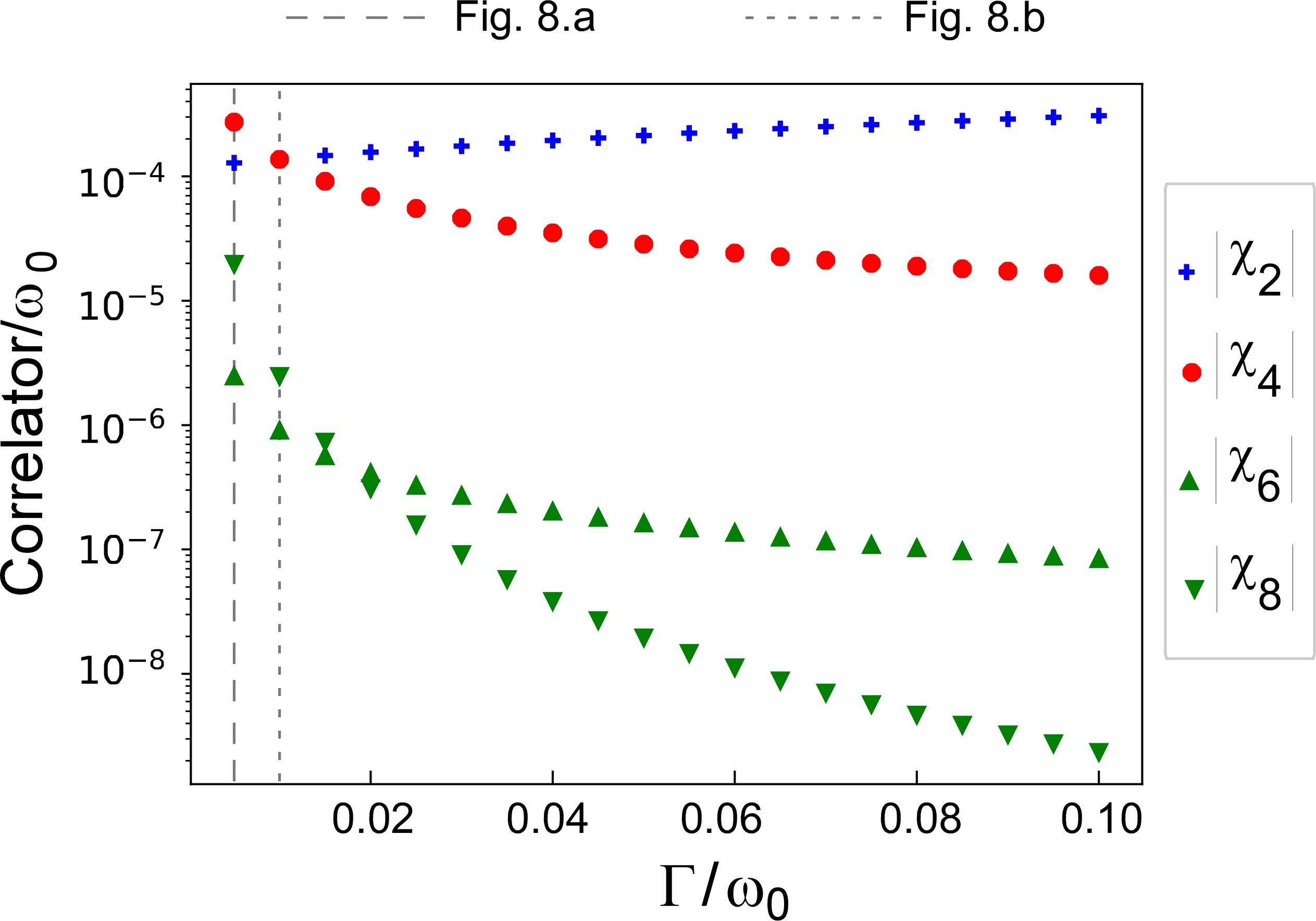}\newline%
\caption{Charge susceptibility coefficients $\chi_{2}$, $\chi_{4}$,
$\chi_{6}$ and $\chi_{8}$ of the DQD of section \ref{DOUBLEDOT} versus
$\Gamma$ for the parameters of Figs.\ref{Fig9} and \ref{Fig8}. The vertical
dashed lines indicate the values of $\Gamma$ used in Fig.\ref{Fig8}a and
Fig.\ref{Fig8}b. }%
\label{HigherCorrelators}%
\end{figure}Figure \ref{HigherCorrelators} shows the absolute values of
$\chi_{2}$, $\chi_{4}$, $\chi_{6}$ and $\chi_{8}$ versus $\Gamma$ for the
parameters of Figs.\ref{Fig9} and \ref{Fig8}. For $\Gamma=0.0025\omega_{0}$,
$\left\vert \chi_{8}\right\vert $ has the same order of magnitude as
$\left\vert \chi_{2}\right\vert $, and $\left\vert \chi_{4}\right\vert $. This
illustrates the fact that our development in $g^{4}$ is not valid for too
small values of $\ \Gamma$. However, for the values $\Gamma=0.005\omega_{0}$
and $\Gamma=0.01\omega_{0}$ corresponding to Figs.\ref{Fig8}a and \ref{Fig8}b
(indicated by vertical dashed lines), one starts to have $\chi_{2}$,$\chi
_{4}\gg\chi_{6}$, $\chi_{8}$ so that our development at fourth order in $g$
seems reasonable. It turns out that we have worked at the limit of the allowed
range of $\Gamma$ in order to maximize the two-photons effects in $K_{loss}$
which decrease for higher values of $\Gamma$ (one has $K_{loss}%
=\operatorname{Im}[\chi_{4}]\simeq-i\chi_{4}$ for the parameters of Fig.
\ref{Fig8}). We have checked that $\chi_{2}$,$\chi_{4}\gg\chi_{6}$, $\chi_{8}$
is also satisfied for the parameters used in section \ref{GENERAL} and
Appendix F. In principle, terms at higher orders in $\beta_{p}$ should also
contribute to Eq.(\ref{Shigher}). However, the next order contribution after
the term in $\beta_{p}g^{3}$ of the main text should be in $\beta_{p}^{2}%
g^{6}$ and since it is also regularized by $\Gamma$, we expect this term to be negligible.


\begin{thebibliography}{99}
\bibitem {Coh}More mathematically speaking, $\varphi_{\pm}(t)$ and $\psi
_{\pm,d}(t)$ represent eigenvalues of the operators $\hat{a}$ and $\hat{c}%
_{d}$ in a coherent eigenstates picture where one can use $\hat{a}\left\vert
\varphi_{\pm}(t)\right\rangle =\varphi_{\pm}(t)\left\vert \varphi_{\pm
}(t)\right\rangle $ and $\hat{c}_{d}\left\vert \psi_{\pm,d}(t)\right\rangle
=\psi_{\pm,d}(t)\left\vert \psi_{\pm,d}(t)\right\rangle $. Details on the
contruction of this representation are given in Refs. \cite{Kamenev,ZJ}.
\bibitem {NoteBias}This is possible provided the capacitances to the source
and drain contacts are equal and the capacitance between the two dots is very
large. In the general case, our results will remain valid provided
$\omega_{L(R)}$ are understood as $V$-shifted orbital energies.
\end{thebibliography}
\end{document}